\documentstyle[epsf,fleqn]{ar209}

\def\LQCD{\Lambda_{\rm QCD}}
\def\LQCDfo{\Lambda_{\rm QCD}^{(4)}}
\def\LQCDfi{\Lambda_{\rm QCD}^{(5)}}
\def\mQ{m_Q}
\def\fpi{f_\pi}
\def\fb{f_B}
\def\fbd{f_{B_d}}
\def\fbs{f_{B_s}}

\def\fds{f_{D_s}}
\def\bbd{B_{B_d}}
\def\bb{B_B}
\def\bnlo{\hat{B}_B^{\rm NLO}}
\def\bnlod{\hat{B}_{B_d}^{\rm NLO}}
\def\bbs{B_{B_s}}

\def\Lb{\Lambda_b}
\def\delm{\Delta m}
\def\delmd{\Delta m_d}
\def\delms{\Delta m_s}
\def\delg{\Delta \Gamma}
\def\Am{{\cal A}_m}
\def\Lxy{L_{xy}}
\def\ptr{p_t^{{\rm rel}}}
\def\Qt{Q_{\rm tag}}
\def\Ppol{P_{\rm pol}}

\def\order{{\cal O}}

\def\bbar{\overline{b}}
\def\dbar{\overline{d}}

\def\ubar{\overline{u}}
\def\cbar{\overline{c}}

\def\qbar{\overline{q}}
\def\pbar{\overline{p}}

\def\Bzbar{\overline{B^0}}

\def\Dsp{D^{*+}}
\def\Dzbar{\overline{D^0}}

\def\Bs{B_s^0}
\def\Bsbar{\overline{\Bs}}

\def\Bbar{\overline{B}}

\def\Bh{|B_H\!>}
\def\Bl{|B_L\!>}
\def\Bht{|B_H(t)\!>}
\def\Blt{|B_L(t)\!>}

\def\mh{m_H}
\def\ml{m_L}

\def\Vqd{V_{qd}}

\def\Vqb{V_{qb}}

\def\Vud{V_{ud}}
\def\Vcd{V_{cd}}
\def\Vtd{V_{td}}

\def\Vcs{V_{cs}}
\def\Vts{V_{ts}}
\def\Vub{V_{ub}}
\def\Vcb{V_{cb}}
\def\Vtb{V_{tb}}
\def\Vqb{V_{qb}}
\def\VQb{V_{Qb}}
\def\Vqd{V_{qd}}
\def\VQd{V_{Qd}}

\def\micron{\mu\mbox{m}}
\def\mQa{m_Q a}

\def\Rxi{R_\xi}

\begin{document}
\input epsf

\title{$B$ Mixing\footnote{This paper is dedicated to the memory of
Elizabeth Bishop Martin, who was originally intended to co-author this paper, but
sadly passed away on March 16, 1999.}}
\author{Colin Gay\affiliation{Department of Physics, Yale University,
New Haven, Connecticut 06511; \\ email: colin.gay@yale.edu}}
\markboth{C. GAY}{$B$ MIXING}
\begin{abstract}
The neutral $B$ mesons, $B^0$ and $\Bs$, can oscillate between their
particle and antiparticle states owing to flavor-changing weak interactions.
In recent years, techniques to detect these
oscillations as a function of the meson's decay time have been developed.
In this article the physics of flavor oscillations is reviewed and
theoretical predictions are summarized.
The many observations that demonstrate
the time-dependence of $B^0-\Bzbar$ oscillations are presented
along with a combined measurement of its frequency,
$\delmd = 0.484\pm0.015$ ps$^{-1}$.
The attempts to measure the $\Bs$ oscillation frequency, both
directly and indirectly, are then summarized, currently resulting in a limit of
$\delms > 14.6$ ps$^{-1}$ (95\% CL).
Finally, values for the CKM elements
$|\Vtd| = (3.6\pm0.4)\times 10^{-3}$ and 
$|\Vts/\Vtd| > 4.7$ (95\% CL) are extracted.
\end{abstract}
\maketitle

\section{INTRODUCTION}
\label{intro}

The ability of a very few neutral mesons to change from their
particle to their antiparticle state is a remarkable consequence of basic
quantum mechanics and the structure of the weak interaction.  This
oscillation from matter to antimatter can be used to measure fundamental
parameters of the standard model; in addition, it might have far reaching
effects, such as breaking the matter/antimatter symmetry of the universe.

In 1955, Gell-Mann \& Pais \cite{GM55} showed that if a
 $\overline{K}^0$ meson with strangeness
$S$=$-1$, as well as the $K^0$ meson with strangeness $S$=$1$ exists,
then a quantum-mechanical mixing due to $K^0-\overline{K}^0$ interactions
takes place. It produces two physical
particles, $K^0_1$ and $K^0_2$,
that are a mixture of these states of well-defined
strangeness, or flavor.
Gell-Mann \& Pais predicted the existence of a long-lived neutral kaon,
$K^0_2$, which decays to three pions, as the companion to the
shorter-lived $K^0_1$
that had already been observed.
Lande \cite{Lande56} at Brookhaven confirmed the existence of the
longer lived state, $K^0_L$, in 1956.

If the physical particles are a mixture of states of well-defined flavor, then
these flavor eigenstates can be considered mixtures of the physical particles.
These physical states must have slightly different masses, as discussed 
in the following section, and so they develop a phase
difference as they evolve in time.  Therefore the physical particle content
of a flavor eigenstate evolves with time -- an initially pure flavor
eigenstate develops a component of the opposite flavor.
The mixing of flavor eigenstates to
form the physical particles, then, is equivalent to the oscillations
of flavor eigenstates into one another.

The only hadrons that can undergo these oscillations are the following
mesons: $K^0, D^0,
B^0$ and $\Bs$.  The $\pi^0$ is its own antiparticle, the top quark is
so heavy that it decays before forming stable hadrons, and excited meson
states decay strongly or electromagnetically before any mixing can occur.

Such particle-antiparticle mixing has since been
seen for $B$ mesons, first in an admixture of $B^0$ and $\Bs$
by UA1 \cite{UA187} and then in $B^0$ mesons by ARGUS \cite{Argus87} and later
CLEO \cite{CLEO89}.
$\Bs$ mixing was established by 
comparing the time-integrated oscillation probability for $B^0$ mesons,
measured by ARGUS and CLEO, to that measured
at LEP, which contains both $B^0$ and $\Bs$ contributions \cite{PDG98}.  
Mixing is
expected to be a very small effect in $D^0$ mesons
 and has not been observed.  The experimental part of this
review focuses specifically on recent analyses that attempt to measure the
time-structure of $B^0$ and $\Bs$ oscillations directly, rather than
those that studied the oscillations in a time-integrated manner. The latter
include
those from ARGUS and CLEO \cite{Chid94} and earlier analyses from the
experiments presented in this paper.  See \cite{PDG98} for a list of these
measurements.

The outline of this paper is as follows.  Section~\ref{formal} collects
the various elements required to 
derive the mixing formalism; Section~\ref{fbSection} presents results on
quantities needed to extract Cabbibo-Kobayashi-Maskawa (CKM) matrix elements
from mixing measurements;
Section~\ref{AnatomySection} presents an overview of experimental techniques
used to measure the time dependence of the flavor oscillations;
Section~\ref{BdSection} presents measurements of the $B^0$ oscillation
frequency;
Section~\ref{BsSection}
presents limits on direct searches for $\Bs$ flavor oscillations; and
Section~\ref{DelGammaSection}
deals with measuring the $\Bs$ oscillation frequency through lifetime
differences in the $\Bs$ system.

Since this paper's focus is $B$-meson mixing, the formalism is presented
in terms
of $b$ quarks coupled with $d,s$ quarks, though it was first derived for the
kaon system.  $B$ is used generically to mean either $B^0$ or $\Bs$ in cases
where the result applies to both; however for clarity and brevity
only the $B^0$ Feynman diagrams are shown.  Also, the charge conjugate of
listed decay modes is implied unless explicitly stated otherwise.  For
convenience we
have set $\hbar=c=1$ and supress the $c$ in the units of momentum and mass.

\section{MIXING FORMALISM}
\label{formal}
Particle-antiparticle oscillations are possible because of the flavor-changing
term of the standard model Lagrangian,
\begin{eqnarray*}
{\cal L} & = & 
\frac{g}{\sqrt{2}} (\overline{u},\overline{c},\overline{t})_L 
V_{\rm CKM}
\gamma_\mu \left( \begin{array}{c} d \\ s \\ b \end{array} \right)_L W^\mu 
+ h.c.,
\end{eqnarray*}
\noindent where $V_{\rm CKM}$ is the CKM matrix \cite{CKM}.
A popular parameterization of this matrix is that of
Wolfenstein \cite{Wolf84}, which
expands each term in powers of the Cabibbo angle $\lambda\approx 0.22$,
shown below to $\order(\lambda^4)$:
\begin{eqnarray*}
\left( \begin{array}{ccc}
V_{ud} & V_{us} & V_{ub} \\
V_{cd} & V_{cs} & V_{cb} \\ 
V_{td} & V_{ts} & V_{tb} \end{array} \right)
  & \approx &  \left( \begin{array}{ccc}
1-\frac{1}{2}\lambda^2 & \lambda & A\lambda^3(\rho-i\eta) \\
-\lambda(1+iA^2\lambda^4\eta) & 1-\frac{1}{2}\lambda^2 & A\lambda^2 \\ 
A\lambda^3(1-\rho-i\eta) & -A\lambda^2 & 1 \end{array} \right).
\end{eqnarray*}
This term in the Lagrangian engenders box diagrams involving an internal
loop with two $W$ bosons,
of the sort shown in Figure~\ref{box}, which result
in nonzero transition matrix elements between $B^0-\Bbar^0$ and $\Bs-\Bsbar$.

Before calculating these transition amplitudes within the standard model,
the relevant general formalism is developed, i.e. the quantum mechanics
of a two-state, particle-antiparticle system weakly coupled to a continuum.

\subsection{Introduction to Mixing Formalism}
The oscillation effect follows from a simple perturbative solution to
Schr\"{o}dinger's equation.
Let ${\cal H}_0$ be the Hamiltonian of the strong interaction.
If this were the only
force, there would be stable states, $|B> = |\bbar q>$ and
$|\Bbar> = |b\qbar>$ ($q=d,s$), that are eigenvectors of ${\cal H}_0$.
By the CPT theorem, the masses are equal,
$m_{B} = m_{\Bbar} = m_0$, and the Hamiltonian is
\begin{eqnarray*}
{\cal H}_0 & = & \left( \begin{array}{cc}
m_0 & 0 \\
0 & m_0 \end{array} \right).
\end{eqnarray*}

\begin{figure}[htb]
\centerline{\epsfxsize=3.5in \epsfbox{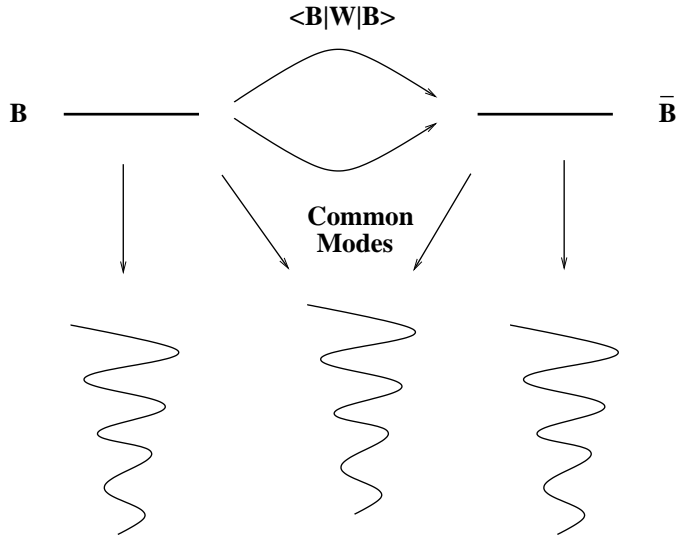}}
\caption{Schematic of a two-state system coupled weakly to a continuum
(indicated by the wiggles)}
\label{twoState}
\end{figure}

When the weak interaction
${\cal H}_W$ is added, the simple two-state system
 becomes much more
complicated, as shown in Figure~\ref{twoState}.  The weak force is responsible
for nonzero matrix elements between the two states
to a continuum of states (i.e. possible decay modes), 
which can be split into three groups:
 ({\em a}) states accessible
only to $|B\!>$, i.e. states $|\alpha>$ with
$<\!\alpha|{\cal H}_W|\Bbar\!>=0$;
 ({\em b}) states accessible only to $|\Bbar\!>$;
and ({\em c}) states coupled to both
 $|B\!>$ and $|\Bbar\!>$.  In addition, the two discrete states
are connected by a direct matrix element $W_{12}=<\Bbar|{\cal H}_W|B>$ or
via off-shell (i.e. with $E\ne m_0$) continuum states accessible to both.
The Hamiltonian
${\cal H} = {\cal H}_0 + {\cal H}_W$ has infinite dimensions.

The time evolution of a wavefunction that is a pure $|B\!>$ at
$t=0$ is easily calculated using standard first-order perturbation theory,
and in general the wavefunction develops both
 $|\Bbar\!>$ and continuum components.
Typically, one limits the state space
to that spanned by the $|B\!>$ and $|\Bbar\!>$, however, 
and derives a matrix ${\cal H}$
(still referred to as a Hamiltonian even though it is
clearly not hermitian) which allows one to use the Schr\"{o}dinger equation in
this subspace.
It is easy to show it has the form:
\begin{eqnarray}
{\cal H} & = & \left( \begin{array}{cc}
m_0 + {\delta E} & W_{12} + {\delta E_{12}} \\
W_{12}^* + {\delta E_{12}^*} & m_0 + {\delta E} \end{array} \right)
-\frac{i}{2} \left( \begin{array}{cc}
{\Gamma} & {\Gamma_{12}} \\
{\Gamma_{12}^*} & {\Gamma} \end{array} \right).
\end{eqnarray}
The introduction of the continuum states has several effects:
\begin{enumerate}
\item It introduces a slight mass shift $\delta E$ in both $|B>$ and $|\Bbar>$,
which can be absorbed into the mass $M=m_0+\delta E$.
\item It introduces off-diagonal elements $M_{12}=W_{12}+\delta E_{12}$
in the real part of ${\cal H}$ due
to the coupling of the particle and antiparticle via ${\cal H}_W$ and
via {\em off-shell} continuum states accessible to both
$|B\!>$ and $|\Bbar\!>$ ($\delta E_{12}$).
\item It introduces an imaginary part to ${\cal H}$, with diagonal elements
given by
the matrix elements of $|B>$ and $|\Bbar>$ to the on-shell
continuum states unique to each.
\item It introduces off-diagonal elements $\Gamma_{12}$ into the
imaginary part of ${\cal H}$  given by
the matrix elements of $|B>$ and $|\Bbar>$ to the on-shell
continuum states common to both.
\end{enumerate}

Diagonalizing this ${\cal H}$ gives us the eigenvectors and eigenvalues.
The eigenstates are conventionally denoted $H,L$,
for heavy and light, rather than short and long as customary for the kaon
system.  If
\begin{eqnarray}
\label{HLeigen}
\Bl & = & p |B\!> + q |\Bbar\!> \hspace{1cm}\mbox{and} \nonumber \\ 
\Bh & = & p |B\!> - q |\Bbar\!>,
\end{eqnarray}
\noindent then
\begin{eqnarray}
\frac{q}{p} & = & \sqrt{\frac{M_{12}^* -\frac{i}{2} \Gamma_{12}^*}
{M_{12} -\frac{i}{2} \Gamma_{12}}},
\end{eqnarray}
and the eigenvalues are
\begin{eqnarray}
\lambda_{H,L} & = & m_{H,L} -\frac{i}{2} \Gamma_{H,L},
\end{eqnarray}
\noindent where the masses and widths of these states are
\begin{eqnarray}
\mh,\ml & = & M \pm Re\sqrt{|M_{12}|^2 - \frac{|\Gamma_{12}|^2}{4} -
i Re(M_{12}\Gamma_{12}^*)} \equiv M \pm \delm/2, \nonumber \\ 
\Gamma_H,\Gamma_L & = & \Gamma \pm 
2 Im\sqrt{|M_{12}|^2 - \frac{|\Gamma_{12}|^2}{4} -
i Re(M_{12}\Gamma_{12}^*)} \equiv \Gamma \pm \delg/2,
\label{eigvals}
\end{eqnarray}
which satisfy
\begin{eqnarray}
\delm^2 - \frac{\delg^2}{4} & = & 4(|M_{12}|^2 - \frac{|\Gamma_{12}|^2}{4})
\hspace{1cm}\mbox{and} \nonumber \\
\delm \delg & = & 4Re(M_{12}\Gamma_{12}^*).
\label{eigvalRelation}
\end{eqnarray}

Equation~\ref{HLeigen} can be
rearranged to give the flavor eigenstate $|B\!>$ in terms
of $\Bh,\Bl$. The standard derivation uses their simple
time dependence, 
[$\Bht = e^{-i\lambda_H t}\Bh$], to obtain the time
evolution of a state which was a pure $|B\!>$ at $t=0$,
expressed in terms of
the flavor eigenstates.  (This procedure implies
the existence of a reference frame with both the $\Bl$ and $\Bh$
components of the $|B\!>$ at rest, and such a frame does not, strictly
speaking, exist.  However, setting the relative velocity of the
$\Bl$ and $\Bh$ states to zero is an excellent approximation.
A comment on its validity and the appropriateness of describing the
system's evolution as a
function of time can be found below.)
\begin{eqnarray}
|B(t)\!> & = & \frac{1}{2p}(\Blt + \Bht) \nonumber \\ 
& = & \frac{1}{2}e^{-iMt}e^{-\frac{\Gamma}{2}t} \left(
(e^{\frac{\delg}{4}t} e^{i\frac{\delm}{2}t} +
e^{-\frac{\delg}{4}t} e^{-i\frac{\delm}{2}t})|B> \right.  \nonumber \\
& &\left.\hskip 1.6cm + \
\frac{q}{p}(e^{\frac{\delg}{4}t} e^{i\frac{\delm}{2}t} -
e^{-\frac{\delg}{4}t} e^{-i\frac{\delm}{2}t})|\Bbar> \right).
\label{Bevolution}
\end{eqnarray}
Similarly, the time evolution of a state which is pure $|\Bbar>$ as $t=0$ is
\begin{eqnarray}
|\Bbar(t)\!> & = & \frac{1}{2q}\left[\Blt - \Bht\right] \nonumber \\ 
& = & \frac{1}{2}e^{-iMt}e^{-\frac{\Gamma}{2}t} \left[
\frac{p}{q}(e^{\frac{\delg}{4}t} e^{i\frac{\delm}{2}t} -
e^{-\frac{\delg}{4}t} e^{-i\frac{\delm}{2}t})|B> \right. \nonumber \\ 
& &\left.\hskip 2.2cm + \
(e^{\frac{\delg}{4}t} e^{i\frac{\delm}{2}t} +
e^{-\frac{\delg}{4}t} e^{-i\frac{\delm}{2}t})|\Bbar> \right].
\label{Bbarevolution}
\end{eqnarray}
The normalizations of these states are given by
\begin{eqnarray}
\eta^2 & = & \int_0^\infty <B(t)|B(t)> dt = \frac{\Gamma}{2}\left[
\frac{1+|q/p|^2}{\Gamma^2-\delg^2/4} +
\frac{1-|q/p|^2}{\Gamma^2+\delm^2}\right] \nonumber \\
\overline{\eta}^2 & = & \int_0^\infty <\Bbar(t)|\Bbar(t)> dt =
\frac{\Gamma}{2}|\frac{p}{q}|^2 \left[
\frac{1+|q/p|^2}{\Gamma^2-\delg^2/4} -
\frac{1-|q/p|^2}{\Gamma^2+\delm^2}\right].
\label{Bnorm}
\end{eqnarray}

Let $P^B_m(t)$ denote the probability that a particle produced as a $B$
oscillated (mixed) and decayed as a $\Bbar$. [i.e.
$P^B_m(t) = \frac{1}{\eta^2}|<\!\Bbar|B(t)\!>|^2$.]  Let $P^B_u(t)$
denote the conjugate probability that this particle did not oscillate, that is,
it remained unmixed (with similar definitions for initial $\Bbar$ states).
Then Equations~\ref{Bevolution}, \ref{Bbarevolution} and \ref{Bnorm} give
the following:
\begin{eqnarray}
P^B_u(t) 
& = & \frac{e^{-\Gamma t}}{\Gamma\left( \frac{1+|q/p|^2}{\Gamma^2-\delg^2/4} +
\frac{1-|q/p|^2}{\Gamma^2+\delm^2} \right)}(\cosh\frac{\delg}{2}t + \cos\delm t), \nonumber \\
P^B_m(t) 
& = & \frac{|q/p|^2e^{-\Gamma t}}{\Gamma\left(
\frac{1+|q/p|^2}{\Gamma^2-\delg^2/4} +
\frac{1-|q/p|^2}{\Gamma^2+\delm^2} \right)}(\cosh\frac{\delg}{2}t - \cos\delm t),
\end{eqnarray}

\begin{eqnarray}
P^{\Bbar}_u(t) & = & 
\frac{|q/p|^2e^{-\Gamma t}}{\Gamma\left( \frac{1+|q/p|^2}{\Gamma^2-\delg^2/4} -
\frac{1-|q/p|^2}{\Gamma^2+\delm^2} \right)}(\cosh\frac{\delg}{2}t + \cos\delm t), \nonumber \\
P^{\Bbar}_m(t) & = &
\frac{e^{-\Gamma t}}{\Gamma\left( \frac{1+|q/p|^2}{\Gamma^2-\delg^2/4} -
\frac{1-|q/p|^2}{\Gamma^2+\delm^2} \right)}(\cosh\frac{\delg}{2}t - \cos\delm t).
\end{eqnarray}
Note that these expressions are not symmetric between $B$ and $\Bbar$
states.

These formulae have two limiting cases: neglecting
CP violation in the mixing, and neglecting the lifetime difference $\delg$
(which also in general implies there is no CP violation in the mixing).

Equation~\ref{HLeigen} can be rewritten as (similarly for the $\Bh$)
\begin{eqnarray*}
\Bl & = & \frac{p+q}{2}\left[(|B>+|\Bbar>)
+ \frac{1-q/p}{1+q/p}(|B>-|\Bbar>)\right],
\end{eqnarray*}
\noindent and thus $(1-q/p)/(1+q/p) \equiv \epsilon_B$ is a measure of
the amount
by which $\Bh$ and $\Bl$ differ from CP eigenstates.
$\epsilon_B$ is expected to be very small
in the standard model, $\order(10^{-3})$. The current world average for
the $B^0$ system is Re($\epsilon_B$)=$0.002\pm 0.007$ \cite{PDG98}.
No measurement or
limit exists for the $\Bs$ system.
The limit of no CP violation in mixing is thus $q/p = 1$.  In this limit
the $B$ and $\Bbar$ symmetry is regained, and we obtain unmixed
and mixed decay probabilities for both $B$ and $\Bbar$ of:
\begin{eqnarray}
P_{u,m}(t) & = & 
\frac{1}{2} \Gamma e^{-\Gamma t} \left(1-\frac{\delg^2}{4\Gamma^2}\right)
(\cosh\frac{\delg}{2}t \pm \cos\delm t),
\label{PmixDelg}
\end{eqnarray}
\noindent where the + sign corresponds to $P_u$.  This form is appropriate for $\Bs$ mesons, which are not
expected to be subject to large CP-violating effects.

On the other hand, even in the presence of CP violation, a simple form
can be obtained.  The lifetime difference between the heavy and light states
 is expected to be
small, $\delg/\Gamma \leq 1\%$ for the $B^0$ and perhaps as large as
25\% for the $\Bs$ \cite{Beneke96}.  From Equation~\ref{eigvalRelation},
$\delg = 0$ in general only if
$\Gamma_{12} = 0$.
In this case,
\begin{eqnarray*}
\frac{q}{p} & = & \sqrt{\frac{M_{12}^* -\frac{i}{2} \Gamma_{12}^*}
{M_{12} -\frac{i}{2} \Gamma_{12}}} =
\sqrt{\frac{M_{12}^*}
{M_{12}}} = e^{-i\phi},
\end{eqnarray*}
thus $|q/p|=1$.
In this $\delg=0$ limit, the time evolutions from Equations~\ref{Bevolution}
and \ref{Bbarevolution} become
\begin{eqnarray}
|B(t)> & = & 
e^{-iMt}e^{-\frac{\Gamma}{2}t} \left( \cos\frac{\delm}{2}t|B> +
ie^{-i\phi}\sin\frac{\delm}{2}t|\Bbar> \right) \nonumber \\
|\Bbar(t)> & = &
e^{-iMt}e^{-\frac{\Gamma}{2}t} \left(
\cos\frac{\delm}{2}t|\Bbar> + ie^{+i\phi}\sin\frac{\delm}{2}t|B> \right).
\end{eqnarray}
The mixed and unmixed decay probabilities again become equal for the $B$ and
$\Bbar$ mesons:
\begin{eqnarray}
P_{u,m}(t) & = &
\frac{1}{2} \Gamma e^{-\Gamma t} (1 \pm \cos\delm t).
\label{Pmix}
\end{eqnarray}
This form is expected to be appropriate for $B^0$ mesons, for which a large
phase $\phi$ (the source of mixing-induced CP violation) is possible.

Equations~\ref{PmixDelg} and \ref{Pmix} will be the expressions used
throughout this
paper.  Their time-integrated versions, expressing the probability that a
$B$ decays as a $\Bbar$, are (with $x\equiv \delm/\Gamma$)
\begin{eqnarray}
\chi & = & \int_0^\infty P_m(t) = \frac{1}{2}
\frac{x^2+\frac{1}{4}\frac{\delg^2}{\Gamma^2}}{1+x^2}
\end{eqnarray}
and in the $\delg=0$ limit,
\begin{eqnarray}
\chi & = & \frac{1}{2}\frac{x^2}{1+x^2}.
\label{chiDefn}
\end{eqnarray}

As has been pointed out in the context of neutrino oscillation experiments
\cite{Lipkin97}, the oscillations observed by any
experiment are oscillations in space not in time. That is, one
has a source creating a pure $B$ or $\Bbar$ meson, which may have oscillated
by the time it reaches a distant detector.  In this spatial picture, we have
a source, very small compared to the oscillation wavelength, which emits
a pure $B$ meson.  The boundary condition that must be imposed, then, is
that the probability of finding a $\Bbar$ meson at the source must vanish
for all time, otherwise a pure $B$ would not be emanating.
The $\Bl$ and $\Bh$ components propagate 
with phase $e^{i(E_{L,H}t-p_{L,H}x)}$, where $x$ denotes the direction of
motion.  At the origin, the only way to
ensure the wavefuntion does not change the relative $\Bl-\Bh$ phase and
develop
a $\Bbar$ component is the condition $E_L=E_H$.  That is, the $B$ meson has a definite
energy.  The components $\Bl$ and $\Bh$ will have the same energy, but
different momenta $p_{L,H}=\sqrt{E^2-m_{L,H}^2}$ respectively.  This induces
spatial oscillations that go as $e^{i(p_H-p_L)x}$.  Thus, if the
light and heavy components have the same energy but different momenta,
they must be moving with different velocities, and there is no frame in which
both the $\Bl$ and $\Bh$ are at rest.  In the derivation above, however, we
put $E_{L,H}=m_{L,H}$ for both states.

At the time $t$, the centers of the $\Bl$ and $\Bh$ wave packets have
separated by
a distance $d=(v_L-v_H)t = (p_L-p_H)t/E$.  As $\delm$ is very
small compared to the mean mass $M$ ($3\times 10^{-13}$ GeV vs. 5.3 GeV), this
distance is $d \approx \frac{2M\delm}{EP} ct$,
where $P^2=E^2-M^2$.  Setting $t$ as large as 10 ps (the $B$
lifetime is approximately 1.5 ps), and noting that the typical $B$-meson energy
in the current experiments is $10-30$ GeV, we see that the
wavepacket separation is only $\order(10^{-10}\micron)$. This justifies
treating the system as moving with a single group velocity given by the
mean velocity $v=P/E$.  Thus the spatial dependence
of the oscillations, $(p_L-p_H)x$, can be converted to a time dependence using
$x=vt=P/E\approx(p_L+p_H)/2E$ to obtain
$(p_L^2-p_H^2)t/2E = (m_L^2-m_H^2)t/2E$.
If we substitute $E=\gamma M$, and the
proper time $\tau=t/\gamma$,
the time dependence goes as $(m_L^2-m_H^2)\tau/2M = \delm\tau$,
recovering
the form used in the standard treatment above.


\subsection{Mixing in the Standard Model}
The time-evolution and decay probabilities presented in the previous
section are valid for any perturbation to the strong Hamiltonian.  This
section
evaluates the matrix elements $M_{12}$ and $\Gamma_{12}$, in the context
of the standard model.

\begin{figure}[htb]
\centerline{\epsfxsize=\hsize \epsfbox[80 600 550 710]{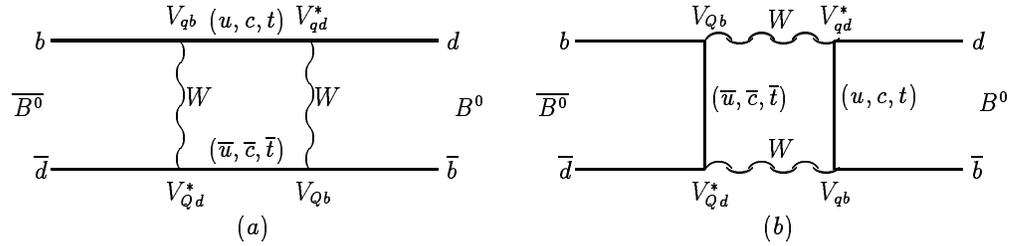}}
\caption{Lowest order box diagrams responsible for $B^0-\Bzbar$ oscillations.}
\label{box}
\end{figure}

To lowest order, the matrix element coupling a $B^0$ and $\Bzbar$
is given by
the Feynman diagrams shown in Figure~\ref{box}{\em a,b}, with similar
diagrams for the $B^0_s$.  With three generations,
the internal quark lines $Q,q$ can be $u, c$, or $t$.  Of course we must
take into account that the quark lines coming in and going
out are not free -- the $b$ and light quark $d$ or $s$ are bound into a hadron.
These diagrams are evaluated using a method first employed by Gaillard \&
 Lee \cite{GaLee74} in 1974 for the $K^0$ system.  The 
Hamiltonian corresponding to the box diagrams is sandwiched between the
bound-state hadrons, ${\cal M} = <B|H_W|\Bbar>$, (so that
$M_{12}={\cal M}/(2m_B)$, including the normalization factor),
the momenta and masses of
the incoming and outgoing quarks are set to zero (since
they are small compared with
$M_W$), and the internal loop momentum integral is performed.

Some care must be taken in evaluating these diagrams.  A good choice of gauge
to perform the calculation is the generalized renormalizable gauge,
$\Rxi$, with $\xi$ finite, introduced by Fujikawa et al
\cite{Fujik72}.  In the standard model, there is an unphysical 
charged scalar $\phi^\pm$,
which is a remnant of the charged Higgs that is absorbed to give the $W$ boson
its mass.  In fact, such particles will generally exist in any model which
generates masses for the vector bosons through spontaneous symmetry breaking.
\begin{figure}[htb]
\centerline{\epsfxsize=\hsize \epsfbox[70 480 550 710]{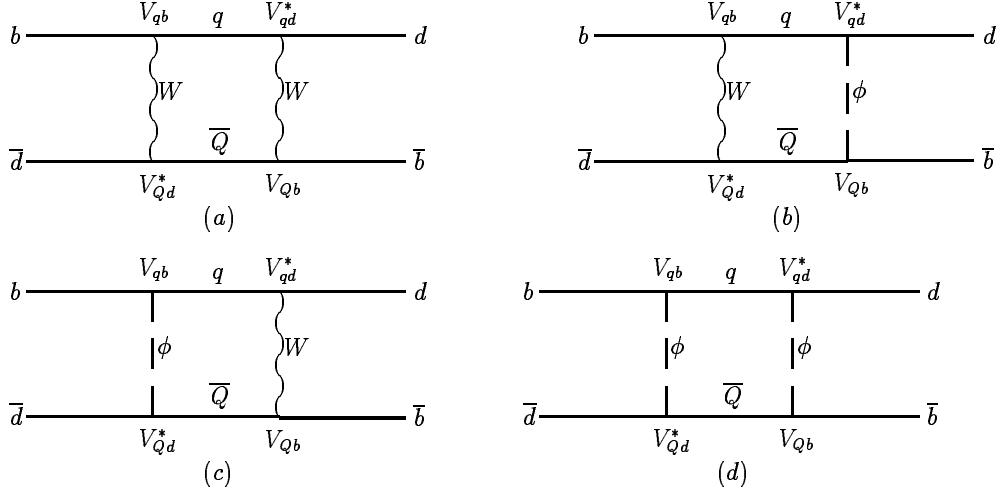}}
\caption{Examples of box diagrams for $b-\bbar$ oscillations,
including unphysical scalar contributions.}
\label{boxunphys}
\end{figure}
These scalars couple to quarks with a strength proportional to the masses.
For an incoming $d$-type quark of generation $j$ and outgoing $u$-type of
generation $i$, the vertex factor is
$\frac{-ig}{2\sqrt{2}M_W} [m_{d_j} (1+\gamma_5)-m_{u_j}(1-\gamma_5)]
V_{{u_i}{d_j}}$.


The $W$ and $\phi$ propagators in the $\Rxi$ gauge are
\begin{eqnarray*}
W^\pm:  & & \frac{-i}{k^2-M_W^2+i\epsilon} \left[ g_{\mu\nu} +
\frac{(\xi-1)k_\mu k_\nu}{k^2-\xi M_W^2} \right], \\
\phi^\pm:  & & \frac{i}{k^2- \xi M_W^2+i\epsilon}.
\end{eqnarray*}

While the unitary gauge, $R_\infty$, is convenient for evaluating
tree-level diagrams (the unphysical scalars drop out), in this gauge there are
extraneous singularities generated in the box diagram from the
$W$-propagator terms, which go as $k_\mu k_\nu/M_W^2$.  In fact, the Green's
functions are unrenormalizable, though one can show that the full S-matrix
is finite \cite{Fujik72}, so extreme
care must be taken in performing the $\xi \rightarrow \infty$ limit.

In the discussion below we work in the 't Hooft-Feynman gauge, $\xi=1$,
in which the propagators are particularly simple.
Note that the relative minus sign in the propagators causes the
diagrams in Figures \ref{boxunphys}$b,c$ to add to,
rather than cancel, the diagrams in Figures \ref{boxunphys}$a,d$.

For example, the basic matrix element for $B^0-\Bzbar$ transitions through
the diagram in Figure \ref{boxunphys}$a$, involving internal quarks of
type $Q$ and $q$ is 
\begin{eqnarray*}
i{\cal M} & = & <B^0| \frac{g^4}{64}
\xi_Q\xi_q \int \frac{d^4k}{(2\pi)^4}
\overline{d}(p_3)\gamma^\mu(1-\gamma_5){ \frac{1}{\not \!k-m_Q}}
\gamma^\nu(1-\gamma_5)b(p_1) \\
& \times & \overline{d}(p_2)\gamma_\nu(1-\gamma_5)
{ \frac{1}{\not \!k-\not \!p_1-\not \!p_2-m_q}}\gamma_\mu(1-\gamma_5)b(p_4) \\
& \times & { \frac{1}{(k-p_1)^2-M_W^2}\frac{1}{(k-p_3)^2-M_W^2}} |\Bzbar>.
\end{eqnarray*}
Here ($p_1$, $p_2$) are the incoming $(b,\dbar)$ momenta;  
($p_3$, $p_4$) are the outgoing $(d, \bbar)$ momenta, and $\xi_Q=\VQd^*\VQb$
and $\xi_q=\Vqd^*\Vqb$ are the CKM factors.

We now invoke the approximation that the external $b$ and $d$ quark
momenta and masses are zero, $p_i=0$.  After a reduction of the Dirac
matrices, the matrix element becomes
\begin{eqnarray*}
i{\cal M}_{WW} & = & \frac{g^4}{16} \xi_Q\xi_q
[<B^0|\overline{d}\gamma^\mu(1-\gamma_5)b
\overline{d}\gamma_\mu(1-\gamma_5)b |\Bzbar>]\\
& \times & { \int \frac{d^4k}{(2\pi)^4}
\frac{k^2}
{(k^2-m_Q^2)(k^2-m_q^2)(k^2-M_W^2)^2}}.
\end{eqnarray*}

Applying the same assumptions to the diagrams with the scalars, one finds that
all diagrams reduce to having a common operator structure,
$\hat{O}_{LL} = (\overline{d}_L\gamma^\mu b_L)^2$, multiplied by a
convergent integral over the loop momentum.  The result is
\begin{eqnarray*}
{\cal M}_{AB}(x_q,x_Q) & = & \frac{g^4}{64 \pi^2 M_W^2} E_{AB}(x_q,x_Q)
<B^0|(\overline{d}_L\gamma^\mu b_L)^2|\Bzbar>
\end{eqnarray*}
\noindent with $x_q \equiv m_q^2/m_W^2$ and
\begin{eqnarray*}
E_{WW}(x_q,x_Q) & = & -\frac{1}{2}\frac{1}{x_q-x_Q}
\left[ \frac{x_q^2 \log x_q}{(1-x_q)^2} - \frac{x_Q^2 \log x_Q}{(1-x_Q)^2}
+ \frac{1}{1-x_q} - \frac{1}{1-x_Q} \right] \\
E_{W\phi}(x_q,x_Q) & = &
\frac{1}{2}\frac{x_q x_Q }{x_q-x_Q}
\left[ \frac{x_q \log x_q}{(1-x_q)^2} - \frac{x_Q \log x_Q}{(1-x_Q)^2}
+ \frac{1}{1-x_q} - \frac{1}{1-x_Q} \right] \\
& = & E_{\phi W}(x_q,x_Q) \\
E_{\phi\phi}(x_q,x_Q) & = & \frac{x_q x_Q}{4} E_{WW}(x_q,x_Q).
\end{eqnarray*}

Table~\ref{tab:boxnumerical} provides values of these functions for
$m_u=0, m_c=1.5$ GeV, and $m_t=170$ GeV.

\begin{table}[htb]
\caption{Contributions of the $W$-$W$, $W$-scalar and scalar-scalar diagrams
with all combinations of internal quark lines.}
\begin{tabular}{@{}lccccccc@{}}
\hline \hline
\multicolumn{2}{c}{Diagram} & {\bf u, u} & {\bf u, c} & {\bf u, t}
& {\bf c, c} & {\bf c, t} & {\bf t, t} \\ \hline
(a) & {\bf $W-W$} & 1 & 1.9952 & 0.5353 & 0.9955 & 0.5352 & 0.1339 \\
(b)+(c) & {\bf $W-\phi$} & 0 & 0 & 0 & $1.45\times 10^{-6}$ &
$1.02\times 10^{-3}$ & 1.196 \\ 
(d) & {\bf $\phi-\phi$} & 0 & 0 & 0 & $3.02\times 10^{-8}$ &
$2.08\times 10^{-4}$ & 0.6690 \\ \hline
\end{tabular}
\label{tab:boxnumerical}
\end{table}

Because the scalar couples to the quarks with a strength proportional to
$m_q/M_W$, only the diagrams with two top quarks contribute substantially
to the
diagrams in Figures~\ref{boxunphys}{\em b,c,d}.
Although the $W-W$ diagrams with $u$ quarks and $c$ quarks
have large
matrix elements, they very nearly cancel when combined with their CKM factors
because both the $u$ and $c$ quark are light compared to the $W$.
Traditionally, the unitarity of the CKM matrix is assumed, and is used to
express the $u$-quark terms $\Vud\Vub^* = -\Vcd\Vcb^* - \Vtd\Vtb^*$.  For
example, it is not the case that the $ut$ matrix element is small, but rather
its contribution almost exactly cancels the $ct$ contribution via this
GIM-like mechanism \cite{GIM70} if the CKM matrix is unitary.

Thus the full matrix element, summing over all internal quark types, becomes
simply a sum over $c$ and $t$ quarks and different bosons:
\begin{eqnarray*}
{\cal M} & = & <B^0|(\overline{d}_L\gamma^\mu b_L)^2|\Bzbar>
\frac{g^4}{64 \pi^2 M_W^2} 2 \sum_{q,Q=c,t} \xi_Q \xi_q \times \\
& & \sum_{A,B=W,\phi} \left[ E_{AB}(x_q,x_Q) -  E_{AB}(x_q,0) -
E_{AB}(0,x_Q) + E_{AB}(0,0) \right].
\end{eqnarray*}

To summarize, we have made the following assumptions:
({\em a}) The $W$ mass is generated
by a spontaneous symmetry-breaking mechanism, resulting in scalar remnants;
({\em b}) the momenta and masses of the $b$ and $d$ quarks inside a bound meson
can be neglected; and ({\em c}) the CKM matrix is unitary.

This program was first carried out, with the important scalar terms included,
by Inami \& Lim
\cite{inamilim} in 1981.  The result of the sum over bosons is
\begin{eqnarray}
f_3(x_Q, x_q) & = & \frac{x_Q x_q}{x_Q - x_q} \left(\ln\frac{x_Q}{x_q}
- \frac{3}{4}\frac{x_Q^2\ln x_Q}{(1-x_Q)^2}
+ \frac{3}{4}\frac{x_q^2\ln x_q}{(1-x_q)^2} \right) -
 \nonumber \\
&  & \frac{3}{4}\frac{x_Q x_q}{(1-x_Q)(1-x_q)}.
\end{eqnarray}
For the same quark type on the two internal lines ($q=Q$), this reduces to
\begin{eqnarray}
f_2(x_q) & = & x_q \left( \frac{1}{4}+\frac{9}{4}\frac{1}{1-x_q} - 
\frac{3}{2}\frac{1}{(1-x_q)^2} \right)
- \frac{3}{2}\frac{x_q^3\log x_q}{(1-x_q)^3}.
\end{eqnarray}

Table~\ref{ilfactor} shows the size of the Inami-Lim functions $f_2(x_c)$,
$f_2(x_t)$ and $f_3(x_t, x_c)$ ($=f_3(x_c, x_t)$),
and the CKM factors that multiply each term.
For the cases of $B^0$ and $\Bs$ mesons, it turns out that these CKM factors
are almost constant, independent of the type of quarks
in the internal loop.  For reference, the CKM coefficients for the
$K^0$ system are shown, which vary by five orders of magnitude.
\begin{table}[htb]
\caption{Factors entering the matrix element, which is proportional to the
product of the Inami-Lim function and the CKM term.}
\begin{tabular}{@{}ccccc@{}}
\hline \hline
{\bf \begin{tabular}{@{}c@{}} Internal \\ quarks\end{tabular}} & {\bf I-L factor} &
{\bf $B^0$ CKM} & {\bf $\Bs$ CKM} &
{\bf $K^0$ CKM} \\ \hline
{\bf c,c} & $3.5\times 10^{-4}$ &
 $A^2\lambda^6$ &
 $A^2\lambda^4$ & 
 $\lambda^2$ \\
& &
 $(7.4\times 10^{-5})$ &
 $(1.4\times 10^{-3})$ & 
 $(2.7\times 10^{-2})$ \\
{\bf c,t} & $3.0\times 10^{-3}$ &
 $A^2\lambda^6|1-\rho-i\eta|$ &
 $A^2\lambda^4$ &
 $A^2\lambda^6|1-\rho-i\eta|$ \\
& &
 $(7.3\times 10^{-5})$ &
 $(1.5\times 10^{-3})$ &
 $(8.8\times 10^{-6})$ \\
{\bf t,t} & 2.5 &
 $A^2\lambda^6|1-\rho-i\eta|^2$ &
 $A^2\lambda^4$ &
 $A^4\lambda^{10}|1-\rho-i\eta|^2$ \\
& &
 $(7.2\times 10^{-5})$ &
 $(1.5\times 10^{-3})$ &
 $(1.1\times 10^{-7})$ \\ \hline
\end{tabular}
\label{ilfactor}
\end{table}

Note that the function $f_2(x_q)$, for which both internal quarks are of
the same type, is proportional to $x_q$, i.e. the square of the mass of
the internal quark, and so the top quark loop completely dominates the charm
loop's contribution.  The function $f_3(x_t, x_c)$ is proportional to $x_c$
and is relatively small, an $\approx 0.1\%$ correction.

In the approximation that only the top quark contributes, the matrix element
can thus be written (with a factor for all permutations) as
\begin{eqnarray}
{\cal M} & = & \frac{G_F^2}{8\pi^2} |\Vtd^*\Vtb|^2 m_t^2 f_2(m_t^2/M_W^2)
<\!B^0|(\overline{d}_L\gamma^\mu b_L)^2|\Bzbar\!>.
\label{eqn:boxmatrix}
\end{eqnarray}

Hence, we need to evaluate the effective matrix element
\begin{eqnarray}
<\!B|(\overline{d}\gamma^\mu&(1-\gamma_5)&b)^2|\overline{B}\!>  = \nonumber \\
& & \sum_n <B|\overline{d}\gamma^\mu(1-\gamma_5)b|n> 
<n|\overline{d}\gamma_\mu(1-\gamma_5)b|\overline{B}> \nonumber \\
& & \equiv { B_B} <B|\overline{d}\gamma^\mu(1-\gamma_5)b|0>
<0|\overline{d}\gamma_\mu(1-\gamma_5)b|\overline{B}> \nonumber \\
& & =  {B_B} |<0|\overline{d}\gamma^\mu\gamma_5 b|B>|^2,
\label{eqn:BagDefn}
\end{eqnarray}
\noindent where $B_B$ is the bag factor, which is 1 if the vacuum insertion
$|0><0|$ saturates the sum over all intermediate states.

Substituting 
$<0|\overline{d}\gamma^\mu\gamma_5 b|B> = if_B p^\mu$,
where $f_B$ is the $B$ decay constant into this expression,
and taking into account the three colours of the quarks gives
\begin{eqnarray}
<B|(\overline{d}\gamma^\mu(1-\gamma_5)b)^2|\overline{B}> & = &
\frac{8}{3} {B_B} {f_B^2} m_B^2.
\label{eqn:boxoperators}
\end{eqnarray}

Finally, the diagrams in Figure~\ref{boxunphys} can have arbitrary numbers
of gluons running between any of the quark lines.
The effect of this on ${\cal M}$ is encoded in the QCD correction
factor $\eta_B$, which multiplies the ${\cal M}$ above.

The size of this correction, including next-to-leading-order (NLO) diagrams,
has been estimated by
Buras et al \cite{Buras90},
\begin{eqnarray}
\eta_B & = & 0.55 \pm 0.01.
\label{eqn:eta}
\end{eqnarray}
These authors found that $\eta_B$ depends upon the definition
of the top quark mass and that only the product $\eta_B f_2(x_t)$ is
insensitive to this choice.  To be consistent with this estimate of $\eta_B$,
the top quark mass must be chosen to be $\overline{m_t}(m_t^{\rm pole})$, the
mass calculated using the $\overline{MS}$ renormalization scheme evaluated
at the measured (pole) mass of $174\pm 5$ GeV \cite{PDG98}.  This gives
a mass 6-7 GeV lower than the measured mass.
Some confusion over $\eta_B$ exists in the
literature; however
$\eta_B$, by definition, is the same for $B^0$ and $\Bs$ mesons.

Putting together
Equations~\ref{eqn:boxmatrix}, \ref{eqn:boxoperators}, \ref{eqn:eta}
results in
\begin{eqnarray*}
|M_{12}| = \frac{|{\cal M}|}{2m_B} & = &\frac{G_F^2}{6\pi^2}
m_B {f_B^2 B_B \eta_B} m_t^2 f_2(m_t^2/M_W^2) {|{\Vtd^*}\Vtb|^2}.
\end{eqnarray*}

To relate this to the heavy-light mass difference $\delm$, we in principle
also need the absorptive part of the matrix element, $\Gamma_{12}$, as per
Equation~\ref{eigvals}.  However, $\Gamma_{12} \ll M_{12}$ as argued below, and
can be neglected.

The matrix element $\Gamma_{12}$ involves decays to modes common to both
 $|B>$ and $|\Bbar>$
that are on-shell and thus energetically allowed, meaning the top quark
loops do not contribute.
Since the matrix element is given approximately by the square of the
mass of the state in the loop, and this is $\approx m_b^2$ for on-shell
transitions, we obtain
$\Gamma_{12}/M_{12} \approx m_b^2/m_t^2 \ll 1$.

Alternatively, this can be seen dirctly from the data.
The fractional lifetime difference $\delg/\Gamma$ is
expected to be small, as previously mentioned -- less than
$0.01$ for the $B^0$ and up to $0.25$ for the $\Bs$.
As shown in Sections~\ref{BdSection} and
\ref{BsSection},
current experiments demonstrate that 
$\delm \approx 0.7 \Gamma$ for the $B^0$ system, and $\delm_s \ge 20 \Gamma$
for the $\Bs$.  Combining this information gives $\delg_d = (0.01/0.7) \delm_d$
and $\delg_s = (0.25/20) \delm_s$, so in both cases
$\delg \sim \order(10^{-2}) \delm$.  Equation~\ref{eigvalRelation} then
implies, in general, that $|\Gamma_{12}| = \order(10^{-2})|M_{12}|$, leaving
\begin{eqnarray}
\delm & = & 2|M_{12}| \hspace{1cm}\mbox{and} \nonumber \\
\delg & = & 2 Re(M_{12}\Gamma_{12}^*)/|M_{12}|.
\end{eqnarray}

Hence, the final formulae are as follows:
\begin{eqnarray}
\Delta m_d & = &\frac{G_F^2}{6\pi^2}
m_{B_d} f_{B_d}^2 B_{B_d} \eta_B m_t^2 f_2(m_t^2/M_W^2) |{\Vtd^*}\Vtb|^2,
 \nonumber \\
\Delta m_s & = &\frac{G_F^2}{6\pi^2}
m_{B_s} f_{B_s}^2 B_{B_s} \eta_B m_t^2 f_2(m_t^2/M_W^2)
|{\Vts^*}\Vtb|^2
\label{delmFormula}
\end{eqnarray}
and
\begin{eqnarray}
\frac{\delm_s}{\delm_d} & = & 
\frac{m_{B_s} f_{B_s}^2 B_{B_s}}{m_{B_d} f_{B_d}^2 B_{B_d}}
\left|\frac{\Vts}{\Vtd}\right|^2.
\end{eqnarray}

Assuming unitarity of the CKM matrix, $|\Vtb| \approx 1$.  Since the
top quark mass has now been measured, the only unknowns in these formulae
are the magnitudes of the CKM
elements $\Vtd$ and $\Vts$, the decay constants $f_B$ (which in principle
can be measured) and
the bag factors, $B_B$, which are theoretical constructs.

Estimating $f_B$ and $B_B$ for the $B^0$ and $\Bs$, then, is essential
for extraction of the CKM elements.  Since these mesons differ only by their
light quarks, it is reasonable to expect that model dependences in
predictions of, for instance, $\fbd$ and $\fbs$ are significantly reduced in the
ratio $\fbd/\fbs$.  These issues are addressed the next section.

\subsection{The $B$ Decay Constant and Bag Parameter}
\label{fbSection}

The $B$ decay constant $\fb$ is necessary to relate the $B$
 oscillation frequency
to the CKM matrix element $\Vtd$, as described in Section~\ref{formal}.
Prospects for measuring this decay constant directly are rather bleak.
In principle, it can be measured just as $\fpi$ is measured, using leptonic
decays.  The partial width of the decay $B\rightarrow l \nu_l$ is
\begin{eqnarray}
\Gamma(B^+\rightarrow l^+ \nu_l) & = &
\frac{G_F^2 |\Vub|^2}{8\pi}\fb^2 m_B m_l^2
\left( 1-\frac{m_l^2}{m_B^2} \right) ^2,
\label{BtauEqn}
\end{eqnarray}
\noindent where $l = e, \mu, \tau$.
The decays into $e$ and $\mu$ are helicity suppressed, leaving the
experimentally challenging $B\rightarrow\tau\nu_\tau$ as the dominant leptonic
decay.  The branching ratio into this mode, however, is expected to be
very small.  Using the range for $\Vud$ of 0.002-0.005 \cite{PDG98}
and $\fbd=200$ MeV (see below) results in a branching ratio
between $1.8 \times 10^{-4}$ and $3 \times 10^{-5}$.

ALEPH \cite{Alephbtau}, DELPHI \cite{Delphibtau} and L3 \cite{L3btau}
have placed 90\% CL upper limits on
this branching ratio, of $1.8\times 10^{-3}$, $1.1 \times 10^{-3}$
and $5.7\times 10^{-4}$ respectively.
Since the LEP data is played out, no further improvement is likely.
CLEO \cite{CLEObtau} searches for
a $B\rightarrow \tau\nu_\tau$ signal by reconstructing the mass of
the $B$ meson, using the beam constraint to account for the missing neutrinos.
The background to this analysis is very high -- the
signal is $-9 \pm 36$ events.  
The CLEO group reports a 90\% CL limit of $2.2\times 10^{-3}$
based on a sample of $2.2\times 10^6$ $B$ decays.

The most promising method to extract $\fb$ from experimental data had seemed to
be via the $B^+-B^0$ lifetime difference.  It was recently estimated
\cite{bigi95} that the lifetime ratio is
\begin{eqnarray}
\label{eqn:liferatio}
\frac{\tau^-}{\tau^0} & = & 1 + 0.05\frac{f_B^2}{(200 \mbox{ MeV})^2},
\end{eqnarray}
\noindent with an extra $\order(15\%)$ error added to account for terms
of order $m_b^4$ that are not included in the calculations.
Thus, a measurement of the lifetime ratio to 1\% corresponds to a measurement
of $\fb$ to $\approx 20$ MeV, with a 30 MeV theoretical error.
The current value of this ratio is $1.04\pm 0.04$
\cite{PDG98}. CDF has measured it as $1.06\pm 0.07\pm 0.01$
in exclusive $B\rightarrow J/\Psi K$ decays \cite{CDFexclusive98} and as
$1.11\pm 0.06 \pm 0.03$ in semileptonic
decays \cite{CDFsl98}.  The 
$J/\Psi$ sample in CDF's upcoming Run II is expected to be more than
20 times larger than the current one, with
a similar increase in the inclusive sample. Along with modest
improvements in the systematic errors, this leads to an expected
precision on this ratio of $\approx$ 1\%.
The BaBar group expects to measure this ratio to close
to 1\% as well \cite{BaBartdr}.
Recently, however, our ability to predict lifetimes and lifetime ratios has been called
into question because none of the models account for the short $\Lb$
lifetime.
This failure has led some authors \cite{neubert95} to suggest that
some of the approximations used in deriving Equation~\ref{eqn:liferatio} are
not valid
-- specifically there are large nonperturbative effects that should not
be ignored.
Until the situation is clarified,
extracting physical quantities from heavy quark lifetimes is suspect.

This leaves us with two ways to obtain $\fb$: lattice QCD and
QCD sum rules.

In the past year or two, lattice calculations of $\fb$ have matured
considerably.
Our understanding of lattice discretation effects has grown, and the advent
of nonrelativistic actions have provided a means to greatly reduce the
uncertainty introduced by the relatively large lattice spacing.

One of the main problems with calculating $\fb$ on the lattice is that the $b$
quark is very heavy, and the wavelength of the $b$ quark is small compared
to the
lattice spacings currently computationally feasible.
Typical lattice spacings, $a$, available today are $\order(0.1 - 0.2)$ fm,
so one has $\mQ a \simeq 25 a > 1$ (where $a$ is in fm).
Thus, in extrapolating to the continuum limit,
$\mQa = 0$, one has a long way to extrapolate; further,
one must contend with lattice-size artifacts which are $\order(\mQa)$,
$\order[(\mQa)^2]$ and so on, and each may be sizeable.

The original Wilson lattice action was developed for light quarks, for which
$ma \ll 1$ holds.  This discretization procedure introduces lattice spacing
artifacts of $\order(a)$, $\order(a^2)$ and so on, which makes the
extrapolation to the continuum, $a=0$, a major source of systematic
uncertainty.  The Sheikholeslami-Wohlert (SW or clover) action \cite{SW85}
was introduced as an improvement on the Wilson action for light quarks;
it removes the leading $\order(a)$ lattice effects, resulting in a
much-improved behaviour.  Lattice calculations continue to use both of these
actions for the light quarks in the $B$ and $D$ meson systems, along with
a variety of approaches in handling the heavy quark.

In placing a heavy-light system on the lattice, different
techniques were developed for handling the heavy quark.
Taking the straight Wilson action is problematic because $\mQ a > 1$.
One common method of dealing with the $b$ quark,
known as the static approximation, is to set its mass to be
infinite.  In the first six or so years of
lattice calculations this method was used extensively, and it
produced results for
$\fb$ which varied over an extremely wide range.  
By 1994, these variations were understood to be due to large contamination
from excited $B^*$ states in the
lattice version of the $B$ meson.  Once this was corrected for, the results
from simulations which used static $b$ quarks stabilized, with values
comparable to those from other methods.

In the meantime, propagating $b$ quarks
were introduced by performing the lattice calculations with finite quark
masses, but closer to $m_c$, where $m_Q a < 1$.
Typically, several calculations with different masses were performed, and then
used to extrapolate up to $\mQ = m_b$ using a linear extrapolation
in $1/\mQ$.  Often the static limit was included to help anchor this process.
In these methods, systematic errors are introduced both by finite
lattice spacing artifacts and the uncertainties in the mass extrapolation.

More recently the heavy $b$-quark problem has been addressed by using an
effective nonrelativistic action for the $b$ quark.
In the rest frame of a $B$ meson, the light valence quark has
a momentum $\order(\LQCD)$.  By momentum conservation, the heavy $b$ quark must
also have momentum of this order.  If the $b$ quark is nonrelativistic,
$p = \mQ v = \LQCD$, and so the velocity $v \approx \LQCD/\mQ \approx 0.05$,
justifying the nonrelativistic approximation.  Thus, one can effectively
expand quantities in powers of $1/\mQ$ and expect reasonable behaviour.
Further, in a nonrelativistic effective theory, the rest mass of the $b$ quark
does not appear, so we sidestep the problem of having a large $\mQ a$.
Since the first use of this approach in 1994 \cite{Hash94}, which
included the $\order(1/\mQ)$ operators, various groups have
continued to add more
and more of the $\order(1/\mQ^2)$ terms.

In such nonrelativistic QCD (NRQCD) calculations, $1/\mQ$ effects and lattice spacing $a$ effects
end up intertwined, so one of the challenges of such work is to 
effectively evaluate
systematics due to the extrapolations.
The uncertainties are much reduced if the next-order $1/\mQ^2$ terms
are included.
In early 1998 the MILC collaboration presented results with a comprehensive
treatment of both these effects, with improvements through
$\order[(\LQCD/\mQ)^2]$ and $\order(a)$ \cite{MILC98}.

A third technique, developed by the Fermilab group (KLM) \cite{KLM},
allows the use of either Wilson or clover actions,
with suitable normalisation and
nonrelativistic interpretation, for the quarks at
any $\mQ a$.  Simulations
based on this heavy-quark action are run directly at $m=m_b$, so no mass
extrapolation is required.

Until recently, all calculations of $\fb$ have been performed within
the quenched approximation, i.e. no virtual quark loops were simulated.
Estimating the size of the shift in $\fb$ due to this omission has been one of
the great outstanding issues for the lattice community.  There are now
three groups reporting partially unquenched simulations, with two flavors of
light fermions:
the MILC collaboration \cite{MILC99,MILC97},
Collins et al \cite{Collins99}, and CP-PACS \cite{CPPACS99-1,CPPACS99-2}.
The results are summarized in Table~\ref{unquenched}.
All three groups found
that unquenching the lattice raised all the $B$ and $D$ decay
constants by 15-20\%, although the results are still preliminary.

\begin{table}[htb]
\caption{Results for the decay constant $\fbd$
from several lattice simulations with two
dynamical sea quarks.}
\begin{tabular}{@{}lc@{}}
\hline\hline
Reference & $\fbd$ (MeV) \\ \hline
MILC 99 \cite{MILC99} & $194\pm 22^{+20}_{-0}$ \\
CP-PACS 99 \cite{CPPACS99-1,CPPACS99-2} & $210^{+9}_{-16}\pm20$  \\
Collins 99 \cite{Collins99} & $186\pm 25^{+50}_{-0}$ \\ \hline
Average & $200\pm30$ \\ \hline
\end{tabular}
\label{unquenched}
\end{table}

\begin{figure}[htb]
\epsfxsize=11cm
\centerline{\epsfbox{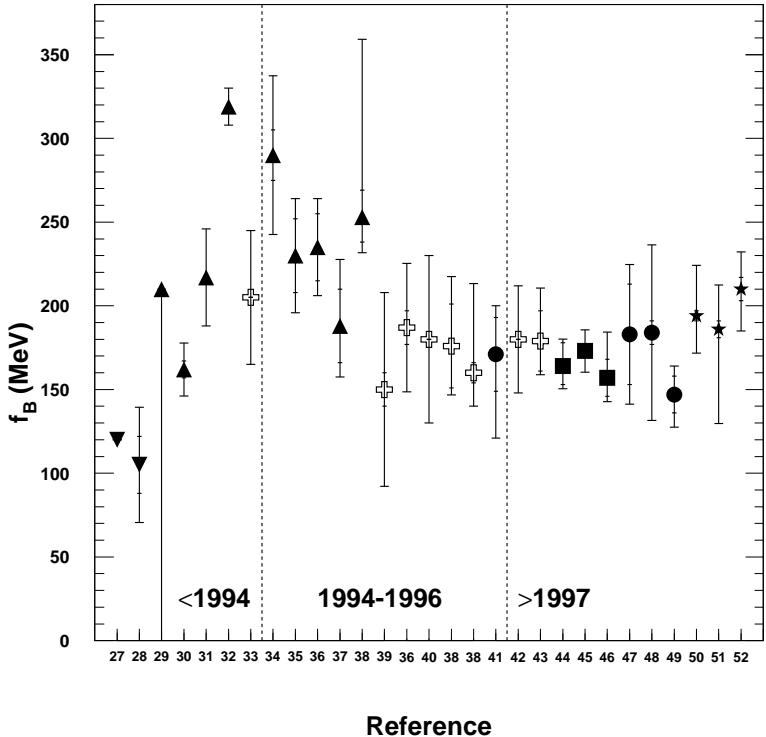}}
\caption
{Summary of lattice calculations of the decay constant $\fbd$.
The results are grouped by the treatment of
the heavy quark:  down-pointing triangles, mass extrapolation from
$m < m_c$; up-pointing triangles, static calculations; open crosses,
mass extraploation from $m > m_c$ to $m_b$; circles, NRQCD;
squares, the KLM action at $m_b$.  The rightmost three results, marked with
a star, are those with two dynamical fermions.}
\label{latSum}
\end{figure}

Figure~\ref{latSum} shows lattice results for $\fbd$
over the 10-year history
of the field.  The results are grouped such that similar methods of dealing
with the $b$ quark (static, mass extrapolation, KLM, NRQCD) are shown
together for easier
comparison.  The points are grouped into three rough time periods:
pre-1994 years, the early years;
1994-1997, during which NRQCD techniques were developed to include
higher-order terms; and post-1997, when fully $\order(1/\mQ^2)$, $\order(a)$
improved simulations were implemented.

Table~\ref{latAvg} shows averages of $\fbd$, $\fds$, and $\fbs/\fbd$,
 using data from the
collaborations shown.  Results with full estimations of systematics were
included in the fit.  The average result, $\fbd = 166\pm 8$ MeV, does not
include a systematic due to quenching.   This is much lower than the
average obtained from the $N_f=2$ calculations, which is $200\pm30$ MeV
(allowing a large systematic as these are still preliminary).

\begin{table}[htb]
\caption{Lattice results from simulations selected as described in the text.
A common systematic error due to quenching is not included.}
\begin{tabular}{@{}lclc@{}}
\hline\hline
Reference & $\fbd$ (MeV) & $\fds$ (MeV) & $\fbs/\fbd$ \\ \hline
MILC 98 \cite{MILC98} & $157^{+27}_{-14}$ & $210^{+27}_{-13}$ & $1.11\pm0.05$ \\
JLQCD 98 \cite{JLQCD98} & $173\pm 13$ & $224\pm19$ & $1.18\pm0.06$\cite{JLQCD98-2} \\
APE 98 \cite{APE98} & $179^{+32}_{-20}$ & $231\pm13$ & $1.14\pm0.03$\\
GLOK 98 \cite{GLOK98} & $147\pm19$ & NA  & $1.20\pm0.04$ \\
FNAL 98 \cite{FNAL98} & $164\pm16$ & $213\pm18$& $1.13\pm0.05$ \\
APE 97 \cite{APE97} & $180\pm32$ & $237\pm16$ & $1.14\pm0.08$ \\ \hline
Average & $166\pm8$ & $226\pm 8$ & $1.15\pm0.02\pm0.06^\dag$\\ \hline
\end{tabular}

$^\dag$ Following a recent review \cite{Dr98} a common systematic of
0.06 is assigned.
\label{latAvg}
\end{table}

\begin{figure}[htb]
\centerline{
\mbox{
\epsfxsize=2.8in \epsfbox{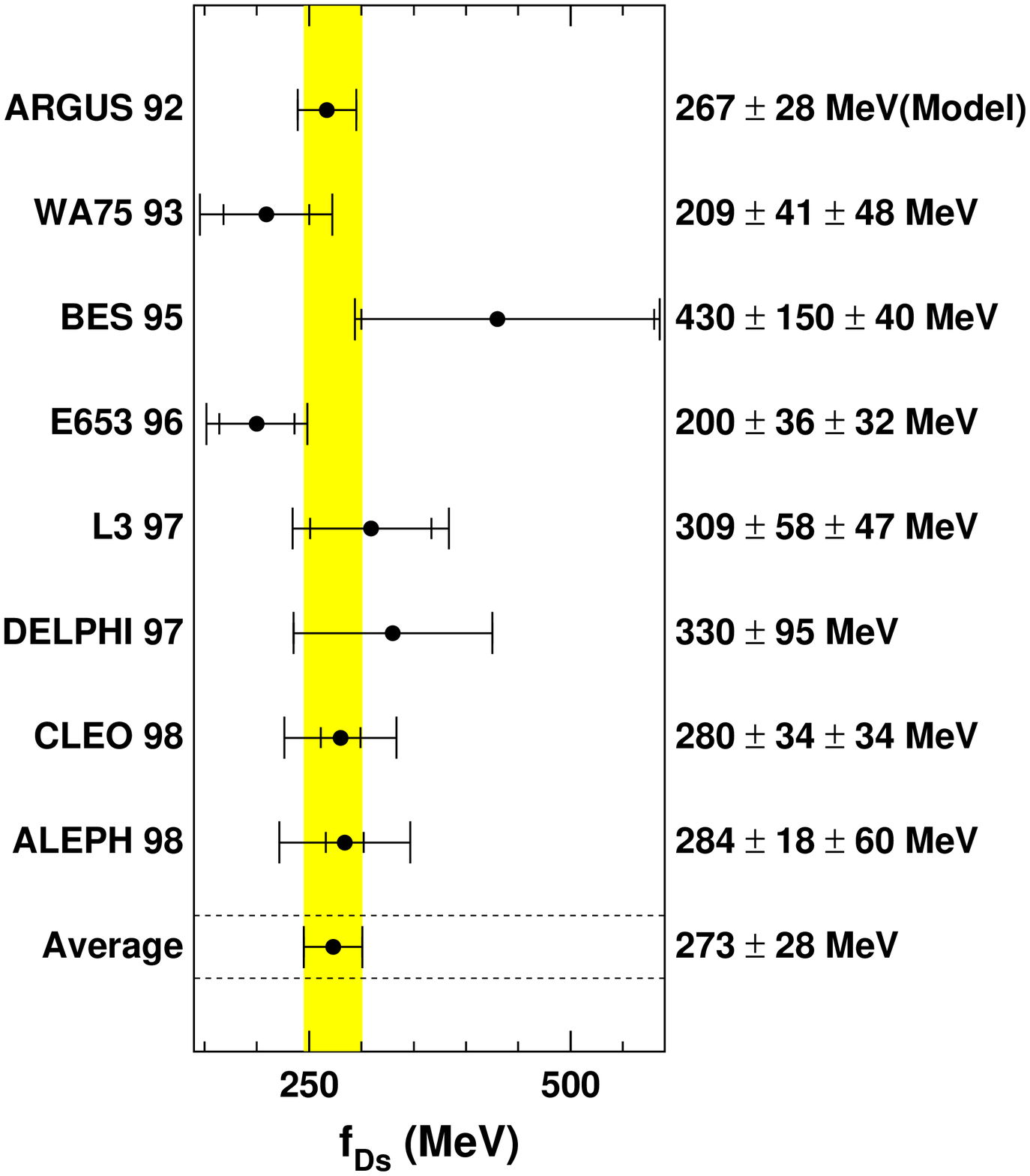}\hspace{0.2in}
\epsfxsize=2.3in \epsfbox{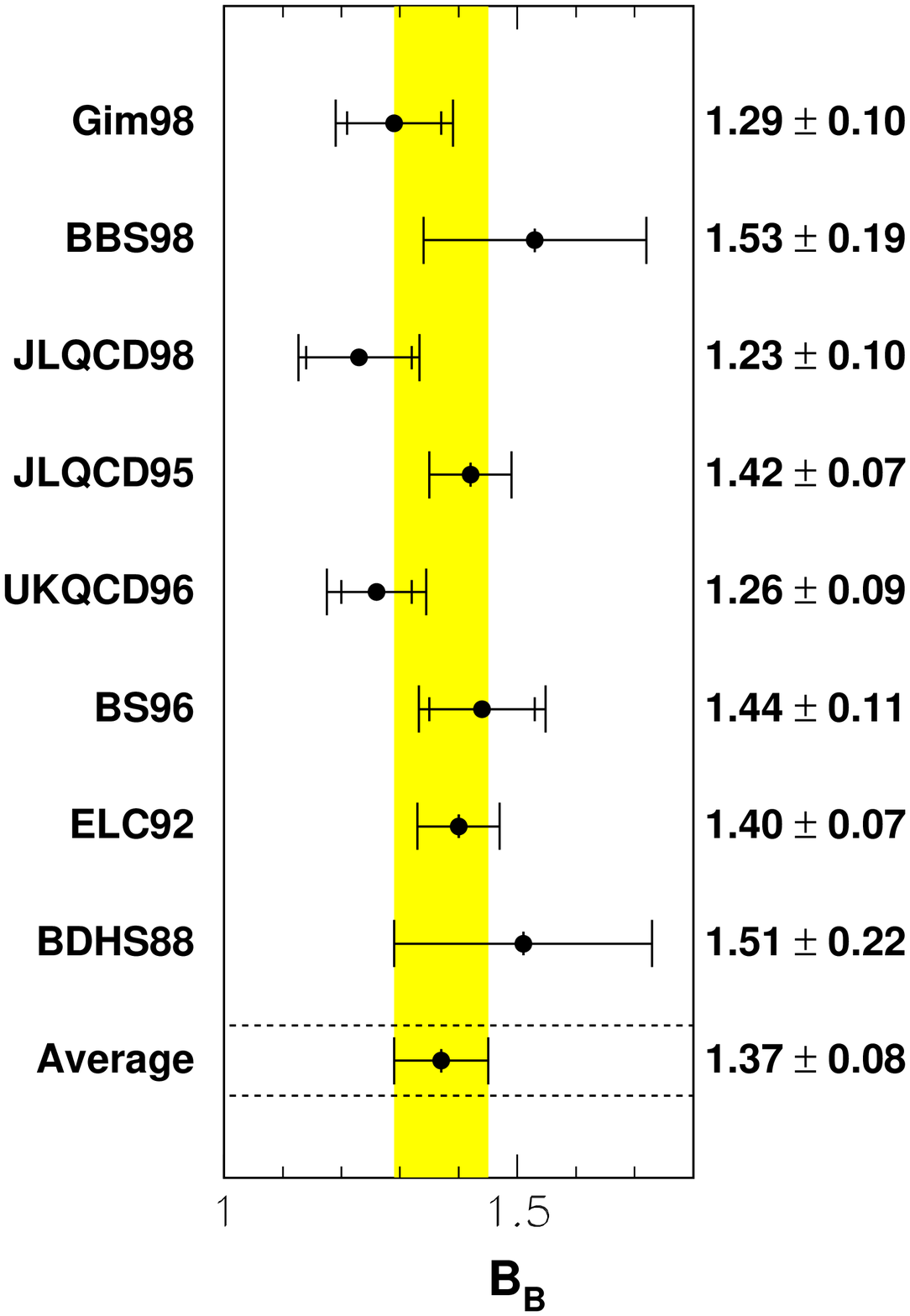}
}
}
\caption{{\em Left:} Average of $\fds$ experimental measurements. The
model-dependent ARGUS result is not included in the average.
{\em Right:} Average of lattice calculations of $\bnlod$.  These have been
adjusted so that the NLO corrections are handled in a uniform manner, and
include a common systematic of 0.07 due to uncertainties in the
renormalization constant.}
\label{fdsAvg}
\end{figure}

Unfortunately, checking the lattice and sum rule results against experiment is
difficult, because experimental data are in short supply as noted earlier.
However, there is one
decay constant prediction which can be correlated with
data, namely
$\fds$.  The branching ratio of $D_s\rightarrow \tau\nu_\tau$ is fairly large,
$7 \pm 4$\% \cite{PDG98}, owing to the relative lack of hadronic decay modes
in contrast to both $B$ and $D$ decays, which have 
ample phase space available for a multitude of hadronic modes.
There are currently eight measurements \cite{fdsMeasures} of the branching
ratio Br($D_s\rightarrow\tau\nu_\tau$), which, along with $\Vcs$, allow us to
extract $\fds$ as per Equation~\ref{BtauEqn}.
Figure~\ref{fdsAvg} summarizes these measurements.

The ARGUS result is model-dependent and omitted from the
average.  Results from WA75 and
E653 have been adjusted to take into account updated $D_s$ and $D^0$ branching
ratios.  Note that the errors have actually
increased as the $D_s \rightarrow \phi \pi$ branching ratio measurement was
downgraded by the PDG from having an error of 0.4\% in 1994 to 0.9\% in 1996.
The measurements were combined, taking into account correlations in the
branching ratios used to derive $\fds$.  The result is
\begin{eqnarray}
\fds^{\rm exp} & = & 273 \pm 28 \hspace{0.5cm} \mbox{MeV}.
\end{eqnarray}

We obtain a lattice average of $\fds = 226 \pm 8$ MeV, not
including quenching errors, from the same group of collaborations
used to derive $\fbd$, as shown in Table~\ref{latAvg}.
There is reasonable agreement between the measurement and the
lattice result, though the agreement would be better using the
unquenched lattices.  More precise experimental data and
a better understanding of the unquenched lattices would be of great help.

Since $\fds$ is likely to be the only experimental input for many years, an
interesting exersize is to bootstrap $\fbd$ and $\fbs$ from it.  The idea is
that many of the largest systematic errors in the lattice calculation of
the heavy-light decay constants either cancel, or are largely ameliorated,
when one takes a ratio.
For example, unquenching the lattice is
expected to have a similar effect on $\fbd$ and $\fds$, as should lattice
size corrections.  Several groups \cite{APE98,MILC98} are now calculating
the ratio
$\fbd/\fds$ directly from the same simulation.  Multiplying the average
value of
$\fbd/\fds = 0.78^{+6}_{-4}$ by the experimental
value for $\fds$ results in:
\begin{eqnarray}
\fbd & = & \frac{\fbd^{\rm lat}}{\fds^{\rm lat}} \fds^{\rm exp} =
213 \pm 22\mbox{(expt.)}
^{+16}_{-11}\mbox{(theor.)}
\end{eqnarray}

The other technique used to calculate $\fb$ is based on QCD sum rules.
(For an
excellent summary of this method, see Reference~\cite{KR98}.)
The sum rules are
derived from the two-point correlation function:
\begin{eqnarray*}
\Pi(q^2) & = & i \int d^4 x e^{iqx}
<0|T{\bar{q}(x)i\gamma_5 b(x), \bar{b}(0)i\gamma_5 q(0)}|0>.
\end{eqnarray*}
Inserting a complete set of states with $B$-meson quantum numbers in the
time-ordered product gives
\begin{eqnarray*}
\Pi(q^2) & = & 
\frac{<0|\bar{q}i\gamma_5 b|B><B|\bar{b}i\gamma_5 q|0>}{m_B^2-q^2} +
 \sum_h \frac{<0|\bar{q}i\gamma_5 b|h><h|\bar{b}i\gamma_5 q|0>}{m_h^2-q^2}.
\label{qcdInsert}
\end{eqnarray*}
Using $m_b <0|\bar{q}i\gamma_5 b|B> = m_B^2 \fb$, we see that the first term
contains $\fb^2$.

The two-point correlation function can also be expanded using an operator
product expansion (OPE):
\begin{eqnarray}
\Pi(q^2) & = & \sum_d C_d(q^2,\mu) \hat{O}_d(\mu),
\label{qcdOPE}
\end{eqnarray}
where $C_d$ are the (calculable) Wilson coefficients of the operators of
various dimension $d$ which appear in the expansion.

Equating these expressions for $\Pi(q^2)$, along
with experimental input on the expectation values of the operators $\hat{O}_d$
(e.g. from bottomonium and charmonium masses), allows the
extraction of $\fb$.

The OPE expansion of $\Pi(q^2)$ is rather sensitive to the value of the
$b$-quark mass chosen, which ends up being the dominant systematic effect.
For example, taking $m_b = 4.7\pm 0.1$ GeV results in $\fb = 180 \pm 30$ MeV
\cite{KR98}.

Given that unquenching the lattice seems to shift $\fbd$ by 15-20\% and that
several groups confirm this result, we choose to use the preliminary
results with $N_f=2$ rather than perpetuate what seems to be a low
$\fbd$ value.  The quenched results on $\fbd$ measured in
a multitude of ways seem to have
stabilized, which suggests that the unquenched simulations will soon have errors
under much better control.

Combining the three values:
\begin{eqnarray*}
\fbd  & = & 200\pm 30, \hspace{100pt}\mbox{(Lattice, $N_f=2$)} \\
\fbd  & = & 180\pm 30, \hspace{100pt}\mbox{(QCD sum rules)} \\
\fbd  & = & 213\pm 22(\mbox{expt.})^{+16}_{-11}(\mbox{theor.}) 
\hspace{20pt}(\fbd^{\rm lat}/\fds^{\rm lat}\times \fds^{\rm exp})
\end{eqnarray*}
\noindent gives our estimate: $\fbd = 200\pm16$ MeV.

\subsubsection{THE BAG PARAMETER}
The second quantity standing between measurements of $\delmd$ and
the clean extraction of
$\Vtd$ is the so-called bag parameter $\bb$.  $\bb$ represents a
``fudge-factor'' introduced because we collapsed the sum over all intermediate
states in Equation~\ref{eqn:BagDefn} to just the vacuum intermediate state.
If the vacuum is the only state which ends up contributing, it saturates the
sum, and $\bb=1$.

As such, $\bb$ is a completely theoretical construct and cannot be measured
experimentally (only the product $\fb^2 \bb$ has physical meaning).  $\bb$ is
also amenable to study on the lattice, and many estimates of its value have
been made by the same groups which calculate $\fb$.

To calculate $\bb$, the ratio
\begin{eqnarray*}
r & = & \frac{<\Bbar|\hat{O}_{LL}|B>}{<\Bbar|\hat{O}_{LL}|B>_{\rm Vac}} =
\frac{<\Bbar|\hat{O}_{LL}|B>}{8/3 \fb^2 m_B^2}
\end{eqnarray*}
is evaluated using standard lattice
techniques (where Vac refers to the vacuum saturation approximation).
As with $\fb$, various groups have used static $b$ quarks,
propagating $b$ quarks with mass extrapolation, and NRQCD actions.
$\bb$ depends on the choice of
scale $\mu$, and groups quote their results at scales ranging from 2 GeV to
5 GeV.  The lattice
and continuum operators must then be matched at this scale.
The physical quantity of interest -- the 
NLO, renormalization-group-invariant quantity, is
\begin{eqnarray}
\bnlo(\mu) & = & \alpha_s(\mu)^{-2/\beta_0}
(1 + \frac{\alpha_s(\mu)}{4\pi}J) \bb(\mu),
\end{eqnarray}
\noindent where \\
$\begin{array}{lcllcl}
J & = & \frac{\gamma_0\beta_1}{2\beta_0^2} -  \frac{\gamma_1}{2\beta_0}, &
\hspace{0.5in} \gamma_0 & = & 4 \\
\beta_0 & = & 11 - \frac{2 n_f}{3}, &
\hspace{0.5in} \gamma_1 & = & -7 +  \frac{4 n_f}{9} \\
\beta_1 & = & 51 - \frac{19 n_f}{3} & & &
\end{array}$

Throughout the literature, a number of different choices are present, e.g.
$n_f=0,4,5$, $\LQCDfo = 200, 239$ MeV, $\LQCDfi = 183, 200$ MeV,
$m_b = 4.33, 5.0$ GeV, as well as scaling with $J=0$.

Following Reference \cite{GR98}, we have tried to rescale the results given
in a consistent manner; namely, $n_f = 4$, $\LQCDfo = 200$ MeV, and
$\mu = 5$ GeV are used
to scale $\alpha_s$ up to the matching point at 5 GeV, with $\alpha_s$ given
by the NLO expression \cite{PDG98}.  We match $\alpha_s$
between $n_f=4$ and $n_f = 5$ at the $b$ mass of 5 GeV, and set $n_f=5$ in 
the exponent of $\alpha_s$, $-2/\beta_0$.  This results in a scaling of about
1.6 between the raw $\bb(m_b)$ and $\bnlo(m_b)$.

The results on $\bbd$ from calculations using static $b$ quarks were
omitted from recent
averages due to an extremely large variation in the results when the NLO
corrections were included in three supposedly equivalent ways
\cite{Flynn97,GM97}.
Recently, several clarifications to this process were
made by Gimenez \& Reyes \cite{GR98}.
First, the calculation of the lattice
operator
$\hat{O}_{LL} = \bbar\gamma^\mu(1-\gamma_5)q\bbar\gamma_\mu(1-\gamma_5)q$
for SW (clover) actions used previously was incorrect,
except for those groups using Wilson actions for the light
quarks.  Second, the
importance of using tadpole-improved light quarks was shown.  Third, some
inconsistencies in the techniques used to match operators to the continuum
were pointed out.
They have reanalyzed their data \cite{GM97} along with data from the UKQCD
group \cite{UKQCD96}, which also used SW light quarks. The results are
now much more stable under variations in the matching scheme, as well as being
in good agreement with the data from simulations that used propagating heavy
quarks \cite{ELC92,JLQCD95,BS}.

Figure~\ref{fdsAvg} shows the average for $\bnlod$ from the analyses in
References \cite{BDHS88,ELC92,GR98,BBS,JLQCD98-2,JLQCD95,UKQCD96,BS}, which is
$\bnlod = 1.37\pm0.08$.  A common systematic error of 0.07, due to
uncertainties in the renormalization, has been included.

The ratio $\bbd/\bbs$ is expected to be very close to one, and has been
estimated as $\bbd/\bbs = 1.01\pm0.01$ \cite{GM97}

\section{ANATOMY OF A TIME-DEPENDENT OSCILLATION MEASUREMENT}
\label{AnatomySection}

We now turn to the experimental side of explicitly measuring the
time dependence of the $B\Bbar$ flavor oscillations.  This section
summarizes the basic analysis elements and techniques that
have been used to produce the results presented in the following two sections.

The basic steps of any time-dependent mixing measurements are evidently as
follows:
\begin{enumerate}
\item Find the interaction point at which the $B$ mesons were produced
(primary vertex) and decay point of one of the mesons (secondary vertex), and
determine the decay distance {$L$}.
\item Measure the (presumed) $B$ meson's momentum to turn the decay distance $L$
into the proper time lived, $t$.
\item Tag the flavor of the (presumed) $B$ at its production point,
and determine the probability the tag was correct, $P_r^P(t)$.
\item Tag the flavor of the $B$ at its decay point, and determine
the probability the tag was correct,  $P_r^D(t)$.
\end{enumerate}

\begin{figure}[htb]
\centerline{\epsfxsize=4.9in \epsfbox{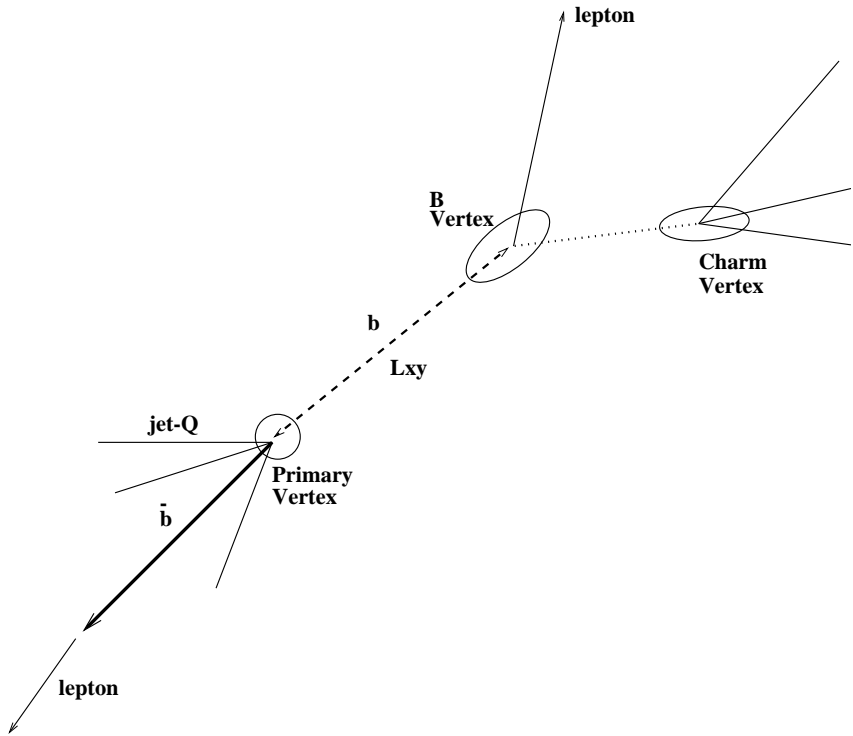}}
\caption{Schematic of a typical $B$ event.}
\label{lxyfig}
\end{figure}

The advent of high-precision silicon tracking systems allows the measurement
of the decay distances of $B$ mesons, which are typically of order
$\gamma\beta c \tau_B \approx \gamma\beta \times 470 \micron$.  With
boost factors 
ranging from $\beta\gamma \sim 3$ at CDF to $\beta\gamma \sim 7$ at LEP and
SLD, decay lengths in
the millimeter range are typical, which are easily discerned by modern detectors.
For detectors with three-dimensional precision tracking (i.e. double-sided silicon or pixel
detectors)
$t = L/\gamma\beta c = L M_B/p_B$. Experiments with single-sided silicon
strip detectors measure the flight distance projected onto the plane perpendicular to
the beamline ($xy$, or $r\phi$ plane),
$\Lxy$, and convert to $L$ using the estimated polar angle of the meson,
$t = \Lxy / (\gamma\beta c \sin\theta) = \Lxy M_B / p_t$.

The uncertainty on the decay time of the meson in its rest frame is
\begin{eqnarray}
\sigma^2(t) & = & \frac{\sigma^2(L)}{\gamma^2\beta^2c^2} +
t^2\frac{\sigma^2(\gamma\beta)}{\gamma^2\beta^2}.
\label{tResolution}
\end{eqnarray}

The resolution on $L$ (or $\Lxy$) is typically $\order(300\micron)$ at LEP
and $\order(100\micron)$ at CDF, so when
divided by $\gamma$ does not contribute significantly to the error on $ct$
compared to the momentum-resolution term $\sigma(\gamma\beta)$,
which grows linearly with the decay length $L$.  This is a critical point
for measuring $\Bs$ oscillations, which have a very rapid oscillation
rate.  Once the smearing induced by the momentum resolution reaches half
the oscillation period, any hint of an oscillation signal is washed out.
With the momentum resolution achievable from
partially reconstructed decays, only the first few oscillations of the
$\Bs$ will be visible, making frequency measurement very difficult.
Fully reconstructed $\Bs$ decays, of course, have optimal momentum
resolution, but current experiments have very small data samples.

The variety of mixing measurements comes mostly from the wide range of
techniques used to identify the $b$-quark flavor at its production and
decay point.  These techniques are breifly described below.

By far the dominant production mode of $b$-type quarks is $b\bbar$
pair production, from $e^+e^-$ annihilation at LEP and SLD, and from
quark-antiquark annihilation and gluon-gluon fusion at CDF.
At collider experiments they
are generally produced roughly back-to-back, so the event can be divided into
two hemispheres (defined by plane
perpendicular to a vector $\hat{a}$ passing through the event origin)
that each contain one of the $b$ quarks.
The analyses are
insensitive to slight changes in the direction of $\hat{a}$,
and several choices for $\hat{a}$ are used, including:
the thrust or sphericity axis of the event;
the direction between the primary and secondary vertices;
or the direction of a jet associated with the secondary vertex,
defined by an appropriate clustering algorithm. The 
hemisphere with
the secondary vertex is called the vertex side, and the other is called
the opposite side.  (This distinction becomes arbitrary if
vertices are reconstructed in both the $b$ and $\bbar$ hemispheres.)

Figure~\ref{lxyfig} is a sketch of an event, showing the $b$ and $\bbar$
quarks originating from the primary vertex, a secondary vertex found
on the $b$ (vertex) side, and some possible flavor tags on the opposite side
(see below).  The hemispheres would be formed by a plane passing through
the primary vertex roughly perpendicular to the $b$-$\bbar$ axis.

\subsection{Decay-Point Flavor Tagging}
\label{decayTagSection}

The $b$-quark flavor at decay is typically identified by one of two methods,
either by the charge of the lepton($e$ or $\mu$) from a semileptonic decay
of the meson, or by
the type of $D$ meson present in a partial reconstruction of the decay.

\begin{table}[htb]
\caption{Sources of leptons from $b$ quark decay  (correct
flavor tags of the $b$ quark are denoted r=right, incorrect tags are denoted
w=wrong).}
\begin{tabular}{@{}lcc@{}}
\hline\hline
Decay chain & Lepton Source & Correct Tag? (r/w) \\ \hline
$b\rightarrow W^- \rightarrow l^-$, & Direct & r \\
$b\rightarrow c \rightarrow W^+ \rightarrow l^+$ & 
Cascade & w \\
$b\rightarrow W^- \rightarrow \overline{c} \rightarrow l^-$ &
Right-sign cascade & r \\
$b\rightarrow W^- \rightarrow \tau^- \rightarrow l^-$ &
Right-sign cascade & r \\
$b\rightarrow J/\psi X \rightarrow l^\pm$ & Both-sign cascade & r/w \\ \hline
\end{tabular}
\label{semilSource}
\end{table}

A lepton from a semileptonic $b$ decay can have the sources given in
Table~\ref{semilSource}.  The main contributions come from the first three
components.  The relative rates of leptons from direct and
wrong-sign sequential decays are approximately equal \cite{PDG98}
The right-sign sequentials $b\rightarrow\overline{c}\rightarrow l^-$
contribute approximately the same amount as the wrong-sign \cite{PDG98},
so the lepton tags the $b$-quark correctly approximately 2/3 of the time.
However, with very simple kinematic cuts, the contribution from the sequential
decays can be greatly reduced.  In particular, the momentum spectrum of the
sequential leptons is softer than that of the direct leptons, as they arise
further down the decay chain.

\begin{figure}[htb]
\centerline{\epsfxsize=\hsize \epsfbox[80 300 550 750]{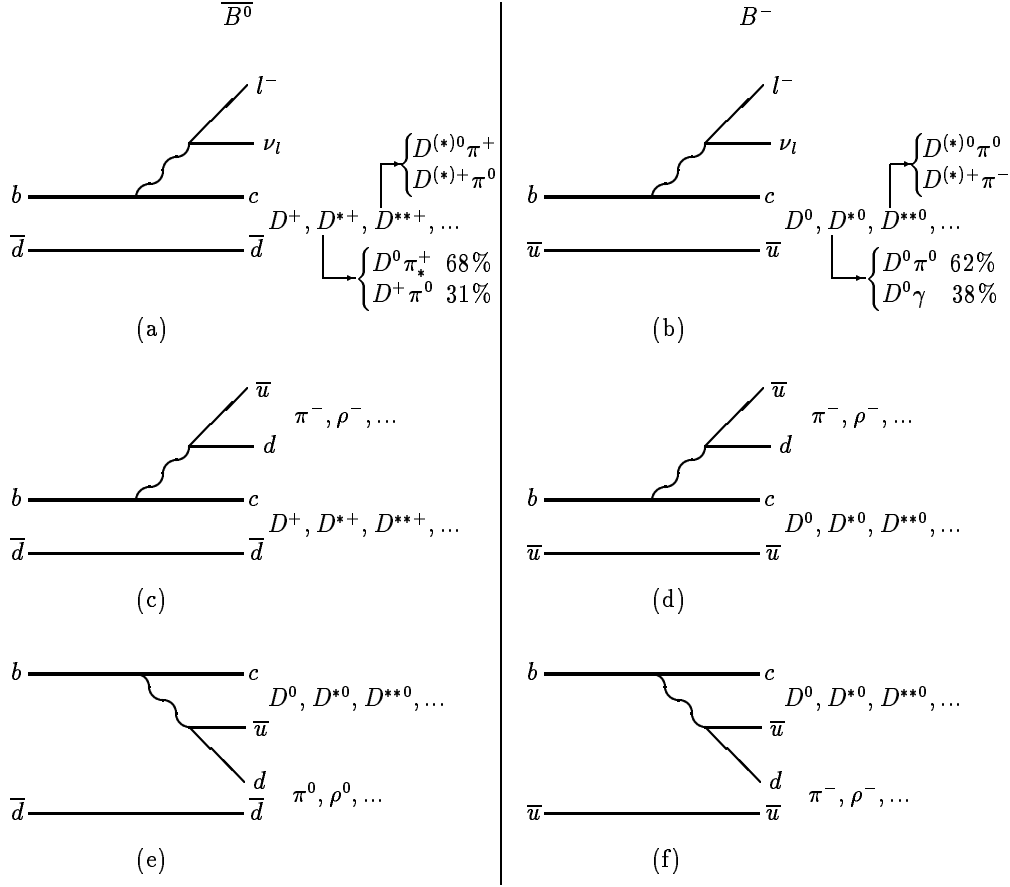}}
\caption{Diagrams for semileptonic and hadronic decay of $\Bzbar$ and $B^-$
mesons (soft gluons are suppressed).}
\label{BtoDstar}
\end{figure}

The second common method of tagging the decay-point $b$ charge is by either
partially or fully reconstructing the charmed meson from the $B$ decay.
In particular, the decays
$\Bzbar\rightarrow \Dsp X$, ($\Dsp\rightarrow D^0 \pi^+_*)$ and 
$\Bzbar\rightarrow D^+ X$ allow the separation of $b$ from $\bbar$ quarks.
($\pi_*$ denotes the soft pion from the $D^*$ decay.)
The sign of the $b$-quark charge is the same as the sign of the kaon charge
in the $D$ meson
decay, and it is opposite that of the $\pi_*$ from the $\Dsp$ decay.

Figure~\ref{BtoDstar} shows the decay modes of the $\Bzbar$ and $B^-$ assuming
final-state interaction effects are small (i.e. factorization approximately
holds). Soft gluons running around the diagrams are not shown.
In the semileptonic modes (Figure~\ref{BtoDstar}{\em a,b}),
note that the $B^-$ produces a $D^0, D^{*0}, D^{**0} ...$, and that the
$D^{*0}$ decays only to $D^0$, not $D^+$.  Thus the only contamination
of $B^-$ in the $\Dsp$ sample is through decays involving $D^{**}$.
In contrast, the $\Bzbar$ decay can produce a $D^+$ directly, or through a
$\Dsp$ with a missing $\pi^0$ and a $\Dsp$ detected in the $D^0\pi^+_*$
configuration.  The decays involving $D^{**}$ will end up scattered around
the various modes, as in the $B^-$ case.

Hadronic decays show the same basic behavior, with the addition of the
internal, color-suppressed diagrams (Figure \ref{BtoDstar}{\em e,f}).
Note that the type of charm
decay product for these diagrams is the same for the $B^-$ decays.  The
$\Bzbar$, however, would tend to produce $D^0, D^{*0}, ...$, identically to
the $B^-$.  This breaks the analogy between the semileptonic and hadronic
modes, resulting in a slightly different fraction of $\Dsp$ coming from
$\Bzbar$ in the two cases, but the
effect is not large.

Diagrams like those in Figure~\ref{BtoDstar}{\em c} and \ref{BtoDstar}{\em d}
can also have the external $W$ decaying into
a $\cbar s$ pair, producing a charm of the same charge as the $b$ quark.
Fortunately, the branching ratio of
$B\rightarrow D^{(*)}\overline{D_s}^{(*)}$ is only
$\order(5\%)$.

Evidently, requiring a reconstructed $\Dsp$ (with or without an associated
lepton) in the final state not only flavor-tags the $b$ quark at the decay
point but also greatly enriches the sample's $B^0$ fraction.  Although
no measurement of the fraction of $\Dsp$ that comes from $\Bzbar$
(as opposed to $B^-$)
is available for hadronic decays, it has been measured to be, e.g.,
$84\pm 9$\% by OPAL \cite{Opal93} in the semileptonic mode. 

The advantages to the $\Dsp$ tag are its excellent purity and its
ability to produce event samples with very high $B^0$ content.  Its obvious
drawback is its efficiency, which is rather low owing to the exclusive
final state.

A third method of tagging the $b$-quark flavor at the decay point is via
the charge of the kaon(s) in the decay chain.  The decay of the $b$ quark
proceeds $b\rightarrow c \rightarrow s$.  Hence a $K^-$ tags a $b$ quark
and a $K^+$ tags a $\bbar$. Using inclusive kaons will greatly increase the
efficiency of this tagging method over the $D^*$ method, but of course there
 are other sources of kaons
besides the $D$-meson decay, which will dilute the
tag's effectiveness.  In addition, the dramatic increase in the $B^0$ content
of the sample is lost.

Finally, a method based on a charge dipole is used by SLD  (see
Section~\ref{InclusiveSection}).

\subsection{Production-Point Flavor Tagging}
\label{prodTagSection}

The flavor tagging of the $b$ quark is more
problematic at its production point than at its decay point.
Broadly speaking, there are two
strategies: same-side techniques, which use information from
the vertex hemisphere to determine the initial flavor of its $b$ quark,
and opposite-side techniques, which measure the
charge of the $b$ quark in the opposite hemisphere (at either its decay
or production point) and infer the vertex-side $b$ charge from that.

Opposite-side tags come in two types: lepton tags and jet charge.
Lepton tagging is essentially the same as described above for decay-point
flavor tagging, although the selection criteria on the lepton may differ for
the two sides.  This type of tagging has high purity 
but suffers from a low efficiency because the semileptonic branching
fraction of $B$ mesons is only $20\%$; in addition, detection and
selection criteria are imposed.

The jet-charge method relies on a weak correlation between the momentum
and charge of tracks in a $b\bbar$ event.  For example, at SLD or LEP,
$b$ and $\bbar$ quarks are generally produced back-to-back each with
momentum of 45 GeV.  Thus the $b$ quark hemisphere starts off with a
net $-1/3$ charge.  As the fragmentation proceeds, 
the $b$ quark gradually loses its momentum
as it radiates off gluons; however,
the $-1/3$ charge continues to
be associated with a higher-momentum particle, whereas the charge required to
keep the hemisphere charge integral is in lower-momentum particles.
Experimentally, a momentum-weighted charge distribution is used to
form a variable
to discriminate between $b$ and $\bbar$ quarks. 
Section~\ref{lJetQSection} provides details on the algorithms used to measure
the jet charge.

\begin{figure}[htb]
\epsfxsize=5cm
\centerline{\epsffile{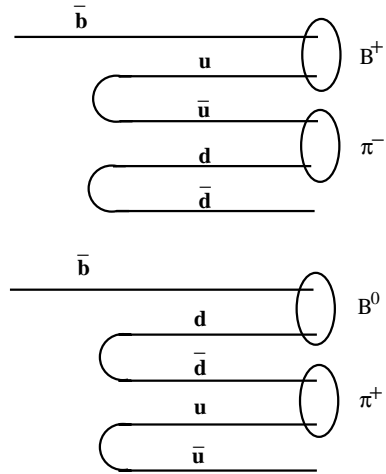}}
\caption{Schematic of $B^0$ and $B^+$ fragmentation chains}
\label{sstpic}
\end{figure}

Two types of same-side tagging (SST) are commonly used.
One method relies on an even
more specific charge correlation than the jet charge \cite{Gronau93}.
Figure~\ref{sstpic} shows
a schematic of a $\bbar$-quark fragmentation chain.  If the first
quark-antiquark
pair radiated is a $u\ubar$, the $\bbar$ becomes a $B^+$, and the first charged
pion in the fragmentation chain is a $\pi^-$.  If the first $q\qbar$ pair is
a $d\dbar$, the meson is a
$B^0$ and the first charged pion is a $\pi^+$.  Hence
for a $B^0/\Bzbar$ sample, a correlated pion $\pi^+$ tags a $B^0$ and $\pi^-$
tags a $\Bzbar$.  The correspondence is opposite for charged $B^\pm$.  This
tagging technique, then, requires a rather clean sample of $B^0/\Bzbar$
events, which rules out applying it to purely leptonically tagged events.
The difficultly lies in selecting the correct pion from all the
fragmentation tracks.  In the case of a fragmentation into a $\Bsbar$ meson,
the first fragmentation track would be a $K^-$.

The second same-side method is unique to SLD, and relies on the large
longitudinal polarization that the SLC was able to achieve.  The 
differential cross section
for producing a $b$ (as opposed to $\bbar$) quark in a
polarized $e^+ e^-$ interaction is
\begin{eqnarray*}
\sigma_b(\cos\theta) \equiv \frac{d\sigma_b(\cos\theta)}{d\cos\theta}&\propto &
(1-A_e P_e) (1+\cos^2\theta) + 2 A_b (A_e - P_e) \cos\theta,
\end{eqnarray*}
\noindent where $\theta$ is the angle between the $b$ quark and the incoming
electron direction. $P_e$ is the longitudinal polarization of the electron
beam ($P_e>0$ for positive helicity, right-handed polarization),
and $A_f = 2 v_f a_f / (v_f^2 + a_f^2)$ where $a_f, v_f$ are the axial
and vector coupling of the fermion $f$ to the $Z^0$.  Thus the probability
a $b$-type quark produced at angle $\theta$ is a $b$, rather than $\bbar$
is simply
\begin{eqnarray*}
\frac{\sigma_b(\cos\theta)}{\sigma_b(\cos\theta)+
\sigma_{\overline{b}}(\cos\theta)} & = & \frac{1}{2} +
\frac{A_b (A_e - P_e) \cos\theta}{(1-A_e P_e) (1+\cos^2\theta)} \equiv
\Ppol(\cos\theta)
\end{eqnarray*}
(Some papers have the sign of $P_e$ swapped and/or measure $\theta$ from the
positron direction.)
The polarization of the electron
beam is randomly chosen between the positive and negative helicity
states on a pulse-by-pulse basis, and the signed polarization
is recorded in the data stream.
The standard model values for the asymmetries are $A_e=0.155$ and $A_b=0.94$
\cite{PDG98}.  Note that the discriminating power 
that survives at $P_e=0$ because of the residual forward-backward $b$ asymmetry
has a maximum amplitude of $A_e A_b \sim 0.1$,
whereas for large $P_e$ this amplitude is approximately $P_e A_b \sim P_e$.
Hence SLD can tag the flavor of the $b$ quark at production simply by
measuring the direction of the $B$-meson system. The probability that the
sign assigned is correct, for a given angle, is simply the
$\Ppol(\cos\theta)$ above.
It is the large (63 and 77\%) polarizations achieved at the SLC that make this
flavor tag effective, with a mean right-tag probability $<\!\Ppol\!>$
of 62\% and 76\% respectively.  
Figure~\ref{SLDtags} shows a typical distribution of $\cos\theta$, along with
the underlying predictions for $b$ and $\bbar$ quarks given by the
Monte Carlo simulations.

Finally, many experiments combine multiple tags, both same-side
and opposite-side, to maximize the probability of tagging correctly.
For example, 
the SLD group uses the event-by-event $\Ppol$
in all their analyses, along with a second flavor-tagging technique
(e.g. jet charge).
The two tags, $P_1$ and $\Ppol$, are combined into the final right-tag
probability
\begin{eqnarray}
P_{\rm comb} & = & \frac{P_1 \Ppol}{P_1 \Ppol + (1-P_1)(1-\Ppol)}.
\end{eqnarray}

\subsection{Fitting Procedure}
\label{mixFit}

The signature for $B$-meson mixing is that the sign of the $b$ quark inferred
at the decay point, $D$, is opposite to that inferred at production,
$P$.  
In the following, the notation $DP+$ refers to events in which the experimental
flavor-tag signs for the vertex-side $b$ quark at production and decay
 are the same,
(i.e. the event looks unmixed) and $DP-$ is used when they are opposite (i.e.
the event looks mixed).  For example, in an analysis that uses leptons
 to tag both the production
and decay charges, $DP+$ events are those with opposite-sign leptons.
The event sample is then split into two components,
with differing time dependences $N_{DP+}(t)$ and $N_{DP-}(t)$.

The basic variable to be measured, then, is the time-dependence of the
decay and production flavor tags' charge correlations,
 $q_c(t)=\mbox{sign}(DP)(t)$.
The mean value of this variable, binned in time $t$, is the standard
measured asymmetry:
\begin{eqnarray*}
<\!q_c(t)\!> \equiv \Am(t) & = &
\frac{N_{DP+}(t)-N_{DP-}(t)}{N_{DP-}(t)+N_{DP-}(t)}.
\label{basicAsymm}
\end{eqnarray*}

Let us first consider this time-dependent charge correlation on a sample
of pure $B^0/\Bzbar$.
The numbers of events in the $DP+$ and $DP-$ samples at a given time are
\begin{eqnarray*}
N_{DP+}(t) & = & N\left(P_u(t) [P_r^P P_r^D + (1-P_r^P)(1- P_r^D)]\right. \\
 & & \left. + \hspace{6.3pt}P_m(t) [P_r^P (1-P_r^D) + (1-P_r^P) P_r^D]\right) \\
N_{DP-}(t) & = & N\left(P_m(t) [P_r^P P_r^D + (1-P_r^P)(1- P_r^D)] \right. \\
& & \left. + \hspace{10.2pt}P_u(t) [P_r^P (1-P_r^D) + (1-P_r^P) P_r^D]\right),
\end{eqnarray*}
\noindent where $N$ is the total number of events in the sample,
$P_u$ and $P_m$ are the unmixed and mixed probabilities from Equation~\ref{Pmix},
and $P_r^P(P_r^D)$ is the
probability that the production (decay) flavor tag got the sign of the
$b$ quark correct.  The asymmetry is
\begin{eqnarray*}
\Am(t) & = &
 (2P_r^P-1)(2P_r^D-1)\frac{P_u(t) - P_m(t)}{P_u(t) + P_m(t)}
 =   {\cal D}_P {\cal D}_D\cos\delm t.
\end{eqnarray*}
That is, the measured asymmetry $A_m$ is reduced from the true asymmetry
$A(t)=\cos\delm t$ by the production and decay tag
dilutions,
${\cal D}_P = 2P_r^P-1$ and ${\cal D}_D = 2P_r^D-1$.

Now suppose we have a variable $\alpha$,
 calculated event-by-event, on which the production tag probability depends.
If we break up the sample into bins of $\alpha$, with $N(\alpha)$ events in
a given bin, then (setting $P_r^D=1$ to reduce algebraic clutter) we have
\begin{eqnarray*}
N_{DP+}(t,\alpha) - N_{DP-}(t,\alpha) & = & N(\alpha) (2P_r^P(\alpha)-1)
\left[P_u(t)-P_m(t)\right]
\end{eqnarray*}
The error on this difference is $\sqrt{N(\alpha)}$, independent of
the dilution $\cal{D}(\alpha)$, while the value of the difference scales with
$\cal{D}(\alpha)$.

If we perform the time-dependent fit separatly for each $\alpha$ bin, the
relative errors on the fits will clearly scale as
$\sigma(\alpha) \sim 1/[(2P_r(\alpha)-1) \sqrt{N(\alpha)}]$.
These (statistically)
independent measurements are optimally combined using the weighted mean.
This is equivalent to weighting each event in the sample by its dilution,
$2P_r^P(\alpha)-1$,
and fitting once on this weighted sample.  
A similar result holds for the case with both $P_r^P$ and $P_r^D$
depending on a set of event-by-event variables $\vec{\alpha}$.

Thus in a time-binned fit to the asymmetry, the optimal procedure is to
weight each event's contribution by the estimated dilution for that event.
Most of the analyses described here use an unbinned likelihood fit, applied
to the $N_{DP+}(t)$ and $N_{DP-}(t)$ samples simultaneously.
In this case,
employing the event-by-event right-tag probability $P_r^P(\alpha)$ is optimal.

The likelihood functions for the $DP+$ and $DP-$ subsamples, or the binned
asymmetry, are constructed based on predicted proper time distributions
of the data sample's components.  These typically include
unmixed $B^0$ decays, mixed $B^0$ decays, $B^+$, $b$ baryons,
unmixed $\Bs$, mixed $\Bs$, cascade decays from all the above $b$ sources,
charm background, and fake or combinatorial background.

The measured decay time in an event differs from the true decay time
because of imperfect reconstruction of the decay length and of the
$B$ momentum, as shown in Equation~\ref{tResolution}.
The probability of reconstructing a decay time $t$ given that
the true decay time was $t^\prime$ is described by the resolution function
$R(t,t^\prime)$.  This is convoluted with the predicted true distribution,
e.g., $P_u(t)$ or an exponential, to get the expected measured distribution.
The vertexing algorithm might also introduce a $t$-dependent efficiency
(actually $L$-dependent, which roughly translates into $t$-dependent), which
is sometimes accounted for implicitly in $R$, or by multiplying the
above convolution with an efficiency curve.

$R$ is typically parameterized as a sum of two or three gaussians, with
separate parameterizations over several bins of decay time to allow for
the increased smearing at longer times.  If the vertex efficiency is included,
the gaussians can differ above and below the mean to fit, e.g., a decrease in
the probability to reconstruct a vertex near the primary vertex.

Alternatively, some analyses fit directly to the predicted distribution
as a function of decay length, in which case the resolution function is
parameterized as a function of decay length rather than time.

\section{$B^0$ MIXING}
\label{BdSection}

The time-dependence of $B^0-\Bzbar$ oscillations was first observed
by ALEPH \cite{Aleph93}.  Since then, a number of analyses have measured
the oscillation frequency, as summarized in Table~\ref{BdTable}.
This section presents
an overview of these measurements
grouped by the techniques used to flavor tag the initial and final charges
of the $b$-quark.  Several preliminary results have been included, which
are indicated on the summary in Figure ~\ref{BdSummary}.

\begin{table}[ht]
\caption{Index of the various flavor-tagging combinations used by each
experiment to measure $B^0$ mixing. See the references indicated for
details on individual analyses.}
\begin{tabular}{@{}l|cc|cc|ccc|cc@{}}
\hline\hline
Vertex tag & \multicolumn{2}{c|}{$l$} & 
\multicolumn{2}{c|}{$D^{(*)}$} & \multicolumn{3}{c|}{$D^{(*)} l$} &
JetQ & $K$/dipole \\ \hline
Prod. Tag & $l$ & JetQ & $l$ & JetQ & $l$ & JetQ & SST & JetQ &
JetQ  \\ \hline
ALEPH  &\cite{Aleph97-1} & \cite{Aleph97-1} & \cite{Aleph97-1} &
\cite{Aleph97-1}&  & & &\cite{Aleph97-2} & \\ 

CDF  & \cite{CDF99-3},\cite{CDF99-2},\cite{CDF3791} & \cite{CDF99-2}& 
\cite{CDF4526} & & \cite{CDF99-5} & & \cite{CDF99-1}& & \\ 

DELPHI  &\cite{Delphi97-1} & \cite{Delphi97-1}& \cite{Delphi97-1}&
&\cite{Delphi97-1} & & & & \\ 

L3  & \cite{L398}& \cite{L398}& & & & & & & \\ 

OPAL & \cite{Opal97-2}& \cite{Opal97-1}& \cite{Opal96}& & &\cite{Opal96} &
 & & \\ 
SLD  & & \cite{SLD96-1},\cite{SLD96-2}& & & & & 
& & \cite{SLD96-3} \\ \hline
\end{tabular}
\label{BdTable}
\end{table}

\subsection{Detector Overviews}
What follows is a brief overview of the six collaborations' detectors,
focussing on the elements which are most relevant to mixing analyses.
More detailed descriptions are available for ALEPH \cite{AlephNIM95},
CDF \cite{CDFNIM88}, DELPHI \cite{DelphiNIM96}, L3 \cite{L3NIM90},
OPAL \cite{OpalNIM91} and SLD \cite{SLDNIM97}.

All of the detectors operate at colliders and have similar cylindrical
geometries with some type of end plug to cover most of the solid angle.
Five of the experiments (ALEPH, DELPHI, L3, OPAL at LEP and SLD at SLC)
operate at $e^+e^-$ machines running at the $Z$-pole, $\sqrt{s} = 91$ GeV.
CDF operates at the Fermilab
$\pbar p$ collider, which has an energy $\sqrt{s} = 1.8$ TeV.

Time-dependent oscillation measurements clearly require precision
tracking detectors, and
it was only after the inclusion of various silicon strip and pixel devices
that each experiment was able to perform these analyses. 
ALEPH, DELPHI and OPAL had silicon detectors in place since
1991.  The ALEPH detector measures both the $r\phi$ and $rz$ views --
OPAL(1993), DELPHI(1994) and L3(1994) soon added vertex detectors which
could do the same.
SLD had a CCD pixel device available since 1994.

Throughout the following, $d$ denotes the impact parameter, or
distance of closest approach,
of a track to the beam direction.  The momentum $p$ of a track
projected onto the $r\phi$ plane is denoted $p_t$.

\subsubsection{PRIMARY VERTEX RECONSTRUCTION}
The position of the interaction point of an event (primary vertex) is an
important part of determining the decay length of the $b$ particle.  The
dimensions of the beam cross section (i.e. in the $x-y$ plane) are small for
all the accelerators used in these analyses: 
$150\mu\mbox{m}\times 10\mu\mbox{m}$ at LEP,
$25\mu\mbox{m}\times 25\mu\mbox{m}$ at Fermilab, and
$1\mu\mbox{m}\times 1\mu\mbox{m}$ at the SLC.  Each experiment tracks the
mean position of the beam centroid, giving an accurate estimate of the
interaction point. In addition, some experiments fit for a vertex
event-by-event, combining this mean position with tracks which have a high
probability of originating there.  This helps reduce the relatively large
uncertainty on the $x$ position of the primary vertex at the LEP experiments.

\subsubsection{PARTICLE IDENTIFICATION}
The analyses presented here rely heavily on detecting and tagging $B$ mesons
through their semileptonic decays to electrons and muons.  These are identified
using standard detector elements such as tracking chambers, electromagnetic
and hadronic calorimeters, and muon chambers (details can be found
in the references).  All experiments remove electron candidates from pair
conversion.

The SLD and LEP experiments have data acquisition systems which can
record virtually all of the relatively low-rate hadronic $Z$ events.
CDF, however, operated with an interaction rate of $\sim 300$ kHz and
simply could not record every event; only those passing a multistage trigger
were selected for analysis.
This trigger could identify those $b$ events that contained a semileptonic
decay by detecting the lepton.  
However, the fake background at low momentum is very large, and a
fairly high lepton $p_t$ threshold of 7.5 GeV was required so that the 
trigger would not saturate the data acquisition bandwidth.
Dedicated dilepton triggers, either $\mu\mu$ or $e\mu$ were also implemented.
For dileptons, the momentum cuts were set much lower (2.0 GeV for the muon and
5.0 GeV for the electron), since requiring two
leptons significantly reduces the fake background.  The data sets collected
with these triggers are referred to as high-$p_t$ single-lepton and low-$p_t$ 
dilepton, respectively.

In addition to leptons, fragmentation kaons and pions are
used by many experiments to
tag the initial $b$-quark charge, and kaons from $B$ decay are used by SLD to
tag the final charge (see Section~\ref{AnatomySection}).
ALEPH, CDF and OPAL identify particles based on
ionization (dE/dx) deposits in their tracking chambers, while
DELPHI and SLD use their ring-imaging Cerenkov detectors.

\subsection{Dilepton Analyses}
\label{dilSection}

\subsubsection{EVENT SELECTION}

The dilepton analyses use a lepton tag as the decay-point tag, as described
in Section~\ref{decayTagSection}, and an opposite-side lepton tag as
the production-flavor tag.

Like all opposite-side-type analyses, the dilepton analyses divide
candidate events
into two hemispheres using one of the methods discussed in
Section~\ref{AnatomySection}.
Events are selected which contain two leptons ($l=e,\mu$)
in opposite hemispheres and
have a secondary vertex associated with at least one of the leptons.
The details of the vertexing algorithms are given below.
The leptons are required to pass the cuts on their total momentum (or momentum
transverse to the beamline, for CDF) listed in Table~\ref{dileptonCuts},
which also shows the number of events and number of secondary vertices
found in each analysis.
In addition, most analyses reject leptons with a $\ptr$ below the threshold
given, where $\ptr$ is the momentum of the lepton transverse to the direction
of the associated $b$ jet (with the lepton removed).  OPAL rejects leptons
based on the output of a neural net, described below.

CDF has three dilepton analyses -- two based on the data recorded with the
$\mu\mu$ and $e\mu$ triggers and the third based on events passing
the single-lepton trigger that have a secondary vertex associated with
the lepton.  From this sample, events
with another so-called soft lepton opposite the trigger lepton are
accepted (the algorithm for selecting the lepton is based on the soft
lepton tag, or SLT, used in the top quark search).

\begin{table}[ht]
\caption{Summary of selection requirements on the
momentum and $\ptr$ of the leptons for the
dilepton analyses, along with the number of events and reconstructed
vertices for each experiment.  $<\!P_r\!>$ is the mean right-tag probability
for the lepton opposite the vertex hemisphere.}
\begin{tabular}{@{}llc@{}c@{}}
\hline\hline
Experiment & Lepton requirements (GeV) &
 Events/Vertices & $<\!P_r\!>$ \\ \hline
ALEPH \cite{Aleph97-1} & $p>2.0(e),3.0(\mu); \ptr>1.25$ & 5957/9710 & 0.83\\ 
CDF($\mu\mu$) \cite{CDF99-3} &  $p_t>3.0, \ptr>1.3$ & 5968/5968 & 0.74 \\ 
CDF($e\mu$) \cite{CDF3791} & $E_t>5.0(e),p_t>2.5(\mu), \ptr>1.25(\mu)$ &
10180/11844 & 0.74 \\ 
CDF(SLT) \cite{CDF99-2} & $p>7.5(\mbox{trig}), 2.0(\mbox{tag}), \ptr(\mbox{weight})$
& 12700/12700 & 0.73 \\
DELPHI \cite{Delphi97-1} & $p>3.0,\ptr>1.2$ & 4778/4778 & 0.89 \\ 
L3 \cite{L398} & $p>3.0(e),4.0(\mu),\ptr>1.0$ & 1490/1928 & 0.84 \\ 
L3(IP) \cite{L398} & same & 2596/$-$ & N/A \\ 
OPAL \cite{Opal97-2} & $p>2.0,\mbox{neural net}$ &  5357/8544 & 0.85\\ \hline
\end{tabular}
\label{dileptonCuts}
\end{table}

\subsubsection{SECONDARY VERTEX RECONSTRUCTION}
Because the lepton flavor-tagging method identifies only one possible
track from the
$B$ decay, all experiments use an inclusive vertex-finding technique
to locate the $B$ decay point. 

  ALEPH \cite{Aleph97-1} approximates that all of the
$B$-decay products except the lepton come from the tertiary charm vertex
(commonly loosely referred to as the $D$ vertex).
They form a grid of possible charm vertex positions.
If a track's
impact parameter (excluding
the lepton) is within $3\sigma$ of a candidate vertex, it is assigned to the
vertex.  ALEPH calculates the $\chi^2$ difference between assigning all tracks to the primary
vertex and allowing some to come from the charm vertex, and the
point which maximizes this difference is chosen as the charm vertex.
A charm pseudotrack is formed, which passes through this vertex, and its
direction is given by the sum of the momentum of the charged particles in the
vertex.  This pseudotrack is then vertexed with the lepton to give the
position of the secondary, $B$ vertex, which is finally projected onto the
$b$-jet direction to give the decay length $L$.
This technique allows a $B$ decay length to be determined for every event and
is appropriate for samples with high initial $b$ purity, which do not need to
rely upon the presence of a well-separated secondary vertex to increase the
$b$ purity.

On the other hand, at CDF, the presence of a secondary vertex is a powerful
means of rejecting large non-$b$ backgrounds.
CDF uses two techniques, both of which involve the projection of the tracks
 onto
the $r\phi$ plane.  The dimuon analysis \cite{CDF99-3} exploits a
correlation between the
impact parameter, $d$, of a track, and its azimuthal ($\phi$) angle.
Tracks coming from a secondary vertex form a line in the $d\phi$ plane,
whereas tracks from the primary vertex have both small $d$ and no $d\phi$ correlation.
A cluster of tracks from a secondary vertex is formed if at least three tracks
with significant separation from the primary vertex ($d/\sigma_d>2$) forming a line
in the $d\phi$ plane can be found.  All tracks in this cluster except the muon
are vertexed to form a presumed charm vertex, and a pseudotrack is formed,
as above, from the summed momentum of these tracks.   This is intersected
with the muon track to give the $b$ decay vertex position, and
hence the flight distance projected onto the $r\phi$ plane, $\Lxy$.

The second method is used for the high-$p_t$-SLT \cite{CDF99-2}
and $e\mu$ analyses \cite{CDF3791}.
First, candidate tracks from the $b$ jet are selected if they have significant
impact parameters to the primary vertex (this is not required for the lepton
track which is presumed to come from the $b$ decay).  Combinations of
at least three tracks consistent with coming from the same point are candidate
vertices.  If no such vertex is found, stricter impact parameter and $p_t$ cuts
are imposed.  All tracks are combined to a common vertex and tracks which
contribute too much to the vertex $\chi^2$ are removed. The process is 
repeated until a good fit is obtained.  This technique does not
attempt to separate a tertiary charm vertex from the $b$ vertex, but
instead combines all tracks into one.  The efficiency of these
algorithms necessarily falls off steeply as the
secondary vertex approaches the primary vertex.

DELPHI \cite{Delphi97-1} uses yet another method, effective
on high $b$ purity samples,
which attempts to identify tracks from the
charm vertex by mass rather than position.  Excluding the lepton, they
cluster particles in the jet associated with the lepton using a jet
clustering algorithm optimized for this purpose (all tracks are given the
pion mass).  Within a cluster, particles
are ordered by decreasing values of their pseudorapidity relative to the
cluster direction.  Tracks with the largest pseudorapidity and $p>500$MeV
are kept until the mass of the resulting system exceeds 2.2 GeV,
and any such system with a large angle ($>500$ mrad relative to the $b$ jet)
is discarded.  The remaining tracks in the cluster are vertexed (in the
$r\phi$ plane), and the charm pseudotrack is formed and intersected with
the lepton to give the $r\phi$ projection of the
$b$ decay point.  The decay point with the largest significance of separation
from the primary vertex is chosen.  The procedure is then repeated once,
using the charm vertex tracks as the cluster seed and including
neutral particles in the $b$ jet, to obtain the $b$ decay length, $\Lxy$.

L3 \cite{L398} uses a track-based approach similar to CDF.
They search for a 
secondary vertex using the lepton and tracks that
are not consistent with coming from the primary vertex.  
An acceptable vertex
can be formed in approximately 70\% of the events, with a loss of efficiency
when the secondary vertex is near the primary.  L3 also has a unique analysis
that does not attempt to find a secondary vertex at all; rather it looks at
the charge correlation of the two leptons as a function of the impact
parameters to the beam axis (IP) of the leptons.

OPAL \cite{Opal97-2} uses a technique in which tracks, ordered by decreasing
significance of separation from the primary, are intersected (in the $r\phi$
plane) with the lepton track to form secondary vertex seeds.
Additional tracks that are more consistent with the secondary seed than the
primary vertex are added, and a candidate secondary vertex is chosen based on
its position and the number of associated tracks.  Quality requirements
based on the error on the vertex position, the invariant mass of the
tracks in the vertex, etc., reduce the efficiency to approximately 70\%, though
in a decay-length-independent manner.

\subsubsection{$B$ MOMENTUM DETERMINATION}

ALEPH estimates the $B$ meson momentum as follows.
First, the (vector) sum of the momentum of the lepton and charm vertex
tracks is formed.
 Second, a fraction
of the neutral energy in the hemisphere, determined from Monte Carlo studies
to be 0.68, is added to this momentum.
Finally, the momentum of the neutrino is approximately accounted for by
adding the missing momentum in the lepton hemisphere, estimated
as the difference between the sum of all visible energy
in the hemisphere and the beam momentum.

CDF begins with the total $p_t^{\rm cl}$ in the vertex cluster, and scales
it to $p_t^{\rm cl}/{\cal K}$, where ${\cal K}$ is a
factor that is determined from Monte Carlo and parameterized as a function
of the obsevered $p_t^{\rm cl}$ and the mass of the tracks in the cluster.

DELPHI starts with an estimate of the $B$ momentum given by the total
energy-momentum reconstructed in its hemisphere minus that of all particles
not included as part of the $B$ in the vertex algorithm, which already
contains contributions from neutral energy.  A single neutrino's
energy and direction
is then solved for by assuming it is the only missing particle in the event.
The visible energy and momentum are rescaled by a
factor $\alpha=1.13$, which gives the
optimal resolution. The energy $E_\nu$ is added to the $B$ momentum if the
direction of the neutrino momentum is within 400 mrad of the $l$-charm
direction previously determined.  Finally this estimate is adjusted to
to give the correct mean value, based on the Monte Carlo prediction.

L3 does not estimate the $B$ momentum event-by-event but rather uses a
constant value of $0.85 E_{\rm beam}$.

OPAL uses a very different technique from the other experiments.  OPAL first
calculates the invariant mass of all the objects in the event that
are not in the $b$ jet, assuming all charged tracks are pions. This mass is
corrected slightly based on the inverse of the total reconstructed energy
(which should be $M_Z$) to improve the resolution.  The energy of the $b$ jet,
including the neutrino, is determined, assuming the $Z^0$ undergoes a two-body
decay to particles of mass 5.3 GeV.
Some of this energy is removed as follows.
For charged tracks, a weight is calculated based on: the probability that
the track comes from the secondary, rather than the primary, vertex; the
track's momentum,
and the angle of the track to the estimated charm-hadron direction.
For neutral clusters, the weight is based on the angle between the
cluster direction and the charm direction alone.  The weighted sum of charged
and neutral energies in the jet is then subtracted from the
full $b$ jet energy to give the estimated $B$ meson momentum.
This technique results in an excellent fractional resolution of
approximately 12\%, with
the core being closer to 7\%.
It was found that the resolution in dilepton events, with two
missing neutrinos,
was not degraded significantly.

\subsubsection{FIT METHOD AND RESULTS}

\begin{figure}[htb]
\epsfxsize=5.5in
\epsfbox[100 550 600 720]{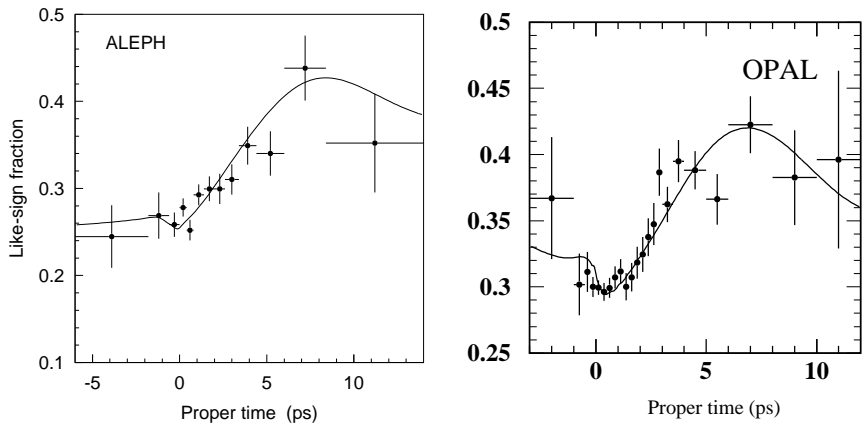}
\caption{Fraction of like-sign events vs. proper time.
{\em Left:} dilepton events from ALEPH {\em Right:} lepton-jet charge events
(selected with $|\Qt|>2$) from OPAL.
The curves represent fits to the data.} 
\label{AlephDilMix}
\end{figure}

The analysis variable for dilepton events is the product of the lepton signs
$q_c(t) = q_l^{\rm opp} \cdot q_l^{\rm vtx}(t)$.
Opposite-sign leptons make up the
"unmixed" $DP+$ sample, and same-sign leptons the $DP-$ sample.
The proper time dependence of these samples is predicted using the
resolution-smeared, efficiency-weighted distributions of the sample components,
weighted by their relative abundance, combined with the
probability of getting the correct charge assignments (as described in
Section~\ref{mixFit}).  If secondary vertices are found for both leptons, the joint
probability of finding decays at their respective decay times is constructed.
The CDF $\mu\mu$ and $e\mu$ analyses use these predictions
in a binned least-squares fit, while the rest of the analyses use
unbinned likelihood fits.

The $b$ purity of these samples is fairly high, ranging from 94\% for
DELPHI, to 98\% for ALEPH, with the CDF samples typically at 80\%.
The remaining events are either $c\cbar$, or fakes.  $c\cbar$ events
represent between
0.5\% of the sample (for ALEPH, L3 and the low-$p_t$ CDF samples) and
8\% (for CDF).  The fake background,
including fake leptons from both $b\bbar$ and non-$b$ sources, constitutes
a few percent -- up to 15\% at CDF.  Approximately 38\% of the leptons in $b\bbar$ events
are estimated to come from $B^0$ mesons.

Many of the following analyses use the opposite-side lepton as a production
flavor tag, so a few words concerning its effectiveness are in order.
The lepton opposite the vertex side can mistag the $b$ in that hemisphere
if the the $b$ hadronizes into a $B^0$ or $\Bs$ and mixes, if the
lepton comes from a cascade $b\rightarrow c\rightarrow l$ decay,
or if the lepton
is a fake.  Denoting
the fractions of the various $B$ hadrons by their light quarks as
$f_d, f_u, f_s$, and $f_{\Lambda_b}$ for all the baryons, the
contribution to the mistag rate from mixed $B^0$ and $\Bs$ mesons is
$\overline{\chi} = f_d \chi_d + f_s \chi_s $ (ingnoring for the moment those
events with vertices on both sides).  Since
$\chi_d = 0.172\pm 0.010$ \cite{PDG98}, $\chi_s\approx0.5$
(see Section~\ref{BsSection}), then
$\overline{\chi} = 0.118\pm0.006$ \cite{PDG98}, with contributions from
$f_d\approx 40\%$ and $f_s\approx 11\%$.  Thus, even though $\Bs$ mesons only
represent 11\% of the $B$ mesons produced, they contribute 
substantially to the overall mistag rate due to
their fast oscillation rate.

The other significant source of mistags are the wrong-sign cascade decays.
Table~\ref{semilSource} lists the sources of leptons which do not come
directly from the $b$ quark.  Among these, the first two make up the bulk
of the events.  The right-sign probability, then, is $P_r = 1 - \overline{\chi}
- f_{bc}^{\rm ws}$,
where $f_{bc}^{\rm ws}$ is the fraction of wrong-sign cascade
decays from the mix of $b$ hadron types in the hemisphere.  In the absence of
any cuts on the lepton, this fraction is comparable to the contribution from
direct $b$ decays.  To reduce this fraction, 
all experiments try to remove, or assign small weights to, events
with a large probability of having a lepton from a cascade decay.  Each
experiment uses the momentum of the lepton referenced to the direction of the
jet (with the lepton removed), $\ptr$.  Leptons from direct $b$ decay can be
produced with significant mometum perpendicular to the $b$ flight direction
due to the large $b$-quark mass, whereas those from the charm decay typically
have less transverse momentum.
For the same reason, cascade leptons also tend to be less well isolated from
the core of the jet, and because they come from further down the
decay chain, their mean momentum is lower.  OPAL feeds these three variables
into a neural net analysis, which is estimated
to be $40\%$ more effective at rejecting cascade decays than simple
cuts on $p$ and $\ptr$.  These cuts reduce the right-sign
cascade component, $b\rightarrow\cbar\rightarrow l$, even more, since these
events are of the $B\rightarrow D\overline{D}$ type, and the momentum available
for the lepton is even smaller than in other cascade decays.  These
end up accounting for $12-25\%$ of the cascade decays.

The SLT analysis of CDF, rather than simply cutting on $\ptr$, weights events
according to the probability that the flavor tag lepton identifies the $b$ charge
correctly (parameterized as a function of $\ptr$).  The other analyses use the
mean probability for each event.

The sample composition on the vertex side need not be the same as on the
opposite side.  For example, in the CDF SLT analysis the vertex-side lepton
has $p_t>7.5$ GeV, while the opposite-side lepton satisfies $p_t>2.0$ GeV.
In addition, vertexing algorithms with efficiencies that decrease near
the primary vertex increase the fraction of the longer-lived sample components.

Figure~\ref{AlephDilMix} shows the like-sign fraction vs. proper time for
the ALEPH analysis.  Table~\ref{domSysDil} lists the results of fits
to the oscillation frequency $\delm_d$,
along with the source and fractional variation of the parameters
which contribute most to the systematic error.  The $\Bs$ and
cascade fractions are important since they dominate the mistag rate.  The
ratio of $B^+$ to $B^0$ lifetime is also important, since this can enhance
or suppress the
fraction of $B^+$ to $B^0$ at long lifetimes, altering the
shape of the like-sign fraction and distorting the $\delm$ measurement.

\begin{table}[ht]
\caption{Fit results and dominant systematic error for dilepton analyses.}
\begin{tabular}{@{}lccc@{}}
\hline\hline
Experiment & Dominant Systematic &
$\delta(\delmd)$ (ps$^{-1}$) & $\delmd$ (ps$^{-1}$) \\ \hline

\vspace{8pt}
ALEPH & \begin{tabular}{@{}l@{}} cascade frac. (15\%) \\
$\Bs$ frac. (9\%) \end{tabular} &
$\begin{array}{c} \pm0.025 \\ \pm0.019 \end{array}$ & 
$0.452\pm0.039\pm0.044$ \\

\vspace{8pt}
CDF($\mu\mu$) & \begin{tabular}{@{}l@{}} cascade frac. (15\%) \\
$\tau_{b}$, sample comp. \end{tabular} &
$\begin{array}{c} \pm0.048 \\ \pm0.043 \end{array}$ &
$0.503\pm0.064\pm0.071$ \\

\vspace{8pt}
CDF($e\mu$) & \begin{tabular}{@{}l@{}} cascade frac. (15\%) \\
$b$-Baryon fract. (40\%)\end{tabular} &
$\begin{array}{c} \pm0.030 \\ \pm0.021 \end{array}$ & $0.450\pm0.045\pm0.051$ \\

\vspace{8pt}
CDF(SLT) & \begin{tabular}{@{}l@{}} cascade frac. (25\%) \\
$\tau_{B^+}/\tau_{B^0} (5\%)$ \\
charm dilut. (100\%) \end{tabular} &
$\begin{array}{c} \pm0.004 \\ \pm0.021 \\ \pm0.032\end{array}$ &
\begin{tabular}{@{}c@{}} $0.500\pm0.052\pm0.043$ \\ (includes Jet Q tags)
\end{tabular} \\

\vspace{8pt}
DELPHI & \begin{tabular}{@{}l@{}} cascade frac. (6\%) \\
$\tau_{B^+}/\tau_{B^0} (4\%)$\end{tabular} & 
$\begin{array}{c} \pm0.032 \\ \pm0.017 \end{array}$ & $0.480\pm0.040\pm0.051$ \\ 

\vspace{8pt}
L3 & \begin{tabular}{@{}l@{}} cascade frac. (15\%) \\ 
$\Bs$ frac. (14\%) \end{tabular} &
$\begin{array}{c} \pm0.022 \\ \pm0.015  \end{array}$ & 
$0.458\pm0.046\pm0.032$ \\ 

\vspace{8pt}
L3(IP) & \begin{tabular}{@{}l@{}} cascade frac. (15\%) \\ 
$\Bs$ frac. (14\%) \end{tabular} &
$\begin{array}{c} \pm0.037 \\ \pm0.026  \end{array}$ & 
$0.472\pm0.049\pm0.053$ \\ 

OPAL & \begin{tabular}{@{}l@{}} cascade frac. (15\%) \\ 
$\tau_{B^+}/\tau_{B^0} (6\%)$ \end{tabular} &
$\begin{array}{c} \pm0.011 \\ \pm0.023  \end{array}$ &
$0.430\pm0.043\pm0.030$ \\ \hline

\end{tabular}
\label{domSysDil}
\end{table}

\subsection{Lepton-Jet-Charge Analyses}
\label{lJetQSection}

This section introduces the second commonly used opposite-side production
 flavor tag, the jet charge, as it is used in events with a lepton-tagged
decay vertex.

\subsubsection{EVENT SELECTION AND PROPER TIME MEASUREMENT}

Events with a single lepton passing each experiment's standard lepton
selection are considered.
The lepton identification procedure follows that of the dilepton analyses
for ALEPH \cite{Aleph97-1}, DELPHI \cite{Delphi97-1}, L3 \cite{L398} and
OPAL \cite{Opal97-1}, except that ALEPH raises the momentum 
cut on the electron to $3$ GeV to match that of the muon.
The CDF \cite{CDF99-2} data sample is the same as that of
the lepton-SLT analysis described
in the previous section.  These experiments 
search for a secondary vertex in the lepton
hemisphere and convert to proper time, 
as described above for the dilepton analyses.

SLD select leptons with $p>2.0$ GeV and $\ptr>0.4$
GeV ($\ptr>0.8$ GeV) for the inclusive lepton \cite{SLD96-1} and
lepton-$D$ \cite{SLD96-2} samples, defined below.

SLD's vertexing algorithm involves combining all the (well-measured)
tracks in a hemisphere.  The position of the $b$ vertex relative to the
primary is estimated by $\vec{L} = \frac{\sum W_i \vec{X_i}}{\sum W_i}$, where
$\vec{X_i}$ is the vector from the primary vertex to the point on the
lepton track that is closest to the track of particle $i$, and
$W_i$ is a weighting function that is the product of three weights.
The first function suppresses tracks which
are consistent with the primary vertex; it depends upon the
impact parameter (three-dimensional) of the track. The function approaches zero for small 
impact parameters and becomes roughly constant for large impact parameters.
The second function suppresses tracks that have a poorly measured intersection
with the lepton, and the third gives more significance to tracks for which this
intersection is close to the intersection of the lepton track and
jet direction.  This method results in a vertex resolution with a core
distribution of width $170\mu$m.

SLD uses a second method in which a $D$-like vertex is searched for inclusively
in the lepton hemisphere.  Tracks are first classified as non-primary if
they have an impact parameter (three-dimensional) to the primary vertex
of $>3.5\sigma$
and a momentum greater than $0.8$ GeV.  The non-primary tracks
(excluding the lepton, which
is presumed to come from the $B$ decay) are vertexed,
and the resultant pseudotrack defined by the $D$ vertex and momentum is
intersected with the lepton.  If this system forms a valid vertex, an attempt
is made to add a primary track first to the $B$ then $D$ vertex.  If this
whole procedure fails, SLD searches for a single non-primary track that forms
a valid vertex with the lepton.  If this succeeds, the remaining tracks
are vertexed to form the $D$ as above.  Although this is not an exclusive $D$
reconstruction, we denote this analysis by $lD$ to distinguish it from the
above method.

Apart from the typical vertex quality requirements, SLD also requires that
the total charge of the tracks in both the $D$ and $B$ vertices be -1, 0 or 1.
Because of the high probability of
correctly assigning tracks to secondary and tertiary vertices from the
excellent tracking detectors, the SLD group can use the total charge at the $B$ vertex
to select $B^0$ events with a high efficiency, while excluding a large
fraction of the $B^+$ events.  They estimate the sample contains 60.8\% $B^0$
and only 19.6\% $B^+$.

SLD obtains the $B$ momentum for the $l$ hemisphere as follows.
An attempt is made to include charged tracks only from the $B$ decay by
ordering the tracks in the jet based on the significance of their impact
parameter to the primary vertex (in three dimensions).
These tracks are combined
until the invariant mass of the combination exceeds 2.0 GeV.  This set of
tracks, along with the lepton, give the charged particle contribution to the
$B$ momentum.
The contribution from the neutrino is given by 
the difference between the beam energy and the total visible energy in the
lepton hemisphere.  Neutral energy is added based on the parameterization
$E_{B^0} = 0.7 E_0 + 0.01 E_0^2$, where $E_{B^0}$ is the neutral energy to
be assigned to the $B$ meson, and $E_0$ is the total visible neutral energy
in the $b$ jet.  This method results in a core resolution of 8\% on the
$B$ momentum.

The SLD $lD$-jet charge analysis fits to the decay
length distribution directly, and so
does not use event-by-event momentum information.

The event selection is summarized in Table~\ref{JetQDefn}.

\begin{table}[htbp]
\caption{Jet charge tag definitions for the inclusive lepton analyses, along
with the right-sign probability parameterization as a function of $\Qt$.
The mean value of $P_r$ is quoted for rough comparison.}
\begin{tabular}{@{}llcrc@{}} \hline\hline
Experiment & $\Qt$ Tag Defn. &
\begin{tabular}{@{}c@{}} $P_r(\Qt)$ \\ (or Event Weight)
\end{tabular} & Events & $<\!P_r\!>$ \\  \hline

ALEPH \cite{Aleph97-1} & $Q_H^{\rm opp}(0.5)$ &
$\begin{array}{c} w=|\Qt| \\ (P_r = \frac{1+|\Qt|}{2})\end{array}$ &
62,320 & 0.53 \\ 

CDF \cite{CDF99-2} & $Q_J^{\rm opp}(1)$ &
$\frac{1+N_D|\Qt| D_{\rm max}}{2}$
 &  120,700 & 0.56 \\ 

DELPHI \cite{Delphi97-1} & $Q_H^{\rm opp}(0.6)$ & 
$<\!P_r\!>,\hspace{0.3cm}(|\Qt|>0.1)$ &
60,381 & 0.69 \\ 

L3 \cite{L398} & $Q_H^{\rm opp}(0.4)-Q_H^{\rm vtx}(0)$ & 
$<\!P_r\!>,\hspace{0.3cm} (|\Qt|>0.12)$ & 8,707 &0.72 \\ 

OPAL \cite{Opal97-1} 
& $Q_J^{\rm vtx}(0)- 10 Q_J^{\rm opp}(1)$ & From MC & 94,843 & \\ 

SLD \cite{SLD96-1} & 
$Q_H^{\rm opp}(0.5)$ & ${\displaystyle \frac{1}{1+e^{-0.26\Qt}}}$ & 2,609 
& 0.68 \\

SLD \cite{SLD96-2}(lD) & 
$Q_H^{\rm opp}(0.5)$ & ${\displaystyle \frac{1}{1+e^{-0.32\Qt}}}$ & 584 
& 0.68 \\ \hline
\end{tabular}
\label{JetQDefn}
\end{table}

\subsubsection{JET-CHARGE FLAVOR TAG}

The basic building block for the jet charge tag $\Qt$ is either the
weighted hemisphere or $b$-jet charge, commonly defined as
\begin{eqnarray*}
Q_{H,J}(\kappa) & = & \frac{\sum_{i} q_i p_{l,i}^\kappa}{\sum_{i} p_{l,i}^\kappa},
\hspace{1cm} p_{l,i}=\vec{p_i}\cdot \hat{a}
\end{eqnarray*}
Here $\hat{a}$ is the unit vector that is used to divide the event into
two hemispheres (thrust, sphericity or $b$-jet axis), $p_{l,i}$ is the
$i^{\rm th}$ track's momentum component along this axis, and the sum runs
over all charged tracks in the
hemisphere ($Q_H$, used by ALEPH, DELPHI, L3 and SLD) or jet ($Q_J$ used
by CDF, OPAL).
OPAL normalizes to the beam
energy $E_{\rm beam}^\kappa$ rather than $\sum_{i} p_{l,i}$; SLD has no
normalization in the denominator;
and L3 uses $C(\phi) q_i$ in place of $q_i$, where $C(\phi)$ is a
function that deweights tracks that pass near an anode wire of the
tracking chamber, for which the charge is not well determined.
$\kappa$ is chosen to
optimize the power of $Q_H$ to distinguish $b$ from $\bbar$ quarks.  For all
experiments except SLD, $-1<Q_H<1$.  $Q_H(\kappa=0)$ is simply the mean
track charge in the hemisphere, $\sum_{i=1}^n q_i / n$
(sum of charges for SLD),
and so is typically small.  $Q_H(\kappa=\infty)$ is simply the charge of
the track with the largest longitudinal momentum fraction, so 
$Q_H(\infty)=\pm 1$.

\begin{figure}[htb]
\epsfxsize=3.5in \epsfbox[30 400 400 650]{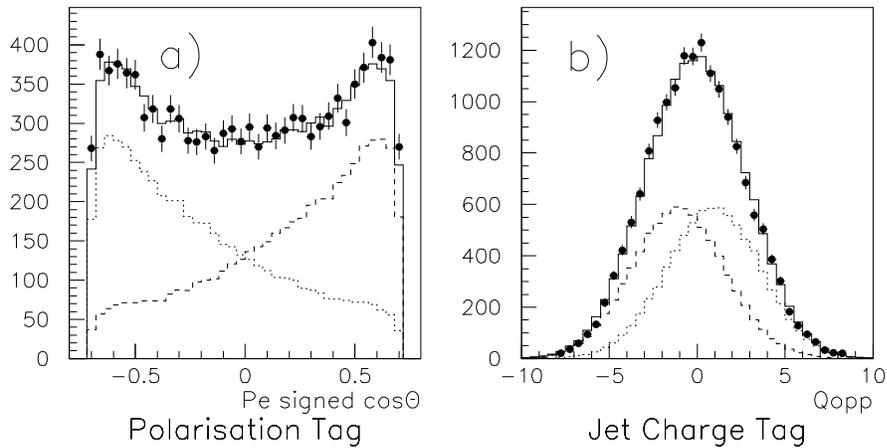}
\caption{$b$-quark production flavor tags for the SLD experiment.
{\em Left:} Polarization tag, based on the polar angle of the event axis.
{\em Right:} Jet charge.  The dashed lines are the distributions for $b$ and $\bbar$
quarks, and the points are the data.}
\label{SLDtags}
\end{figure}

$Q_H$ is the basic building block for the jet-charge $\Qt$; however several
experiments use a linear combination of the two hemisphere (jet) charges
(with different $\kappa$ weights) as the tag variable.
While jet charge is basically an opposite-side tag, the fragmentation tracks
on the (possibly mixed) $B$-vertex side still contain information about the
$b$-quark flavor at its production, although the correlation is weaker than
on the opposite side.  Since one is searching for mixing on this side,
it is desirable to find a combination that only depends weakly on
whether the
$B$ meson has mixed (otherwise the right-tag probability would depend upon
the mixed/unmixed state of the $B$, complicating the fit).
The sum of the charges in the vertex hemisphere, (i.e.
$Q_H^{\rm vtx}(\kappa=0)$) is relatively insensitive to whether the
$B$ meson mixed, as only neutral mesons can mix.
L3, for example, finds that the combination
$\Qt=Q_H^{\rm opp}(\kappa=0.4)-Q_H^{\rm vtx}(\kappa=0)$ has maximal analyzing power.
The combinations used by other experiments in the $l$-$\Qt$ analyses are
given in Table~\ref{JetQDefn}.
SLD combines the jet-charge tag with their polarization tag described in
Section~\ref{prodTagSection}.
Finally, note that for
OPAL, $\Qt$ has been defined with the opposite sign
and a scale factor of approximately ten.
Figure~\ref{SLDtags}{\em b} shows
the measured jet-charge distribution from SLD, along with the
underlying $b$ and $\bbar$ predictions from Monte Carlo studies.

One of the advantages of the jet-charge tag is its high efficiency.  At
LEP and SLD, the $b$ jets are mostly produced back-to-back.  Since the
vertex-side jet must have tracks passing through the silicon detectors,
it is well within the solid-angle covered by each experiment, and the
opposite-side $b$ jet is similarly well contained.  A comparison 
between the $l$-$l$ and $l$-$\Qt$ samples shows the
latter contain approximately ten times more events.

At CDF, however, the initial $b\bbar$ pair is not produced at rest, but with
a potentially large momentum along the beam axis.  A typical rapidity
separation between the two $b$ jets is $\Delta\eta \approx 1$.  The tracking
chambers cover the region $-1<\eta<1$, so quite often the vertexed
$b$ jet is contained within the acceptance of the tracking system while
the other $b$ jet is not.
Even worse, another, non-$b$ jet may be present, diluting
the tag's effectiveness.  These effects result in the average
right-tag probability
of $56$\%, rather than $\sim 67$\% that is typical for the LEP and SLD
experiments.  However, a search for a secondary
vertex in the non-lepton jet is performed.  The subset of events with
such a vertex has a greatly enhanced
right-tag probability, namely 63\%.

\subsubsection{FIT METHOD AND RESULTS}

The sign of the tag variable $\Qt$ is used to categorize the sign of the
$b$-quark charge in the opposite hemisphere. The analyses are based on
the time dependence of
$q_c(t)=\mbox{sign}(\Qt)\cdot \mbox{sign}(q_l(t))$.  
CDF, DELPHI, L3, OPAL and SLD perform likelihood fits to the same- and
opposite-sign data, while ALEPH
performs a binned-$\chi^2$ fit.  The
quantity $<-Q_H \cdot q_l(t)>$ is formed in 0.2 ps bins of proper time.
This is simply the standard charge correlation function $q_c(t)$ with
each event weighted by $|Q_H|$. Although weighting by the dilution
${\cal D}=2P_r-1$ is
optimal, note that CDF finds that for their procedure, in fact,
${\cal D}\sim Q_H$.

Events with $|\Qt|$ near zero clearly have little power to separate opposite
hemisphere $b$ from $\bbar$ quarks.
DELPHI and L3 simply cut out events with
$|\Qt|<0.1$ and $|\Qt|<0.12$ respectively.  

OPAL and SLD, on the other hand, parameterize
$P_r(\Qt)$ based on Monte Carlo studies.
For example, 
the separation between $\Qt$ for $b$ and $\bbar$ jets at SLD is shown
in Figure~\ref{SLDtags}, which can be combined to give the $b$ probability
as a function of measured $\Qt$.
CDF prefers to measure the shape of $P_r(\Qt)$ from data.
They measure the raw dilution,
${\cal D}_{\rm raw}= (N_{SS}-N_{OS})/(N_{SS}+N_{OS})$, where SS refers to
same-sign and OS refers to
opposite-sign lepton and jet-charge.
They find
that the raw dilution can be parameterized by a simple straight line,
$D_{\rm raw}(|\Qt|) = \alpha |\Qt|$.
Since the raw dilution is lower than the
true dilution due to mixing and cascade semileptonic decays,
the final fit includes an overall normalization,
${\cal D} = N_D D_{raw}(|\Qt|)$.  The normalization is $N_D\approx 1.5-2.0$,
depending on the subsample.

In addition to estimating the production tag right-sign probability for each
event, OPAL also estimates the sample composition fractions event-by-event
based on the value of the output of the neural net that is used to select lepton
candidates, as described in Section~\ref{dilSection}.

The quality of the vertex-side lepton tag depends on the parameters
discussed in the dilepton section.  The sample composition on the lepton
side is very similar to the dilepton case. The right-sign probability for this
decay-point tag is typically estimated from the Monte Carlo predictions for
the $\Bs$ and cascade fractions.  The tag purity of the jet charge is
extracted from the data (see Table~\ref{JetQDefn} for a summary and the
number of events in each analysis).  The systematic errors are dominated by
the same effects which distort the relative amount of mixed and unmixed
events as a function of the proper decay time, namely the
sample composition parameters and the $B^+$-$B^0$ lifetime ratio.
The results of the fits are summarized in Figure~\ref{BdSummary} and
Table~\ref{domSyslQjet}.

\begin{table}[ht]
\caption{Fit results and dominant systematic errors for
the lepton-jet charge analyses.}
\begin{tabular}{@{}llcc@{}}
\hline\hline
Experiment & Dominant Systematic & $\delta(\delmd)$ (ps$^{-1}$) &
$\delmd$ (ps$^{-1}$) \\ \hline

\vspace{8pt}
ALEPH & \begin{tabular}{@{}l@{}} $\tau_{B^0} (3\%)$ \\ 
$\tau_{B^+} (3\%)$ \end{tabular} &
$\begin{array}{c} \pm0.012 \\ \pm0.016  \end{array}$ &
$0.404\pm0.045\pm0.027$ \\ 
\vspace{8pt}

CDF & $\tau_{B^+}/\tau_{B^0} (5\%)$ & $0.021$ &
\begin{tabular}{@{}c@{}}$0.500\pm0.052\pm0.043$ \\ (includes $l$ tags)
\end{tabular} \\ 
\vspace{8pt}

DELPHI & \begin{tabular}{@{}l@{}} right-sign frac., fakes (2\%) \\ 
fake frac. (20\%) \\ $b$-Baryon frac. (33\%) \end{tabular} &
$\begin{array}{c} \pm0.014 \\ \pm0.010 \\ \pm0.010  \end{array}$ & 
$0.493\pm0.042\pm0.027$ \\ 
\vspace{8pt}

L3 & $<\!P_r\!>$  (2\%) & $\pm0.037$ & $0.437\pm0.043\pm0.044$ \\ 
\vspace{8pt}

OPAL & \begin{tabular}{@{}l@{}} $\Bs$ fraction (17\%) \\ 
$\tau_{\Lambda_b}/\tau_{B^0} (8\%)$ \end{tabular} &
$\begin{array}{c} \pm0.011 \\ \pm0.007  \end{array}$ &
$0.444\pm0.029\pm0.020$ \\ 
\vspace{8pt}

SLD($l$) & \begin{tabular}{@{}l@{}} $\Bs$ frac. (26\%) \\ 
$b$-Baryon frac. (42\%) \end{tabular} &
$\begin{array}{c} \pm0.014 \\ \pm0.015  \end{array}$ & 
$0.520\pm0.072\pm0.035$ \\ 

SLD($lD$) & \begin{tabular}{@{}l@{}} Fit binning and range \\ 
Tracking efficiency \end{tabular} &
$\begin{array}{c} \pm0.032 \\ \pm0.017  \end{array}$ & 
$0.452\pm0.074\pm0.049$ \\ \hline

\end{tabular}
\label{domSyslQjet}
\end{table}

\subsection{$D^*-l$ and $D^*-$Jet Charge Analyses}
\label{DstarSection}

As described in Section~\ref{decayTagSection},
a reconstructed $\Dsp$ meson can be used
as a very pure tag of the charge of the $b$ quark at its decay point;
moreover, events
with $\Dsp$ are also enriched in $B^0$ content.  Further, the tracks forming
the $\Dsp$ can be vertexed and used to estimate the $B$ decay length.

The experiments search for the following decay modes: $\Dsp\rightarrow D^0 \pi^+_*$, with
$D^0\rightarrow K^-\pi^+, K^-\pi^+\pi^0$, or $K^-\pi^+\pi^+\pi^-$.
All experiments exploit the small Q-value of the $\Dsp$ decay: 
$M_{\Dsp} - M_{D^0} - M_{\pi^+} = 5.8$ MeV, meaning the decay products are
basically at rest in the $\Dsp$ rest frame, which
greatly suppresses the combinatorial background.  The quantity
$\Delta M = M(\mbox{all found decay products
including } \pi^+_*) - M(\mbox{all found decay products excluding } \pi^+_*)$
is very insensitive to the momentum resolution of an individual track,
 since resolution effects mostly cancel in the difference.
This is true even to the extent that the $\pi^0$
can be missing and a clean $D^*$ signal extracted.

ALEPH \cite{Aleph97-1} fully reconstructs the
three $D^0$ modes listed, including the
$\pi^0\rightarrow \gamma\gamma$.
CDF \cite{CDF4526}
reconstructs the $\Dsp$ only in the $D^0\rightarrow K^-\pi^+$ mode, but
adds in the $D^+\rightarrow K^-\pi^+\pi^+$ decay.
DELPHI \cite{Delphi97-1}
does not reconstruct the $\pi^0$, resulting in larger background.
OPAL \cite{Opal96}
finds the signal/background for the $K\pi(\pi^0)$ and $K3\pi$
modes to be too high, and excludes them from the analysis.
Table~\ref{DstarAnal} lists the modes reconstructed, as well as the
signal-to-background ratio for each analysis.

\begin{table}[htb]
\caption{Overview of the decay modes, sample sizes, flavor tag and mean
right-sign tag probability for the $D^*$-$l$ and $D^*$-jet charge analyses.}
\begin{tabular}{@{}llccc@{}}
\hline\hline
Experiment & $D^0/D^+$ Modes & Sig/Bgnd & Prod. Tag & $<\!P_r\!>$  \\ \hline 

\vspace{8pt}
ALEPH \cite{Aleph97-1} &
$\begin{array}{@{}l@{}} K^-\pi^+ \\ K^-\pi^+\pi^0 \\ K^-\pi^+\pi^+\pi^-
\end{array}$ & 840/565 &
$\begin{array}{@{}c@{}} l \\ (p>3.0 \mbox{ GeV} \\ \ptr>0.75 \mbox{ GeV})
\end{array}$ & 0.79 \\

\vspace{8pt}
ALEPH \cite{Aleph97-1} &
$\begin{array}{@{}l@{}} K^-\pi^+ \\ K^-\pi^+\pi^0 \\ K^-\pi^+\pi^+\pi^-
\end{array}$ & 1555/1096 &
$\begin{array}{c} \mbox{Jet Q} \\ Q_H(0.5)-0.08\sum q_i \\ |\Qt|>0.1
\end{array}$ & 0.76 \\ 

\vspace{8pt}
CDF \cite{CDF4526} &
$\begin{array}{@{}l@{}} K^-\pi^+ \\ K^-\pi^+\pi^+\end{array}$
& $\begin{array}{@{}c@{}}358/520 \\ 460/510 \end{array}$ &
$\begin{array}{@{}c@{}} l \\ p_t>8.0 \mbox{ GeV} \\ \ptr>1.5 \mbox{ GeV}
\end{array}$ &  0.79 \\

\vspace{8pt}
DELPHI \cite{Delphi97-1} &
$\begin{array}{@{}l@{}} K^-\pi^+ \\ K^-\pi^+(\pi^0) \\ K^-\pi^+\pi^+\pi^-\end{array}$
& $\begin{array}{@{}c@{}}1554/745 \\ 1370/1121 \\ 1288/3904\end{array}$
& \begin{tabular}{@{}c@{}} Jet Q$^\ddag$ \end{tabular} & $0.65^\dag$ \\ 

OPAL \cite{Opal96} &
$K^-\pi^+$ & 253/95 & \begin{tabular}{@{}c@{}}$l$ \\ Neural net$^\ast$
\end{tabular} &  0.79 \\ \hline
\end{tabular}

$^\dag$Includes wrong-sign $D^*$ mistags.  $^\ddag$See Table~\ref{JetQDefn}.
$^\ast$Approximately equivalent to cuts of $p>3.0$ GeV and $\ptr>0.75$ GeV.
\label{DstarAnal}
\end{table}

\subsubsection{FLAVOR TAGGING}
The vertex-side $b$ quark flavor at its decay point has the opposite sign
of the reconstructed $D^{*\pm}$.  The production flavor tags
are either the opposite-side lepton, or opposite-side jet-charge tags
which were introduced above. 

\subsubsection{DECAY LENGTH AND PROPER TIME DETERMINATION}

These analyses have the advantage of a built-in decay vertex, namely
that given by vertexing the $D^0$ decay products.  Unfortunately, intersecting
this with the $\pi^+_*$ from the $\Dsp$ decay to obtain the $B$ decay point
is not viable; the pion is emitted almost at rest in the $\Dsp$ frame,
and so its direction is the same as the $D^0$ direction in the laboratory frame.
All analyses begin by vertexing the $D$ decay tracks.

Only OPAL attempts to explicitly find the $B$ decay vertex, as follows.
After vertexing the tracks forming the $\Dsp$ and creating
a pseudotrack passing
through this vertex with the $\Dsp$ momentum, OPAL attempts to
combine this pseudotrack with
other tracks in the hemisphere that are likely to have
come from the $B$ decay.  These tracks are selected based on their momentum
and angle relative to the $\Dsp$ pseudotrack direction, and they must have
an intersection point with the pseudotrack that is consistent with
a $b$ decay.
The $B$ momentum is estimated using the same technique as 
in the OPAL dilepton analysis.

CDF estimates the most likely $B$ decay position based on its distance
to the $D$ vertex, the momentum of the $D$, and knowledge of the $D$ and $B$
lifetimes.  The $B$ momentum is estimated from the momentum of the
$D$ meson, using a correspondence determined from Monte Carlo studies.

ALEPH and DELPHI fit the charge correlation to variables described below.

\subsubsection{FIT PROCEDURE AND RESULTS}

ALEPH fits directly to the charge correlation of the $D^*$ sign and opposite
hemisphere sign ($l$ or jet charge) as a function of the
$D^0$ decay distance, $q_c(L_{D^0})$.
This involves combining both $B$ and $D$ decay
time and momentum distributions in Monte Carlo studies to generate the
predicted
decay length distributions used in the likelihood fit.

DELPHI fits to the charge correlation as a function of the
sum of the proper decay times of the $B$ and $D$ meson.
For
$l_B$=flight distance of the $B$, $l_D$=flight distance of the $D$,
and $l\approx l_B+l_D$ = total decay length to the $D$ vertex, one defines
a proper time
$t=t_B + t_D = \frac{m_B}{p_B}l_B + \frac{m_D}{p_D}l_D
\approx \frac{m_B}{p_B}l$. This approximation is good to about a percent.
DELPHI takes the average value $p_B = 0.7 E_{\rm beam}$ to convert $l$ to $t$.

CDF fits to the mean charge correlation in eight bins of proper time while
OPAL performs an unbinned likelihood fit as described previously.

The vertex-side tag (the sign of the $D^{(*)}$), has a very high purity -- the
main source of mistags comes from $B\rightarrow D \overline{D_s}$ events
(see Section~\ref{decayTagSection}).  The $B^0$ fraction in these samples
is high, with a typical value of 83\%.  
None of the experiments measure the production
tag's right-sign probability event-by-event; instead they use the
mean value $<\!P_r\!>$, which is left as a free parameter in the fits.
The values of $<\!P_r\!>$ are
summarized in Table~\ref{DstarAnal}.

Figure~\ref{BdSummary} and
Table~\ref{domSysDstar} summarize the results of the fits.

\begin{table}[ht]
\caption{Fit results and dominant systematic errors
for the $D^*$-$l$ and $D^*$-jet charge analyses.}
\begin{tabular}{@{}llccc@{}}
\hline \hline
Experiment & Dominant Systematic & $\delta(\delmd)$ (ps$^{-1}$) &
$\delmd$ (ps$^{-1}$) \\ \hline

\vspace{8pt}
\begin{tabular}{@{}l@{}} ALEPH \\ (combined)\end{tabular} &
 \begin{tabular}{@{}l@{}} cascade frac. (15\%) \\ 
$\Bs$ frac. (9\%) \end{tabular} &
$\begin{array}{c} \pm0.025 \\ \pm0.019  \end{array}$ &
$0.482\pm0.044\pm0.024$ \\ 

\vspace{8pt}
CDF & \begin{tabular}{@{}l@{}} $B^+$ frac. (35\%) \\ 
$\tau_{c} (25\%)$ \end{tabular} &
$\begin{array}{c} \pm0.037 \\ \pm0.021 \end{array}$ &
$0.562\pm0.068\pm0.050$ \\ 

\vspace{8pt}
DELPHI & \begin{tabular}{@{}l@{}} cascade frac. (6\%) \\ 
$\tau_{B^+}/\tau_{B^0} (4\%)$ \end{tabular} &
$\begin{array}{c} \pm0.032 \\ \pm0.017  \end{array}$ & 
$0.523\pm0.072\pm0.043$ \\ 

OPAL & \begin{tabular}{@{}l@{}} cascade frac. (15\%) \\ 
$\tau_{B^+}/\tau_{B^0} (6\%)$ \end{tabular} &
$\begin{array}{c} \pm0.011 \\ \pm0.023  \end{array}$ &
$0.567\pm0.089\pm0.029$ \\ \hline

\end{tabular}
\label{domSysDstar}
\end{table}


\subsection{$D^*l-l$ and $D^*l-$Jet Charge Analyses}
\label{DstarLepSection}

CDF \cite{CDF99-5}, DELPHI \cite{Delphi97-1} and OPAL \cite{Opal96}
 perform analyses in which the vertex side of the event
has both a $\Dsp$ and a lepton.
A semileptonic $B$ decay involving
a $D^*$ has the lepton sign opposite the $D^*$ sign, e.g. $\Dsp l^-$, whereas
combinatorial background contributes to both opposite-sign and same-sign.
The production flavor
tags used are the opposite-side lepton (CDF), or the opposite-side jet
charge (DELPHI, OPAL).

\subsubsection{EVENT SELECTION}
CDF searches for $\Dsp$ candidates in the low-$p_t$ dilepton samples
described previously, using the same decay modes described in the above section
and similar techniques.  OPAL adds the $D^0\rightarrow K^-\pi^+(\pi^0)$ decay.
Neither experiment reconstructs the $\pi^0$.

DELPHI does not actually reconstruct the $\Dsp$ at all; they only
try to find the $\pi^+_*$ from the decay.  
Tracks likely to belong to the $B$ decay are selected by the
inclusive vertex method in the lepton jet, exactly as described in the
dilepton analysis in Section~\ref{dilSection}.
Just as in the case of the missing $\pi^0$,
forming the mass difference $\Delta M = M(\mbox{All found decay products
including } \pi^+_*) - M(\mbox{All found decay products excluding } \pi^+_*)$
allows a relatively pure $\Dsp$ signal to be extracted even if several
$B$ decay products are missing.  The larger background can be controlled
via the
extra handle of the charge correlation of the lepton with the $\pi_*$.

Table~\ref{DstarLepAnal} summarizes the event selection.

\subsubsection{DECAY LENGTH AND MOMENTUM MEASUREMENTS}

In $D^*l$ analyses, 
the lepton track, which is presumed to come from the
$B$ meson, can be used to reconstruct the $B$ decay point.
The $D^0$ decay products are vertexed
to reconstruct the
$D$ decay point, and a pseudotrack is formed with the $D$ momentum passing
through this point.  This pseudotrack is intersected with the lepton
trajectory to give the $B$ decay position.
OPAL solves for the decay length $L$ using the position of the primary vertex,
this secondary $B$ vertex and a constraint based on the direction of the
$l\Dsp$ system.  CDF performs a similar vertexing procedure
in the $r\phi$ plane and projects
the decay length vector formed by the $B$ and primary vertices onto the
$l-D^*$ momentum direction to give $\Lxy$.

OPAL and CDF estimate the $B$ boost as $\beta\gamma = p^{l\Dsp}/M_B \times
{\cal K}(p^{l\Dsp},M_{l\Dsp})$, and 
$\beta\gamma\sin\theta = p^{l\Dsp}_t/M_B \times <\!{\cal K}\!>$ respectively,
where the correction factor, ${\cal K}$, is determined from
Monte Carlo studies.
The full distribution of ${\cal K}$ is used to derive the
momentum smearing needed for
the likelihood functions.
Typical resolutions on the decay time are 15-20\%.

DELPHI converts the decay length found by the inclusive vertex algorithm to
proper time exactly as in the dilepton analysis.

\begin{table}[htb]
\caption{Overview of the decay modes, sample sizes, flavor tag and mean
right-sign tag probability for the $D^*l$-$l$ and $D^*l$-jet charge analyses.}
\begin{tabular}{@{}llccc@{}}
\hline \hline
Experiment & $D^0$ Modes & Sig/Bgnd & Production Tag & $<\!P_r\!>$ \\ \hline
\vspace{8pt}
CDF \cite{CDF99-4} & $\begin{array}{@{}l@{}}K^-\pi^+ \\
K^-\pi^+(\pi^0) \\
K^-\pi^+\pi^+\pi^-\end{array}$ &
$\begin{array}{@{}l@{}} 167/49 \\ 190/226 \\ 173/83 \end{array}$ &
 $l^\dag$ & 0.67 \\

\vspace{8pt}
DELPHI \cite{Delphi97-1} &
$\pi^+_*-l$ & 4132/1823 & Jet Q$^\ddag$ & 0.69 \\ 

OPAL \cite{Opal97-1} & $\begin{array}{@{}l@{}}K^-\pi^+ \\
K^-\pi^+(\pi^0)\end{array}$ & 
$\begin{array}{@{}l@{}} 406/49 \\ 794/225\end{array}$ &
\begin{tabular}{@{}c@{}}Jet Q$^\ddag$ \\ ($|\Qt|>1$)\end{tabular} & 0.72 \\
\hline
\end{tabular}

$^\dag\!$ See $e\mu$ and $\mu\mu$ in Table~\ref{dileptonCuts}.
$^\ddag\!$ See Table~\ref{JetQDefn}.
\label{DstarLepAnal}
\end{table}

\subsubsection{FIT METHOD AND RESULTS}

Each experiment performs a likelihood fit to the charge correlation of the
charge of the vertex-side lepton and the opposite side lepton or jet-charge
sign, as a function of the proper decay time.

The decay-point tag is of course very pure, having both a $D^*$ and lepton
requirement.
None of the experiments uses an event-by-event measure of the production
tag's right-sign probability; instead, they use the mean
value $<\!P_r\!>$, which  is a free parameter in the fits.
To increase this mean probability, OPAL and DELPHI remove low $|\Qt|$ events.
The fitted values of $<\!P_r\!>$ are given in Table~\ref{DstarLepAnal}.

\begin{figure}[htb]
\mbox{
\epsfxsize=1.9in \epsfbox[70 130 400 700]{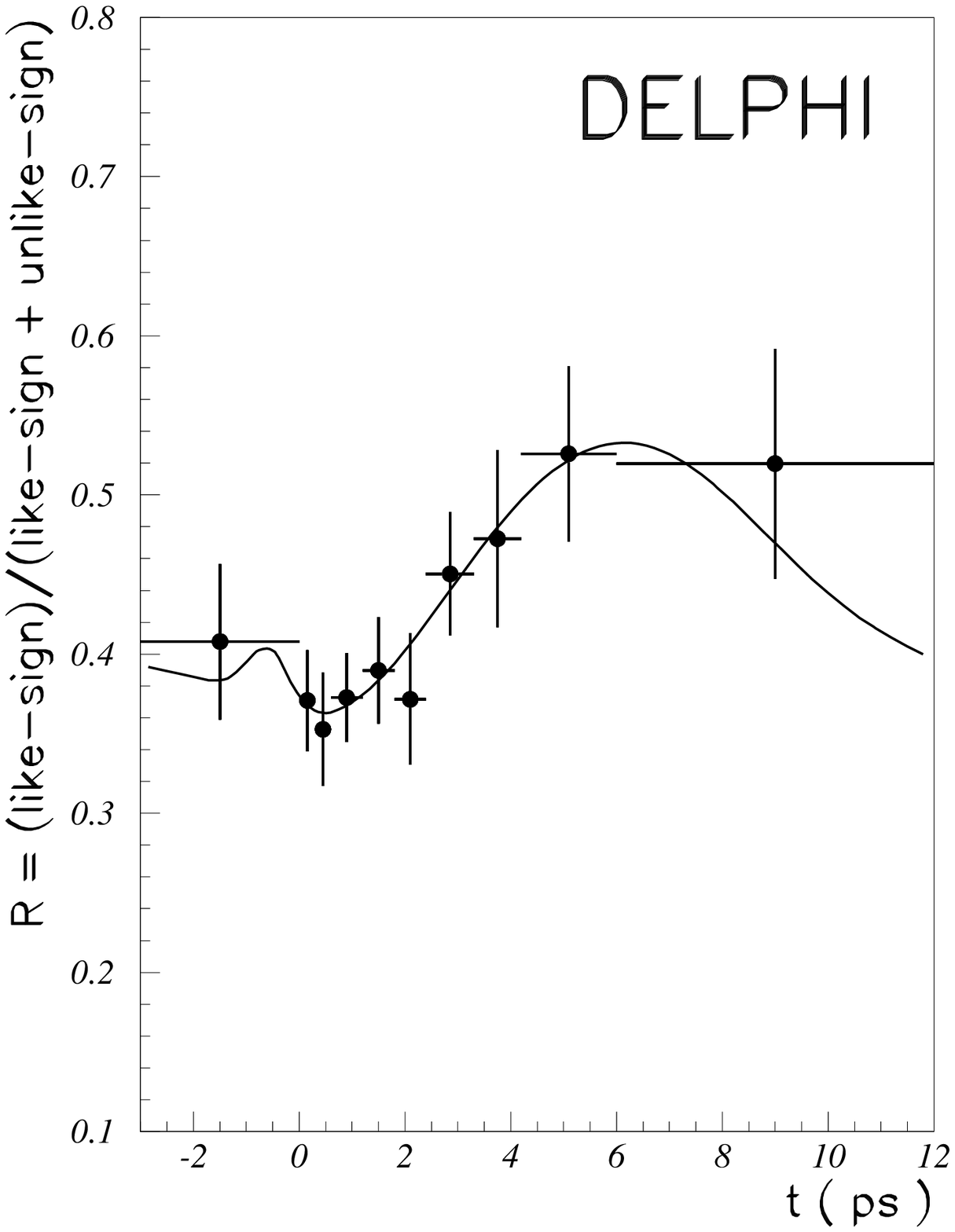} \hspace{40pt}
\epsfxsize=2.2in \epsfbox[0 0 400 600]{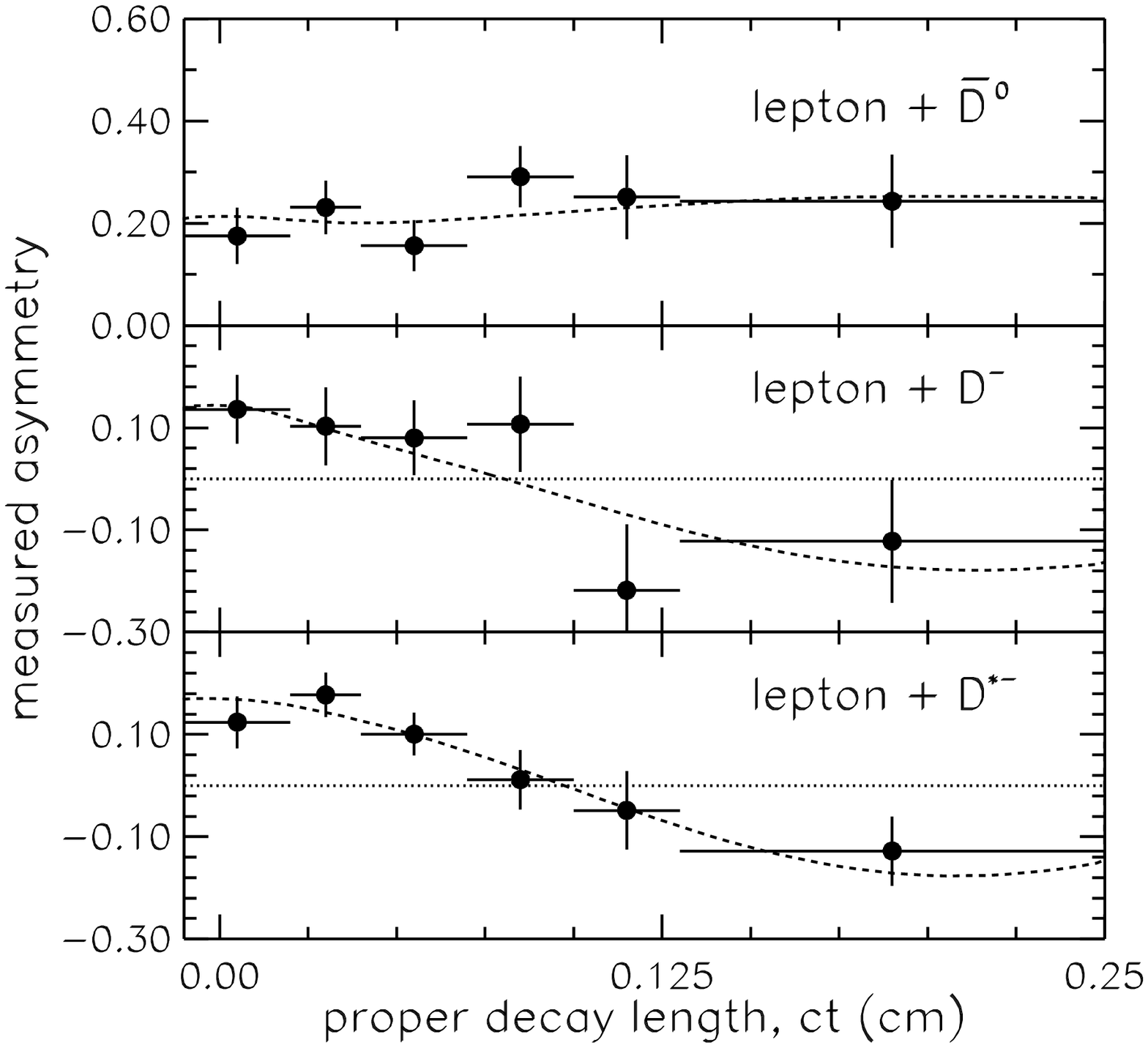}}
\caption{{\em Left}: Fraction of $\pi_*l-l$ events with like-sign (mixed)
leptons, from
the DELPHI analysis, and the fit results.
{\em Right}: Right/wrong sign asymmetry from the CDF same-side tagging analysis
and fit results.}
\label{SSTmix}
\end{figure}

Figure~\ref{SSTmix} shows the $DP-$ (mixed) fraction
for the DELPHI $\pi_*l-l$ analysis.
Figure~\ref{BdSummary} and
Table~\ref{domSysDstarl} summarize the results of the fits.
The systematic errors are dominated by
the uncertainty in the $B^-$ contamination.

\begin{table}[ht]
\caption{Fit results and dominant systematic errors for the $D^*l$ analyses.}
\begin{tabular}{@{}llcc@{}}
\hline\hline
Experiment & Dominant Systematic & $\delta(\delmd)$ (ps$^{-1}$) &
$\delmd$ (ps$^{-1}$) \\ \hline

\vspace{8pt}
CDF & \begin{tabular}{@{}l@{}} $B^-$ frac. (30\%) \\
$D^{**}\rightarrow D^*$ frac. (50\%)\end{tabular} &
$\begin{array}{c} \pm0.020 \\ \pm0.027 \end{array}$ &
$0.516\pm0.099\pm0.035$ \\ 

\vspace{8pt}
DELPHI & $B^-$ frac. (25\%) &
$\pm0.012$ & $0.499\pm0.053\pm0.015$ \\

OPAL & $B^-$ frac. (50\%) & $\pm0.019$ & $0.539\pm0.060\pm0.024$ \\
 \hline

\end{tabular}
\label{domSysDstarl}
\end{table}

\subsection{Other Inclusive Analyses}
\label{InclusiveSection}

There have been several attempts to increase the
sample sizes beyond those involving leptons or $D$ tags by using very
inclusive methods, both in
locating candidate $B$ vertices and in flavor tagging.

Both ALEPH \cite{Aleph97-2} and SLD \cite{SLD96-3}
 have selected events using similar inclusive
topological vertexing methods.
SLD applies kaon and dipole final-state tags, using jet-charge and
polarization for the initial tag, while ALEPH uses a highly efficient
double jet-charge method.

\subsubsection{EVENT SELECTION}

The vertex algorithms used by ALEPH and SLD, rather than being driven by
track intersections, search for viable secondary vertex positions.
The ALEPH vertex algorithm is as described in the dilepton section, and is
based on the $\chi^2$ difference of assigning all tracks to the primary
vertex vs.
assigning some to a secondary vertex.
The SLD algorithm is similar; it searches for points in space from which
several tracks originate with high probability.
In both algorithms, no attempt is made to form separate $B$ and $D$ vertices.

SLD finds a suitable vertex in $\sim 50\%$
of $b$ hemispheres, compared to $\sim 15\%$ for charm and $\sim 3\%$ for
light quarks.  A vertex axis is formed by the direction between the primary
and this secondary vertex.
A final pass is made to assign tracks to the vertex based on
their impact parameter to the vertex axis, and the distance along the vertex
axis of the closest approach position.
Finally, the quantity $M_{p_t}  = \sqrt{M^2+p_t^2} + |p_t|$
is formed, where $M$ is the mass of the tracks in the vertex (
assuming they are all pions) and $p_t$ is their total momentum transverse to
the vertex axis.  Selected events are required to have
$M_{p_t}<2.0$ GeV, which rejects most of the remaining charm and $uds$ events,
leaving 16,803 vertexes with an estimated $b$ purity of 93\%.

ALEPH uses the vertex as a seed and accepts tracks with
$R_{sig} = S_p/(S_p+S_s) > 0.7$, where $S_{p,s}$ is the significance of the
impact parameter of the track to the primary and secondary vertices
respectively.  Hemispheres with a vertex $\chi^2$ probability $>1\%$ are
accepted -- if both hemispheres have a valid vertex, the one with the
highest probability is selected as the vertex side of the event.

\begin{table}[htb]
\caption{Overview of kaon, dipole and double jet-charge analyses.}
\begin{tabular}{@{}lccc@{}}
\hline\hline
Experiment &  Events & Decay Tag & Production Tag \\ \hline

ALEPH \cite{Aleph97-2} & 423,169 & Jet Q & Jet Q$^\dag$ \\ 

SLD \cite{SLD96-3}(Kaon) & 5,694 & Kaon & Jet Q +
Polarization$^\dag$ \\ 

SLD \cite{SLD96-3}(Dipole) & 3,291 & Dipole & Jet Q +
Polarization$^\dag$ \\ \hline

\end{tabular}

$^\dag\!$ See Table~\ref{JetQDefn}.
\label{InclAnal}
\end{table}

\subsubsection{INITIAL- AND FINAL-STATE FLAVOR TAGGING}

SLD determines 
the $b$ flavor of the initial state by using both
the jet charge and polarization methods, as described above.

For the final-state tag, two methods were developed.
First, the standard $b\rightarrow c \rightarrow s$ decay chain suggests that
inclusive kaons should dominantly
have the same sign charge as the $b$ quark.  This correspondence is diluted
by $B\rightarrow D\overline{D}$ decays and by strange quark pairs produced
during fragmentation.

Kaons are required to be separated by more than $2.4\sigma$ from the pion
hypothesis and by $3.2\sigma$ from a proton.  The final-state tag is simply the
sum of the charges of all kaon candidates, and so $\sum Q_K >0$ would tag a
$B^0$, for example.  The 12\% of events with $\sum Q_K=0$ are discarded.  The 
probability of correctly tagging the charge of the $b$ at its decay is 77\%
with this method.

The second method involves constructing a charge dipole from the $B$ and
$D$ meson decay products.  SLD requires the total charge of tracks associated
with the vertex to be zero, to enhance the $B^0$ content of the sample as
described previously.  The direction between the primary and secondary
vertices defines the vertex axis.  A dipole is formed by summing over
positive and negative tracks:
\begin{eqnarray*}
\delta q & = & \frac{\sum_{i=+} w_i L_i}{\sum_{i=+}w_i} - 
\frac{\sum_{i=-} w_i L_i}{\sum_{i=-}w_i},
\end{eqnarray*}
\noindent where $L_i$ is the location of the distance of closest approach
of track $i$ to the vertex axis, and the weight
$w_i=\sin^2\theta_i/\sigma_{Ti}$, where $\theta_i$ is the angle of the track
relative to the vertex axis and $\sigma_{Ti}$ is the uncertainty on the
impact parameter to the vertex axis.
A $B^0\rightarrow l^+\nu D^-$ decay, for example, will thus tend to have 
a negative dipole, whereas the equivalent $\Bzbar$ decay has a positive
dipole.  The probability of a correct tag is parameterized as a function of
$|\delta q|$,
and reaches 68\% for large $|\delta q|$.

ALEPH tags both hemispheres using the jet-charge algorithm previously
described, using $Q_H(\kappa=0.5)$ for the opposite side hemisphere, and
$Q_H(\kappa=1.0)$ for the vertex side.

\subsubsection{FIT METHOD AND RESULTS}

Both SLD analyses perform $\chi^2$ fits to the
mixed/total (i.e. $N_{DP-}/N$)
fraction of the data,
binned in decay length.  Each event's contribution is weighted by its
estimated dilution to maximize the sensitivity (see Section~\ref{mixFit}).

ALEPH converts the decay length to proper time using an
event-by-event estimate of the momentum constructed from a weighted sum of the
momentum of the charged tracks, energy from neutral clusters associated with
the $b$ jet (projected onto the vertex direction) and a correction due to the
missing neutrino.
They then fit to the weighted charge correlation
$<\!-Q_H^{\rm opp}(0.5)Q_H^{\rm vtx}(1.0)\!>(t)$,
i.e. the usual correlation function
given by the signs of the hemisphere jet charges, with an event weight given
by the product of their magnitudes similarly to the $l$-jet charge analysis
described previously.

The results are summarized in Table~\ref{domSysSLD} and Figure~\ref{BdSummary}.

\begin{table}[htb]
\caption{Fit results and dominant systematic error for double jet charge,
dipole and kaon tag analyses.}
\begin{tabular}{@{}llcc@{}}
\hline\hline
Experiment & Dominant Systematic & $\delta(\delmd)$ (ps$^{-1}$) &
$\delmd$ (ps$^{-1}$) \\ \hline

\vspace{8pt}
ALEPH & \begin{tabular}{@{}l@{}} Frag. params. \\
$\Lambda_b$ frac. \end{tabular} &
$\begin{array}{c} \pm0.015 \\ \pm0.012\end{array}$ & 
$0.441\pm0.026\pm0.029$ \\ 

\vspace{8pt}
SLD(Kaon) & \begin{tabular}{@{}l@{}} Fit method \\
Kaon ID efficiency \end{tabular} & $\begin{array}{c} \pm0.042 \\ \pm0.035\end{array}$ & 
$0.580\pm0.066\pm0.075$ \\ 

SLD(Dipole) & \begin{tabular}{@{}l@{}} Fit method \\
MC statistics \end{tabular} & $\begin{array}{c} \pm0.016 \\ \pm0.021\end{array}$ & 
$0.561\pm0.078\pm0.039$ \\ \hline

\end{tabular}
\label{domSysSLD}
\end{table}

\subsection{Same-side Tag Analyses}
\label{SSTSection}

As mentioned above, in an experiment at a hadron collider such as CDF,
it is quite common that one $b$ jet falls within
the tracking acceptance while the other $b$ jet does not.
Hence a flavor tagging method which only relies on finding one $b$ jet
is attractive.  Such methods are generally called same-side tags (SST).

\subsubsection{EVENT SELECTION}

The only $b$ samples available to CDF involve semileptonic 
$b$ decays, selected by either the low-$p_t$ dilepton or high-$p_t$ single
lepton triggers described above.  Since this analysis is designed to only
require one $b$, the single lepton samples are chosen as the basis.
The SST algorithm used by CDF \cite{CDF99-1} requires a relatively pure
 $B^0$ meson sample,
as discussed below.
To obtain this, a reconstructed $\Dsp$ or $D^+$
associated with the lepton is required.  Three $D^0$ decay modes,
$K^-\pi^+, K^-\pi^+(\pi^0)$, and $K^-\pi^+\pi^+\pi^-$, (with the $\pi^0$ not
reconstructed), are used, and the $\Dsp$ reconstruction follows the
standard methods described in Section~\ref{DstarSection}.
The $D^+$ is reconstructed in the
$D^+\rightarrow K^-\pi^+\pi^+$ mode.
In addition to the $B^0\rightarrow D^{(*)+} l^-\nu X$ mode, the
$B^-\rightarrow D^0 l^-\nu X$ mode, with $D^0\rightarrow K^+\pi^-$, is searched
for, where the $D^0$ is required to not come from a $\Dsp$.  This
signal is used to help estimate the amount of $B^+$ contamination in the
(mostly) $B^0$ sample.

\subsubsection{PRODUCTION FLAVOR TAG}
As described by the schematic picture in Figure~\ref{sstpic},
the SST tag used by CDF attempts to find the "first" charged pion in the
fragmentation chain of the $b$ quark.  If the $b$ quark forms a $\Bzbar$,
the first pion should be a $\pi^-$;
conversely, a $\pi^+$ would come from a $\bbar$
quark.  However, if the $b$ quark forms a $B^-$, the pion
should be a $\pi^+$, with a $\pi^-$ associated with a $\bbar$,
exactly the opposite correlation to the $B^0$ case.
Hence the charge of the pion, if it can be identified,
determines the sign of the $b$ quark, as long as it is known whether the
decaying meson is a $B^0/\Bzbar$ or $B^\pm$.

The SST pion is a fragmentation track, and so should originate at the
primary vertex, not the $B$ decay vertex.  SST candidates are thus
searched for near the $B$ direction, defined as the $l+D$ direction, with
impact parameter significance to the primary vertex of less than three.
String fragmentation models suggest that the fragmentation particles have
little momentum transverse to the $b$-quark direction, and so the
track with the smallest $\ptr$ is chosen, where $\ptr$ is the track's momentum
orthogonal to the direction given by the $l+D+$SST pion system.
A suitable candidate can be found in approximately 70\% of the events.

\begin{table}[htb]
\caption{Event selection and same-side tag properties from CDF.}
\begin{tabular}{@{}lc|c@{}}
\hline\hline
& \begin{tabular}{@{}c@{}} $B^0$ sample, $l^-D^{(*)+}$ \\
$(\Dsp\rightarrow D^0\pi^+_*)$\end{tabular} &
\begin{tabular}{@{}c@{}} $B^-$ sample, $l^-D^0$ \end{tabular}
\\ \hline
Modes/Events & 
$\begin{array}{lc} 
D^0\rightarrow K^-\pi^+ & \\
\hspace{1.2cm} K^-\pi^+\pi^+\pi^- & 1754 \\
D^0\rightarrow K^-\pi^+(\pi^0) & 2515 \\
D^+\rightarrow K^-\pi^+\pi^+ & 1997 \\ 
\end{array}$ &
\begin{tabular}{@{}l@{}} $D^0\rightarrow K^-\pi^+$ \\
(not from $\Dsp$)\end{tabular} \hspace{0.6cm}2928 \\ 
Tag Effic. & 70\% & 70\% \\ 
Tag $<\!P_r\!>$ & 0.59 & 0.64 \\ \hline
\end{tabular}
\label{SSTsample}
\end{table}

\subsubsection{FIT METHOD AND RESULTS}
The decay length and conversion to proper decay time is performed as
in the $D^*l-l$ analysis described in Section~\ref{DstarLepSection}.

The sign of the lepton in the $lD^{(*)}$ decay tags the $b$ quark charge at its
decay point.
For the predominantly $B^0$ sample, given by the $l\Dsp$ events, the
"unmixed" $DP+$ sample has same-sign, $l^--\pi^-_{SST}$ and $l^+-\pi^+_{SST}$
events,
whereas the $DP-$ sample has opposite-sign events.
For the $B^+$ sample the assignment is reversed.

Each sample is divided into six bins in proper decay length,
and the mean value of the charge-correlation function
$<\!q_c(ct)\!> = [N_{DP+}(ct)-N_{DP-}(ct)]/[N_{DP+}(ct)+N_{DP-}(ct)]$
is evaluated for each bin.
If the samples were pure $B^0$ and $B^+$, the first would show
the usual $\cos\delmd t$ dependence whereas the second would be flat.  However,
there is some cross contamination between the samples.  First, $B^0$
events with $D^0\rightarrow K^-\pi^+$ decays end up in the $B^+$ sample 
if the $\pi^+_*$ from the $\Dsp$ decay is missed.  CDF determined that
this is the case for $15\pm 7$\% of these $B^0$ events.
The most serious cross contamination, however, comes from decays involving
the P-wave $D^{**}$ resonances
(or non-resonant $D^{(*)}\pi$ pairs), as shown in
Figure~\ref{BtoDstar}.
For example, the decay $B^-\rightarrow D^{**0}l^-\nu$ followed by
$D^{**0}\rightarrow \Dsp\pi^-_{**}$ (denoting the $\pi$ from the $D^{**}$
as $\pi_{**}$) is classified as a $B^0$ decay.  A further complication
is that the $\pi_{**}$ may be selected as the SST pion.  As this example
shows, this pion's charge is always correlated with that of the
lepton such that the event is classified as $DP+$, which biases the
charge-correlation function.  The $\pi_{**}$ particles, however,
originate from the $B$-decay vertex, whereas
the SST pions are selected to come from the primary vertex, which suppresses
this effect.

The fraction of $D^{**}$ in semileptonic $B$ decays, $f^{**}$, and the
branching fractions of the $D^{**}$ mesons are rather poorly known.
CDF assumes $f+f^*+f^{**}=1$, where $f$ and $f^*$ are the
decay fractions into $lD$ and $lD^*$.  CDF sets $f^{**}=0.36\pm0.12$
\cite{Cleo91} and $f^*/f=2.5\pm0.6$ \cite{PDG96}.  Further, there are
four $D^{**}$
states of differing spin-parity, some of which decay to $D^*\pi$ and some
to $D\pi$.  The relative abundances of the $D^{**}$ states in
the $B$ decay, then, have an effect on the amount of cross contamination
among the samples by changing the fraction involving the $D^*$ state.
This fraction is left as a free parameter in the fit and is found to be $0.3\pm0.3$.

\begin{table}[ht]
\caption{Fit result and dominant systematic errors for the
same-side tag analysis.}
\begin{tabular}{@{}lccc@{}}
\hline\hline
Experiment & Dominant Systematic & $\delta(\delmd)$ (ps$^{-1}$) &
$\delmd$ (ps$^{-1}$) \\ \hline

CDF \cite{CDF99-1} & \begin{tabular}{@{}l@{}} $D^{**}$ fraction(30\%) \\
$D^{**}\rightarrow D^*$ fraction (100\%)\end{tabular} & $\pm0.031$ & 
$0.471\pm0.078\pm0.034$ \\ \hline

\end{tabular}
\label{domSysSST}
\end{table}

\subsection{Summary of $B^0$ Mixing Results}

As this exhaustive summary of methods and results shows, a large number
of different techniques are used to extract the oscillation frequency of
$B^0$ mixing.  Within each single technique, each experiment typically has
unique methods of measuring the decay length and momentum of the $B$ meson.
Further, in evaluating systematics, the range of variation of variables,
even those common to all analyses, is often not consistent.

Therefore, combining all these results must be done with care.  Since there are
so many measurements, the final error can quickly become dominated by
systematic effects, and the handling of common systematics
is especially important.  A $B$ Oscillation Working Group was
formed \cite{BOSC} to combine the measurements in as consistent a
manner as possible.

\begin{figure}[htb]
\epsfxsize=5.5in \centerline{\epsfbox{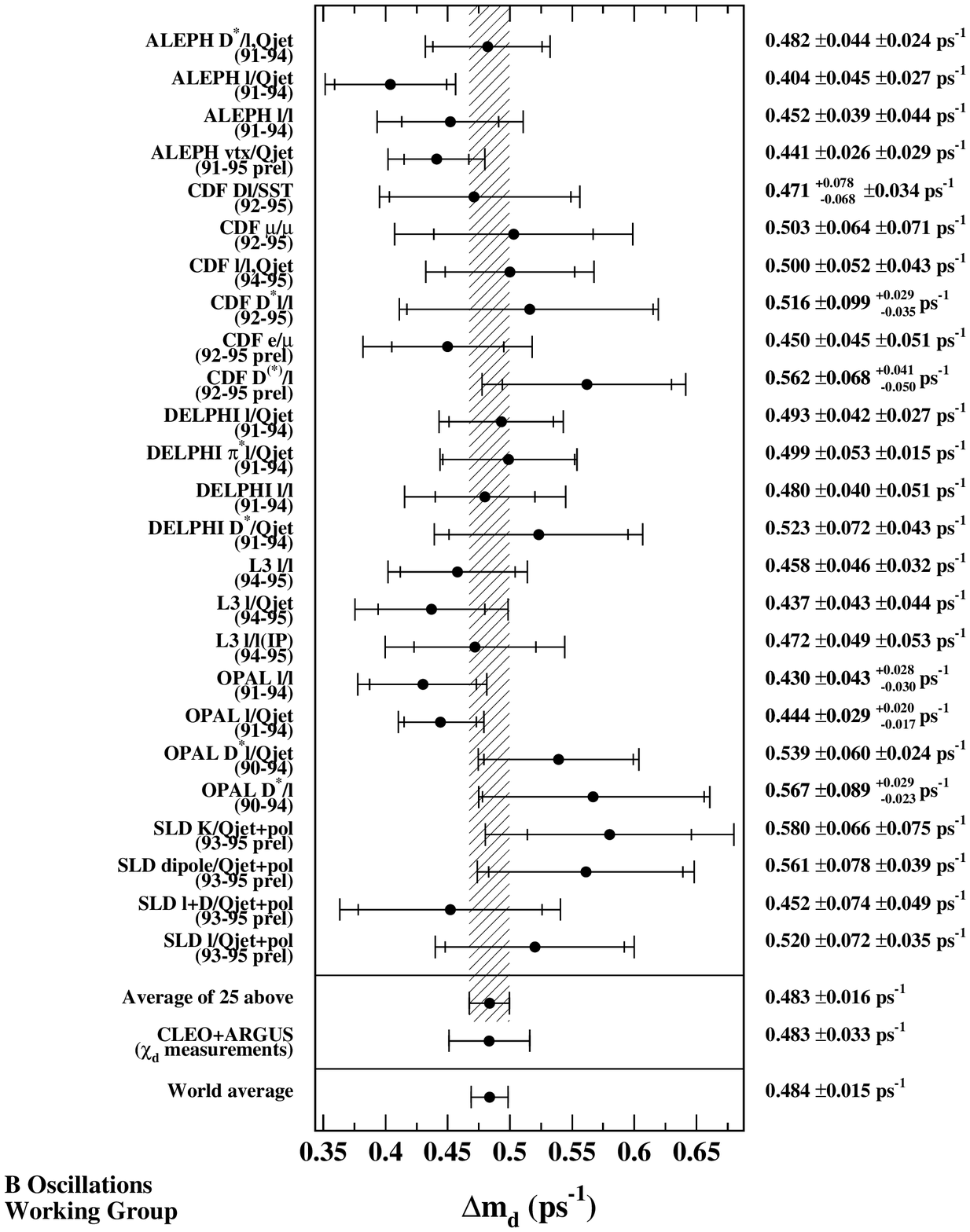}}
\caption{Summary of $B^0$ mixing results from the $B$ Oscillation
Working Group \cite{BOSC}.}
\label{BdSummary}
\end{figure}

Their results, shown in Figure~\ref{BdSummary}, give a best fit of
$\delmd = 0.484\pm0.015$ ps$^{-1}$, including the time-integrated
ARGUS and CLEO measurements.

\section{$\Bs$ MIXING}
\label{BsSection}

That $\Bs$ mesons undergo oscillations is clear.  The
time-integrated probability that a $B^0$ mixes, $\chi_d$, is measured
by the ARGUS and CLEO experiments,
which operate at $e^+e^-$ colliders tuned to the
$\Upsilon(4S)$ resonance.  The $\Upsilon(4S)$ is not sufficiently heavy to
decay to $\Bs \Bsbar$ pairs, and so the only contribution to a mixing signal
is from $B^0$ mesons.  At LEP, SLD and CDF, however, both $B^0$ and $\Bs$
are produced, so time-integrated mixing measurements determine
$\overline{\chi} = f_d\chi_d + f_s\chi_s$, where $f_d\approx 40\%$ is
the $B^0$ fraction in the $b$ sample, and $f_s\approx 11\%$ is the $\Bs$
fraction.
Since $\chi=0.172\pm0.010$ and $\overline{\chi} = 0.118\pm0.006$ \cite{PDG98},
a large value for $\chi_s$ is required, indicating that $\Bs$ mesons do
indeed mix.

The problem is that $\Bs$ mesons oscillate at a high frequency, so
$\chi_s \approx 0.5$ and $\chi_s$ has no power to resolve $\delms$
near this limit.
Unfortunately, no experiment has yet succeeded in directly measuring the
frequency of $\Bs$ oscillations; only lower limits have been determined.

For several reasons, including the notorious difficulty in combining
exlusion regions from several experiments, an alternate method of fitting
for $\delms$ was proposed by Moser and Roussarie \cite{Moser97}.  This method,
commonly called the amplitude method, provides a consistent fitting
procedure for all experiments, making it much easier to combine the results.
Essentially this method entails searching for a peak in the power spectrum of the data
as a function of frequency.  The likelihood function for a given data sample
is constructed in the usual way, by estimating the amounts of
various components of the sample.  Each component's expected
proper-time distribution is formed based on its true distribution,
convoluted with the
decay-length and momentum-resolution functions, and combined in the
appropriate fractions.  The amplitude method introduces one difference
in the construction of the likelihood function:
the true probability densities of the $\Bs$ and $\Bsbar$ mesons are replaced by
\begin{eqnarray}
P(t) & = & \frac{\Gamma_s}{2}e^{-\Gamma_st}(1\pm{\cal A}\cos\delms t),
\label{amplDefn}
\end{eqnarray}
\noindent i.e. an amplitude ${\cal A}$ is introduced
in front of the oscillation term.

The new likelihood function is then maximized as a function of ${\cal A}$
alone, i.e. all parameters, such as sample composition parameters,
$B$ hadron lifetimes, and even $\delms$, are fixed.  This gives a measure of
${\cal A}(\delms)$ and an error $\sigma_{\cal A}(\delms)$ for that value of
$\delms$.  This fit is repeated for many different values of $\delms$ to 
sketch out the shape of the ${\cal A}(\delms)$ curve.  If
$\delms=\delms^{\rm true}$, the fit should
return ${\cal A}=1$, within its errors.
For $\delms$ far from its true value, ${\cal A}$ should fluctuate around 0,
within its errors.

A given $\delms$ can thus be excluded at greater than 95\% CL if
${\cal A} + 1.645\sigma_{\cal A} \le 1$, assuming $\sigma$ represents a
gaussian error.  An experiment's 95\% CL lower limit on $\delms$ is
defined as the largest $\delms$ such that all smaller values have measurement
probabilities below 5\%.

Of course, as with all limits, for a given error $\sigma_{\cal A}$,
fluctuations in the central value ${\cal A}$
can result in more or less favourable exlusion regions.
The sensitivity of an experiment is therefore defined as the
largest $\delms$ excluded if ${\cal A}=0$ for all $\delms$,
i.e. the point at which $1.645 \sigma_{\cal A}(\delms)=1$.

The lower limits on $\delms$ and sensitivities of the analyses presented here
are dominated by their statistical errors.  The values quoted include a
degradation due to systematic effects, but since these change the result very
little, the major systematic errors are not listed here.

Searches for $\Bs$ mixing fall into two rough categories: high
statistics but
low $\Bs$-purity analyses, based primarily on inclusive lepton
samples, with typical
$\Bs$ purities of $\sim 10\%$; and low statistics, high-purity ($\sim 60\%$)
samples in which the $D_s$ daughter of the $\Bs$ has been reconstructed.
In order to obtain the maximal sensitivity to $\delms$, these experiments
combine several production flavor tags.

\subsection{Inclusive Analyses}
\label{BsInclSection}
These analyses closely follow the form of their $B^0$ counterpart --
the inclusive $l$-jet charge analyses described in Section~\ref{lJetQSection}.

\subsubsection{EVENT SELECTION AND VERTEXING}

ALEPH \cite{Aleph99}, DELPHI \cite{Delphi98-2}, OPAL \cite{Opal99} and
SLD \cite{SLD99-1,SLD99-2} (with two analyses, referred to here as
$l$-track and $lD$) search for $B_s$ oscillations in the inclusive
lepton samples described in Section~\ref{lJetQSection}.
A third SLD analysis (referred to as ``dipole'') is based on an inclusive
vertex sample similar to the
one used for the dipole-based $B^0$ analysis described in
Section~\ref{InclusiveSection}.  Table~\ref{BsInclLepCuts} shows an overview
of the event selection for these analyses.

\begin{table}[ht]
\caption{Event selection for the inclusive-type $\Bs$ mixing measurements.
All analyses except the SLD dipole are based on single lepton or dilepton
samples.}
\begin{tabular}{@{}lccrc@{}}
\hline\hline
Experiment & \begin{tabular}{@{}c@{}}$p(l)$\\ (GeV)\end{tabular} &
\begin{tabular}{@{}c@{}}$\ptr(l)$\\ (GeV) \end{tabular}& Events &
$\Bs$ fraction \\ \hline 
ALEPH \cite{Aleph99} & 3.0 & 1.25 & 33,023 & 10.4\% \\
DELPHI \cite{Delphi98-2} & 3.0 & 1.0 & 78,476 & 10.5\% \\
OPAL \cite{Opal99} & 2.0 & 0.7 & 53,050 & 10.5\%   \\ 
SLD \cite{SLD99-1} ($l$+track) & $1.0(e),2.0(\mu)$ & 0.8 & 9,691 & 8.5\%  \\ 
SLD \cite{SLD99-2} ($l$+D) &  & 0.9 & 2,009 & 15.9\%  \\ 
SLD \cite{SLD99-2} (dipole) & N/A & N/A & 7,547 & 15.6\%  \\ \hline
\end{tabular}
\label{BsInclLepCuts}
\end{table}

ALEPH, DELPHI and SLD ($l$-track and $lD$ analyses)
search for a vertex by applying the same inclusive algorithms
used in the $B^0$ $l$-jet charge analyses, described in
Section~\ref{dilSection} (ALEPH and DELPHI)
and Section~\ref{lJetQSection} (SLD).  
For the SLD $lD$ sample, a $D$-like vertex is required in the events.
Demanding the total charge of tracks coming from the
$B$ and $D$ vertices to be zero enhances the $\Bs$ (and $B^0$) content of this
sample to 15.9\%.  DELPHI and SLD reconstruct the $B$ decay
length and momentum
following the procedure used for the $B^0$ analyses;
in addition, SLD adds a few extra requirements designed to
improve the decay length resolution and supress backgrounds from
cascade $b\rightarrow c \rightarrow l$ decays (such as the increased
$\ptr$ requirement).

ALEPH adds
requirements on the angle that the charm pseudotrack (produced by the vertexing
algorithm) makes with the lepton and
with the jet, and on the mass of the $B$ decay products.  These cuts,
which are not present in the $B^0$ analysis, reduce the number of
events by $\sim 65\%$, but decrease the non-$b$ background by a factor of
$\sim 4$
and significantly increase both the decay-length and momentum resolution
of the remaining events.
The energy of the charm particle is estimated by clustering charged and
neutral particles with energy $>0.5$ GeV (to reduce fragmentation clutter)
with the charm vertex's tracks until the mass exceeds 2.7 GeV.  The neutrino
energy is estimated from the missing energy in the lepton hemisphere using the
beam constraint as described above.  The $B$ momentum is then estimated
by $p_B = \sqrt{(E_c+E_\nu+E_l)^2-m_B^2}$.  The core resolutions given
by these methods are $280\mu$m on the decay length and 7\% on the fractional
momentum resolution.

OPAL uses a different vertexing algorithm than that used in the $B^0$ inclusive
lepton analysis.  They attempt to find both the $b$ and charm decay vertex
simultaneously.  They form a likelihood to which  each track contributes
$(w/2)\times (P_b+P_c) + (1-w)\times P_p)$, where $P_p, P_b$ and $P_c$ are the
probabilities that the track originates from the primary, $b$ or charm vertex,
as determined from the track's (three-dimensional) impact parameters to 
these vertex positions, and
$w$ is the probability that the track originates from the $b$ vertex,
determined from the track momentum and angle relative to the $b$-jet direction.
For the lepton track, $w$ is set to 1.

The inclusive vertexing algorithm used in SLD's dipole analysis differs from
that used in the $B^0$ analysis.  In brief, it relies upon the fact that the
$D$ and $B$ flight directions are almost identical.  This implies that
a single straight line (in three-dimensional space) can be found such that
all the
$B$ and $D$ decay tracks intersect it at one of two points.
The algorithm first
tries to find this line by minimizing a $\chi^2$ based on the
intersections of all charged tracks with the line.  Once this
line is found, it is assigned a finite width and considered a "ghost" track.
Tracks are again vertexed with this track or the primary vertex to build up
the primary, secondary ($B$) and tertiary ($D$) vertices (see 
Reference~\cite{SLD99-2} for the  details of this
method).
The $B$ momentum is estimated as in the $lD$ analysis.

\subsubsection{FLAVOR TAGGING}

All the inclusive lepton analyses use the sign of the
lepton to tag the $b$ flavor at the decay point .  The SLD dipole analysis defines
a charge dipole $\delta Q \equiv l_{BD} \times \mbox{sign}(Q_D-Q_B)$, where
$Q_B$ and $Q_D$ are the charges of the $B$ and $D$ vertices, and $l_{BD}$ is the
distance between them.  As in the SLD $B^0$ dipole analysis, $\delta Q >0$
tags a $\Bbar$.

ALEPH uses a combination of three production flavor tags:
\begin{itemize}
\item Opposite side lepton ($p>3.0$ GeV) and
opposite-side jet charge ($Q_H(0.5)$).  See
Sections~\ref{dilSection} and \ref{lJetQSection}.
No $\ptr$ cut is imposed on the lepton.
\item Fragmentation (SST) kaon: The kaon candidate, selected by
dE/dx, is required to be more consistent with originating from the primary
vertex
than the secondary vertex and have direction
within $45^\circ$ of the $\Bs$ direction.
\end{itemize}

The data are categorized according to the presence of these tags.
The lepton tag always takes precedence if it exists.  If not, the
kaon tag then used, if present; otherwise the sign of the jet-charge
serves as the tag.  The average right-sign probability of this tag is 71\%.

DELPHI uses the following combination of production tag variables in the
hemisphere opposite the lepton:
\begin{itemize}
\item The opposite-side jet charge, $Q_H(0.6)$ (the hemisphere
jet charge is defined in Section~\ref{lJetQSection}); 
the opposite-side kaon charge, $Q_K$ (the jet charge evaluated for
kaon candidates alone); and the opposite-side lepton, weighted by its
$\ptr$ (see Section~\ref{dilSection}).
\item The sum of charges of tracks not compatible with the primary event vertex
and the sum of charges of tracks compatible with the primary event vertex
\end{itemize}
\noindent In the lepton hemisphere, DELPHI uses:
\begin{itemize}
\item The vertex-side jet charge: $Q_H(0.6)$
\item The fragmentation (SST) kaon or fragmentation (SST) $\Lambda^0$
(the rapidity of the highest-momentum kaon or $\Lambda^0$ candidate
relative to the
thrust axis that is compatible with being a fragmentation track, signed
by its charge.)
\end{itemize}
All this information, along with a tag based on the polar angle of the
thrust axis (see Section~\ref{prodTagSection}), is combined
into one variable.  The
average right-sign probability of this tag is 69\%.

OPAL also uses a combination of tags:
\begin{itemize}
\item The opposite-side lepton -- a lepton passing the standard
OPAL requirements.
\item The modified jet charge --  a combination of  jet charges from both the
opposite and vertex hemisphere, as well as fragmentation kaon
information and opposite-side vertex charge.  On the vertex side, $Q_{\rm vtx}$
combines the unweighted 
$b$-jet charge [$Q_H(0)$, with $Q_H$ given in Table~\ref{JetQDefn}]
and the hemisphere charge
[$Q_H(0)$ summed over all tracks in the hemisphere], as well as
two other jet charges
(with track weights given by neural nets which
were fed the track's kaon and pion probabilities and momentum, direction and
impact parameter information.

The opposite hemisphere's tag, $Q_{\rm opp}$, combines
the standard jet-charges $Q_H(1.0)$ and $Q_H(0)$, the output of
a neural net similar to
that used in the vertex hemisphere, and, if available, the total
charge of tracks in a viable secondary vertex.

The tag variables $Q_{\rm vtx}$ and $Q_{\rm opp}$ are used to form the
final variable
\begin{eqnarray*}
\Qt & = & \frac{2 Q_{\rm vtx} Q_{\rm opp}}{ Q_{\rm vtx} Q_{\rm opp} +
(1- Q_{\rm vtx} Q_{\rm opp})}-1.
\end{eqnarray*}
\end{itemize}
OPAL estimates this tag variable is 40\% more effective than the combined
jet charge used in the $B^0$ $l$-jet charge analysis.

All three SLD analyses use the same combination of
jet-charge and polarization (polar angle) initial-state
tags described in the $B^0$ $l$-jet charge section.

\subsubsection{FIT METHOD AND RESULTS}

The sample composition of the lepton hemisphere is determined event-by-event.
It is similar to that of the
inclusive lepton samples described in Section~\ref{lJetQSection}.  The
resulting increase in sensitivity is
equivalent to having approximately 30\%
more events.
All experiments perform amplitude fits to the data.  The individual
lower limits on $\delms$ and analysis sensitivities are listed in
Table~\ref{dmsLimits}.

\subsection{Analyses with Reconstructed $D_s$ Decays}
\label{DsSection}

The remaining analyses take a more exlusive approach, ranging from
partially reconstructing the $D_s$ decay in $\Bs\rightarrow D_s l$ decays,
 to fully reconstructing the
$D_s$ decay but partially reconstructing the $\Bs$ in
$\Bs \rightarrow D_s l X, D_s h X$ decays, to fully
reconstructing the $\Bs$ decay.  Clearly the latter will give ideal vertex
and momentum determinations, since all the $\Bs$ decay products are
found; however
the very small exclusive $\Bs$ branching ratios mean that the event sample
is rather small.

ALEPH and DELPHI reconstruct many $D_s$ decay modes,
summarized in Table~\ref{DsModes}.
The strange mesons are reconstructed in their usual accessible modes:
$\phi\rightarrow K^+K^-$, 
$\overline{K}^{*0}\rightarrow K^-\pi^+$,
$\overline{K}^{*+}\rightarrow K^0_s\pi^+$, and the $f$ resonance as
$f(980)\rightarrow\pi^+\pi^-$.  For details on the selection criteria,
see the references indicated in Table~\ref{DsSamples}.

\begin{table}[ht]
\caption{$D_s$ decay modes searched for by various experiments.}
\begin{tabular}{@{}lcc|ccc|c@{}} \hline\hline
& \multicolumn{2}{c|}{ALEPH} & \multicolumn{3}{c|}{DELPHI} & CDF \\ \hline
$D_s$ mode & $D_s l$ & $D_s h$ & $D_s l/\phi l$ &
$D_s h$ & Full $\Bs$ & $\phi l$  \\ \hline
$\phi \pi^+$ & Y & Y & Y & Y & Y & {} \\ 
$\phi \pi^+\pi^0$ & Y & {} & Y & {} & {} & {} \\ 
$\phi \pi^+ \pi^-\pi^+$ & Y & {} & Y & {} & Y & {}\\ 
$\overline{K}^{*0} K^+$ & Y & Y & Y & Y & Y & {} \\ 
$\overline{K}^{*0} K^{*+}$ & Y & {} & Y & {} & Y & {}\\ 
$K^0_s K^+$ & Y & Y & Y & {} & Y & {} \\ 
$f(980)\pi^+$ & {} & {} & {} & {} & Y & {} \\ 
$\phi l \nu_l (l=e,\mu)$ & Y & Y & Y & {} & {} & {} \\  \hline
$\phi h^+$ & {} & {} & Y & {} & {} & Y \\ \hline
\end{tabular}
\label{DsModes}
\end{table}

\subsubsection{EVENT SELECTION}

ALEPH \cite{Aleph96-1,Aleph98}
and DELPHI \cite{Delphi98-1,Delphi99-1} reconstruct
$\Bs\rightarrow D_s^- l^+ X$ and 
$\Bs\rightarrow D_s^- h^+ X$, with the $D_s$ decaying in the modes given
in Table~\ref{DsModes}.  In addition, DELPHI searches for fully
 reconstructed $\Bs$ (up to a missing $\gamma$ or $\pi^0$) in the following
modes: $D_s^-\pi^+, D_s^-a_1^+, \overline{D^0}K^-\pi^+$, and
$\overline{D^0}K^-a_1^+$, with
$a_1^+\rightarrow \rho^0\pi^+\rightarrow \pi^+\pi^-\pi^+$ and
$\Dzbar\rightarrow K^+\pi^-, K^+\pi^-\pi^+\pi^-$.

Note that for $\Bs\rightarrow D_s^*\rightarrow D_s\gamma,
D_s\pi^0$ transitions, the photon or $\pi^0$ may not be reconstructed.
Thus the mass
spectrum of the $\Bs$ candidates is expected to
have a sharp peak at the $\Bs$ mass, and a wider satellite peak at lower mass
from $D_S^*$ decays.  DELPHI finds $8\pm 4$ signal events out of the
11 events in the $\Bs$ mass region, and
$15\pm 8$ signal events out of 33 in the satellite region.

\begin{table}[ht]
\caption{Summary of the $B_s$ decay modes, sample sizes and purity, and
right-sign tag probability for analyses with a partially or fully reconstructed
$D_s$.}
\begin{tabular}{@{}lcrcc@{}}
\hline\hline
Experiment & $B_s$ Mode & Events & $\Bs$ frac. & $<\!P_r\!>$ \\ \hline
\vspace{8pt}
ALEPH \cite{Aleph96-1,Aleph98} & $\begin{array}{c}D_s l \\ D_s h\end{array}$ &
$\begin{array}{@{}r@{}}277\\ 1,620 \end{array}$ &
$\begin{array}{@{}c@{}}66\% \\ 22\% \end{array}$ &
$\begin{array}{@{}c@{}} 0.73 \\ 0.74 \end{array}$ \\

\vspace{8pt}
CDF \cite{CDF99-6} & $\phi h l$ & 1,068 & 61\% & 0.76 \\ 

DELPHI \cite{Delphi98-1,Delphi99-1} & $\begin{array}{c} D_s l \\ \phi h l \\ D_s h \\
\Bs \mbox{ (full)}\end{array}$ &
$\begin{array}{@{}r@{}} 436 \\ 441 \\ 2,953 \\ 44 \end{array}$
& $\begin{array}{@{}c@{}}53\% \\ 9\% \\ 20\% \\ 51\% \end{array}$ &
$\begin{array}{@{}c@{}} 0.78 \\ 0.78 \\ 0.78 \\ 0.78 \end{array}$  \\ \hline
\end{tabular}
\label{DsSamples}
\end{table}

In addition to the exclusive $D_s$ modes listed, for events with a lepton,
DELPHI also searches for a
partially reconstructed $D_s$ of the form $D_s\rightarrow \phi h^+ X$ (which
was not found as one of the above modes).

CDF \cite{CDF99-6} searches
for $\phi\rightarrow K^+K^-$ candidates in the low-$p_t$ dilepton
samples ($\mu\mu$ and $e\mu$) described in Section~\ref{dilSection}.
The lepton
on the $\phi$ side is required to have $\ptr>1.0$ GeV.  A charged hadron
near the $\phi l$ pair is required, and the effective masses are restricted
to be $1.0<m_{\phi h}<2.0$ and
$m_{\phi h l}<5.0$, consistent with decay kinematics for
$\Bs\rightarrow D_s l \nu$ and $D_s\rightarrow \phi h X$.

Table~\ref{DsSamples} gives an overview of these data samples.

\subsubsection{DECAY LENGTH AND PROPER TIME}
A $B$ decay vertex is easily obtained in these events.  First the $D_s$
decay products are vertexed and a pseudotrack formed, which passes through
this vertex and has a momentum given by the $D_s$ momentum.  This track is
intersected with the other track available, $l$ or hadron,
to give the $\Bs$ decay point.

For the $D_s l$ samples, ALEPH and DELPHI estimate the $B$ momentum based
on the momentum of the $D_s l$ system and apply a correction for the neutrino
obtained from the missing energy in the hemisphere.
For the $D_s h$ sample, ALEPH adds charged and neutral particles to the
$D_s h$ system and selects the most likely combination based on the
effective mass.
The momentum of the fully reconstructed decay is simply measured from all the
decay products.
CDF corrects the measured momentum of the $\phi h l$ system for the missing
energy using a mean value derived from Monte Carlo simulation.

\subsubsection{FLAVOR TAGGING}
All experiments use the sign of the lepton on the vertex side as the
final-state flavor tag.
ALEPH and DELPHI use the combination of production flavor
 tags described in the inclusive lepton section above, whereas
CDF uses the sign of the opposite-side lepton.

\subsubsection{FIT METHOD AND RESULTS}

DELPHI estimates the right-tag probability event-by-event by using the shape of
the $P_r(x)$ distributions, which are determined from Monte Carlo simulation.
This improves the
effective mean right-tag probability, e.g., from 0.745 to 0.78 for the
1994-1995 $D_s l$ data.  The effectiveness of the combined tags can be seen
by comparing the right-sign tag probabilities in Table~\ref{DsSamples}
to the individual tags used in the $B^0$ analyses.

All the experiments perform an amplitude style fit, with the $\delms$ exclusion
regions and sensitivities given in Table~\ref{dmsLimits}.

\subsection{Summary of $\Bs$ Mixing}

\begin{table}[ht]
\caption{Excluded regions of $\delms$ and the sensitivity of each analysis.}
\begin{tabular}{@{}lcc@{}}
\hline\hline
Experiment & \begin{tabular}{@{}l@{}}Excluded $\delms$ \\
region (95\% CL)\end{tabular} & Sensitivity \\ \hline
ALEPH $l$ \cite{Aleph99} & $<9.5$ ps$^{-1}$ & $9.5$ ps$^{-1}$ \\ 
ALEPH $D_s l$ \cite{Aleph96-1} &  $<6.6$ ps$^{-1}$ & $6.7$ ps$^{-1}$ \\ 
ALEPH $D_s h$ \cite{Aleph98} & $<3.9$ ps$^{-1}$, $6.5-8.8$ ps$^{-1}$ & $4.1$ ps$^{-1}$ \\ 
CDF $\phi l-l$ \cite{CDF99-6} & $<5.8$ ps$^{-1}$ & $5.1$ ps$^{-1}$ \\ 
DELPHI $l$ \cite{Delphi98-2} & $<4.7$ ps$^{-1}$ & $6.5$ ps$^{-1}$ \\ 
DELPHI $D_s l$ \cite{Delphi98-1} & $<7.5$ ps$^{-1}$ & $8.2$ ps$^{-1}$ \\ 
DELPHI $\Bs + D_s h$ \cite{Delphi99-1} & $<4.0$ ps$^{-1}$ & $3.2$ ps$^{-1}$  \\ 
OPAL $l$ \cite{Opal99} & $<5.2$ ps$^{-1}$ & $7.2$ ps$^{-1}$ \\ 
SLD $l$ \cite{SLD99-1} & $<1.3$ ps$^{-1}$,$2.0-8.6$ ps$^{-1}$,$9.8-12.2$ ps$^{-1}$
& $3.8$ ps$^{-1}$ \\ 
SLD $lD$ \cite{SLD99-2} & $<5.2$ ps$^{-1}$ & $3.5$ ps$^{-1}$ \\ 
SLD dipole \cite{SLD99-2} & $<5.2$ ps$^{-1}$ & $5.4$ ps$^{-1}$ \\ \hline
\end{tabular}
\label{dmsLimits}
\end{table}

\begin{figure}[htb]
\epsfxsize=10cm
\centerline{\epsffile{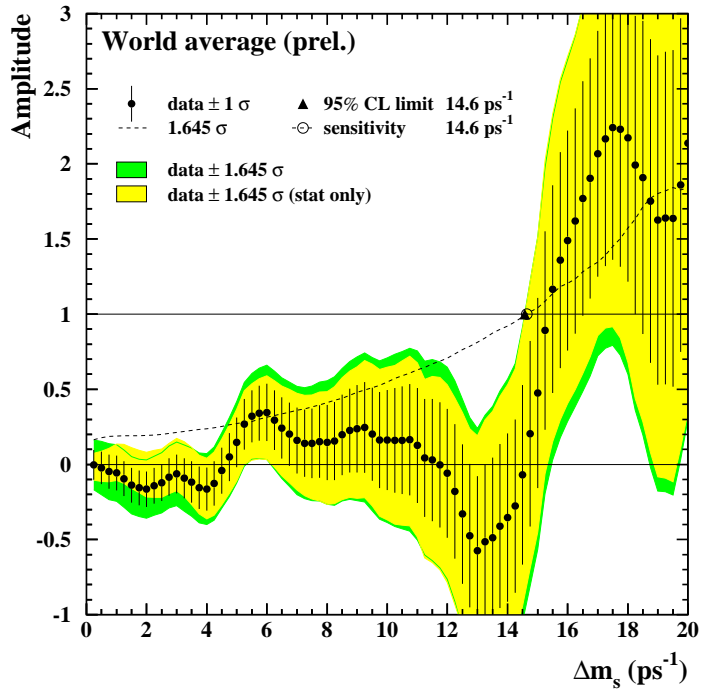}}
\caption{$B$ Oscillation Working Group's \cite{BOSC} combined plot of fitted
amplitude versus $\delms$, for the
$B_s$ mixing analyses presented in this section.}
\label{BsSummary}
\end{figure}

Table~\ref{dmsLimits} summarizes the excluded regions of $\delms$ from
each of the analyses presented, along with their sensitivities.
The $B$ Oscillation Working Group has combined these, and finds
that $\delms > 14.6$ ps$^{-1}$ at the 95\% CL \cite{BOSC}.
Figure~\ref{BsSummary}
contains the combined amplitude plot showing the value of the amplitude
${\cal A}$ in Equation~\ref{amplDefn} versus assumed $\delms$, from which
the lower limit and experimental sensitivity can be read.

\section{$\Bs$ LIFETIME DIFFERENCE}
\label{DelGammaSection}

Another, more indirect way of searching for $\Bs$ mixing utilizes the
relationship between the mass difference of the two eigenstates, $\delms$, and
the decay rate difference $\delg_s$ \cite{Hagelin81},
\begin{eqnarray}
\frac{\delg_s}{\delms} & = & -\frac{3}{2}\pi \frac{m_b^2}{m_t^2}
\frac{\eta^{\delg_s}_{\rm QCD}}{\eta^{\delms}_{\rm QCD}}.
\end{eqnarray}
The QCD correction factors $\eta$ have recently been estimated, 
giving $\delg_s/\delm_s = (5.6\pm2.6)\times 10^{-3}$
\cite{Beneke96}.

A large $\delm_s$
corresponds to a large lifetime difference between the heavy and light states.
Even a moderately large $\delm_s \approx 20$ would result in a 17\%
difference in their lifetimes.

This difference has been searched for in two ways.
First, $\Bs$ decays to final states that are not CP eigenstates
should contain a mixture of both $B_H$ and $B_L$ components.  
Several experiments have fit their $\Bs$ lifetime distributions to a sum
of two exponentials of decay widths $\Gamma_s\pm \delg_s/2$.
From their sample of $\sim 600$ $\Bs\rightarrow l^+ D_s^-\nu X$ events,
CDF \cite{CDF99-4} finds
$\delg_s / \Gamma_s = 0.34^{+0.31}_{-0.34}$ (statistical only),
corresponding to an upper limit of $\delg_s / \Gamma_s <0.83$ (95\% CL).
DELPHI \cite{Delphi99}, using their $D_s^\pm l^\mp$ and $D_s^\pm h^\mp$
samples from the $\Bs$ mixing analysis (Section~\ref{DsSection}),
finds $\delg_s / \Gamma_s <0.42$ (95\% CL).

L3 \cite{L399} fits 445,000 events, selected using an inclusive vertexing
 algorithm, to a sum of 
$B^0$, $B^+$, $\Lambda_b$ exponential decay distributions
and a two-lifetime $\Bs$ component, resulting in
$\delg_s / \Gamma_s = 0.00^{+0.30}_{-0.00}$ (statistical only).
Including systematic effects, L3 estimates an upper limit of
$\delg_s / \Gamma_s <0.67$ (95\% CL).

A second method to search for a lifetime difference is to measure the lifetime
of $\Bs$ decays to modes that are CP eigenstates or are mixtures
dominanted by
one CP state.  These decays arise primarily from either $B_H$ or $B_L$ decays,
depending on the CP eigenvalue of the final state.
The difference between the lifetime
measured in this sample and the mean lifetime measured over a mixture
of final states measures $\Delta \Gamma$.
For example, ALEPH \cite{Aleph99-1} found $32\pm 17$ decays of
$\Bs\rightarrow D_s^{(*)+}D_s^{(*)-} \rightarrow \phi\phi X$, from which
they extracted $\tau(B_s) = 1.42\pm 0.23 \pm 0.16$ ps.  Assuming this is
a pure CP-even decay, comparison to the mean $\Bs$ lifetime \cite{PDG98} of
$\tau_{B_s} = 1.55$ ps gives $\delg_s / \Gamma_s = 0.17\pm 0.35$.

CDF measures the lifetime using $58\pm 12$ decays of
$\Bs\rightarrow J/\psi \phi$
and finds $\tau(B_s) = 1.34\pm 0.23\pm 0.05$ ps \cite{CDF96}.
This final state is not necessarily a pure CP eigenstate, and using a
transversity analysis \cite{CDF4672}
CDF finds
that the CP-odd component represents $23\pm 19$\% of the sample.
The higher statistics samples soon to be available at the Tevatron
will clearly benefit this type of analysis.

The DELPHI result, combined with the mean $\Bs$ lifetime, gives an
upper limit of $\delms < 48$ ps$^{-1}$ (95\% CL).

\section{CONCLUSION}

Mixing between the particle and antiparticle states of the $B^0$ and $\Bs$
mesons has several consequences.  It leads to physical
states that are no longer
eigenstates of flavor, particle-antiparticle oscillations can
generate CP asymmetries between  $B$ and $\Bbar$ decays
and their frequency is relatively directly related to CKM matrix elements.
The $B$ system is unique in having two such particles readily available,
which can be studied and compared.

Several experiments have succeded in observing the time-dependence
of $B^0$ flavor oscillations, using a wide variety of techniques.
The measurement
of the $\Bs$ oscillation frequency will have to wait for the next generation of
experiments -- current experiments are only able to place a lower
limit.
The principle results presented here are:
$\delmd = 0.484\pm0.015$ ps$^{-1}$ and $\delms > 14.6$ ps$^{-1}$.

With the results from Section~\ref{fbSection}, $\fbd=200\pm 16$ MeV,
$\bnlod=1.37\pm 0.08$, $\fbd/\fbs = 1.15\pm0.06$, $\bbd/\bbs=1.01\pm0.01$
and $\overline{m_t}(174) = 167 \pm 5$ GeV, $\eta_B = 0.55\pm 0.01$,
the CKM matrix element $|\Vtd|$ and ratio $|\Vts|/|\Vtd|$ can be extracted
according to Equation~\ref{delmFormula},
\begin{eqnarray*}
\delmd & = & \left( 24.4
\left[\frac{\fbd}{200 \mbox{ MeV}}\right]^2 \left[\frac{\bnlod}{1.37}\right] \pm 5.0 \right)
 |\Vtd|^2 \hspace{0.5cm} \mbox{(eV)} \\
\frac{\delms}{\delmd} & = & 
(1.36\pm0.14) \left|\frac{\Vts}{\Vtd}\right|^2,
\end{eqnarray*}
corresponding to
$|\Vtd| = (3.6\pm0.4) \times 10^{-3}$ and
$|\Vts/\Vtd| > 4.7$ (95\% CL).

The future direction for $B$ mixing measurements is clear.
Measuring $\delms$, either
directly through a time-dependent oscillation analysis, or through a
heavy-light lifetime difference is crucial.  The prospects for this occurring
in the next few years are good, as the upgraded CDF and D\O\ detectors
are expected to provide
significantly improved statistics and sensitivity to measure
$\delms$.

It is equally important to make further
progress in calculating $f_B$ and $B_B$.
The lattice calculations of these constants are maturing rapidly, and with
unquenched simulations becoming available, more precise estimates with
a full evaluation of systematic uncertainties should be within reach.

These measurements are an important part of exploring the unitarity of the
CKM matrix, along with the various $B$-sector CP violation searches which
will take place over the next few years.  The study of $B$ mixing will surely
remain vibrant for years to come.


\begin{thebibliography}{999}


\bibitem{GM55} Gell-Mann M, Pais A.
{\em Phys. Rev.} 97:1387 (1955)

\bibitem{Lande56} Lande K, et al
{\em Phys. Rev.} 103:1901 (1956)

\bibitem{UA187} Albajar C, et al (UA1 Collaboration).
{\em Phys. Lett.} B186:247 (1987) 

\bibitem{Argus87} Albrecht H, et al (ARGUS Collaboration).
{\em Phys. Lett.} B192:245 (1987) 

\bibitem{CLEO89} Artuso M, et al (CLEO Collaboration).
{\em Phys. Rev. Lett.} 62:2233 (1989) 

\bibitem{PDG98} Caso C, et al (Particle Data Group)
{\em Eur. Phys. J. C} 63:1 (1998)

\bibitem{Chid94} Albrecht H, et al (ARGUS Collaboration).
{\em Z. Phys. C} 55:357 (1992);
Albrecht H, et al (ARGUS Collaboration).
{\em Phys. Lett.} B324:249 (1994);
Bartelt J, et al (CLEO Collaboration).
{\em Phys. Rev. Lett.} 71:1680 (1993) 

\bibitem{CKM} Cabbibo N. {\em Phys. Rev. Lett.} 10:531 (1963); 
Kobayashi M, Maskawa K {\em Prog. Theor. Phys.} 49:652 (1973)

\bibitem{Wolf84} Wolfenstein L. {\em Phys. Rev. Lett.} 51:1945 (1984)

\bibitem{Beneke96} Beneke M, Buchalla G, Dunietz I.
{\em Phys. Rev. D} 54:4419 (1996) 

\bibitem{Lipkin97} Grossman L, Lipkin H. {\em Phys. Rev. D} 55:2760 (1997)

\bibitem{GaLee74} Gaillard M, Lee B. {\em Phys. Rev. D} 10:897 (1974)

\bibitem{Fujik72} Fujikawa K, Lee B, Sanda A.
{\em Phys. Rev. D} 6:2923 (1972)

\bibitem{GIM70} Glashow S, Illiopoulos J, Maiani L.
{\em Phys. Rev. D} 2:1285 (1970) 

\bibitem{inamilim} Inami T, Lim CS. {\em Prog. Theor. Phys.}
65:297 (1981)

\bibitem{Buras90} Buras A, Jamin M, Weisz P. {\em Nucl. Phys.}
B347:491 (1990)


\bibitem{Alephbtau} Buskulic D, et al (ALEPH Collaboration).
{\em Phys. Lett.} B343:444(1995)

\bibitem{Delphibtau} Abreu P, et al (DELPHI Collaboration).
CERN-EP-99-162 (1999)

\bibitem{L3btau} Acciarri M, et al (L3 Collaboration).
{\em Phys. Lett.} B396:327 (1997)

\bibitem{CLEObtau} Artuso M, et al (CLEO Collaboration).
{\em Phys. Rev. Lett.} 75:785 (1995)

\bibitem{bigi95} Bigi I. hep-ph/9508408 (1995); 
Bigi I, Uraltsev N. {\em Phys. Lett.} B280:271 (1992)

\bibitem{CDFexclusive98} Abe F, et al (CDF Collaboration).
{\em Phys. Rev. D} 57:5382 (1998)

\bibitem{CDFsl98} Abe F, et al (CDF Collaboration).
{\em Phys. Rev. D} 58:092002 (1998).

\bibitem{BaBartdr} BaBar Collaboration. {\em The BaBar Physics Book}.
SLAC-R-504 (1998)

\bibitem{neubert95} Neubert M, Schrajda CT.
{\em Nucl. Phys.} B438:238 (1995)

\bibitem{SW85} Sheikholeslami B, Wohlert R.
{\em Nucl. Phys.} B259:572 (1985) 


\bibitem{Gavela88} Gavela M, et al {\em Phys. Lett.}
B206:113 (1988)

\bibitem{BDHS88} Bernard C, et al {\em Phys. Rev. D} 38:3540 (1988)

\bibitem{Boucaud89} Boucaud P, et al
 {\em Phys. Lett.} B220:219 (1989)

\bibitem{Allton91} Allton C, et al
{\em Nucl. Phys.} B349:504 (1991)

\bibitem{Alex91} Alexandrou C, et al
{\em Phys. Lett.} B256:60 (1991)

\bibitem{FNAL92} Duncan A, et al {\em Nucl. Phys. Proc. Supplement} 
30:433 (1992)

\bibitem{ELC92} Abada A, et al {\em Nucl. Phys.} B376:172 (1992)

\bibitem{APE94} Allton C, et al (APE Collaboration).
{\em Phys. Lett.} B326:295 (1994)

\bibitem{Alex94} Alexandrou C, et al
 {\em Nucl. Phys.} B414:815 (1994)

\bibitem{BLS94}  Bernard C, Labrenz J, Soni A.
{\em Phys. Rev. D} 49:2536 (1994)

\bibitem{Duncan95} Duncan A, et al {\em Phys. Rev. D} 51:5101 (1995)

\bibitem{UKQCD94} Baxter R, et al (UKQCD Collaboration).
{\em Phys. Rev. D} 49:1594 (1994)

\bibitem{HEMCGC94} Duncan A, et al (HEMCGC Collaboration).
{\em Phys. Rev. D} 49:3546 (1994)

\bibitem{PCW94} Alexandrou C, et al (PCW Collaboration).
{\em Z. Phys. C} 62:659 (1994)

\bibitem{Hash94} Hashimoto S. {\em Phys. Rev. D} 50:4639 (1994)

\bibitem{APE97} Allton C, et al (APE Collaboration).
{\em Phys. Lett.} B405:133 (1997)

\bibitem{APE98} Becerevic D, et al (APE Collaboration).
hep-lat/9811003 (1998)

\bibitem{FNAL98} El-Khadra A, et al
{\em Phys. Rev. D} 58:014506 (1998)

\bibitem{JLQCD98} Aoiki S, et al (JLQCD Collaboration).
{\em Phys. Rev. Lett.} 80:5711 (1998)

\bibitem{MILC98} Bernard C, et al (MILC Collaboration).
{\em Phys. Rev. Lett.} 81:4812 (1998)

\bibitem{SGO97}  Ali Khan A, et al (SGO Collaboration).
{\em Phys. Rev. D} 56:7012 (1997)

\bibitem{Ishi97}  Ishikawa K, et al
{\em Phys. Rev. D} 56:7028 (1997)

\bibitem{GLOK98} Ali Khan A, et al (GLOK Collaboration).
{\em Phys. Lett.} B427:132 (1998)

\bibitem{MILC99} Bernard C, et al (MILC Collaboration).
hep-lat/9909121 (1999) 

\bibitem{Collins99} Collins S, et al
hep-lat/9901001 (1999) 

\bibitem{CPPACS99-1} Ali Khan A, et al (CP-PACS Collaboration).
hep-lat/9909052 (1999) 

\bibitem{CPPACS99-2} Ali Khan A, et al (CP-PACS Collaboration).
hep-lat/9911039 (1999) 


\bibitem{KLM} Lepage G and Mackenzie P.{\em Phys. Rev. D}
48:2250 (1993); 
El-Khadra A, Kronfeld A, Mackenzie P.{\em Phys. Rev. D}
55:3933 (1997) 

\bibitem{MILC97} Bernard C, et al (MILC Collaboration).
hep-lat/9709142 (1997)

\bibitem{fdsMeasures} Albrecht H, et al (Argus Collaboration).
{\em Z. Phys. C} 54:1 (1992); 
Aoki S, et al (WA75 Collaboration).
{\em Prog. Theor. Phys.} 89:131 (1993); 
Kodama K, et al (E653 Collaboration).
{\em Phys. Lett.} B382:299 (1996); 
Bai J, et al (BES Collaboration).
{\em Phys. Rev. Lett.} 74:4599 (1995); 
Acciari M, et al (L3 Collaboration).
{\em Phys. Lett.} B396:327 (1997); 
Chadha M, et al (CLEO Collaboration).
{\em Phys. Rev. D} 58:032002 (1998); 
Abreu P, et al (DELPHI Collaboration).
DELPHI 97-105 CONF 87 (1997);
Buskulic D, et al (ALEPH Collaboration).
ALEPH 98-063 (1998) 

\bibitem{KR98} Khodjamirian A, Ruckl R. hep-ph/9801443 (1998) 

\bibitem{Dr98} Draper T hep-lat/9810065 (1998)

\bibitem{GR98} Gimenez V, Reyes J. hep-lat/9806023 (1998)

\bibitem{Flynn97} Flynn J, Sachrajda CT. hep-lat/97100057 (1997) 

\bibitem{GM97} Gimenez V, Martinelli G.{\em Phys. Lett.} B398:135 (1997)

\bibitem{BBS} Bernard C, Blum T, Soni A. hep-lat/9801039 (1998) 

\bibitem{JLQCD98-2} Aoiki S, et al (JLQCD Collaboration).
hep-lat/9809152 (1998) 

\bibitem{JLQCD95} Aoiki S, et al (JLQCD Collaboration).
hep-lat/9510033 (1995) 

\bibitem{UKQCD96} Ewing A, et al (UKQCD Collaboration).
{\em Phys. Rev. D} 54:3526 (1996) 

\bibitem{BS} Bernard C, Soni A. {\em Nucl. Phys.} B47:43 (1996) 

\bibitem{Opal93} Acton P, et al (OPAL Collaboration).
{\em Phys. Lett.} B307:247 (1993) 

\bibitem{Gronau93} Gronau M, Nippe A, Rosner J.
{\em Phys. Rev. D} 47:1988 (1993) 

\bibitem{Aleph93} Buskulic D, et al (ALEPH Collaboration).
{\em Phys. Lett.} B313:498 (1993);
Buskulic D, et al (ALEPH Collaboration).
{\em Phys. Lett.} B322:441 (1994) 

\bibitem{AlephNIM95} Buskulic D, et al (ALEPH Collaboration).
{\em Nucl. Instr. Meth. A} 360:481 (1995) 

\bibitem{CDFNIM88} Abe F, et al (CDF Collaboration).
{\em Nucl. Instr. Meth. A} 271:387 (1988); 
Abe F, et al (CDF Collaboration).
{\em Phys. Rev. D} 50:2966 (1994) 

\bibitem{DelphiNIM96} Abreu P, et al (DELPHI Collaboration).
{\em Nucl. Instr. Meth. A} 378:57 (1996) 

\bibitem{L3NIM90} Adeva B, et al (L3 Collaboration).
{\em Nucl. Instr. Meth. A} 289:35 (1990); 
Adriani O, et al (L3 Collaboration).
{\em Phys. Rep.} 236:1 (1993) 

\bibitem{OpalNIM91} Ahmet K, et al (OPAL Collaboration).
{\em Nucl. Instr. Meth. A} 305:275 (1991); 
Allport P, et al (OPAL Collaboration).
{\em Nucl. Instr. Meth. A} 324:34 (1993); 
Allport P, et al (OPAL Collaboration).
{\em Nucl. Instr. Meth. A} 346:476 (1994) 

\bibitem{SLDNIM97} Abe K, et al (SLD Collaboration).
{\em Phys. Rev. D} 53:1023 (1996); 
Abe K, et al (SLD Collaboration).
{\em Nucl. Instr. Meth. A} 400:287 (1997) 

\bibitem{Aleph97-1} Buskulic D, et al (ALEPH Collaboration).
{\em Z. Phys. C} 75:397 (1997) 

\bibitem{CDF99-3} Abe F, et al (CDF Collaboration).
{\em Phys. Rev. D} 60:051101 (1999) 

\bibitem{CDF99-2} Abe F, et al (CDF Collaboration).
{\em Phys. Rev. D} 60:072003 (1999) 

\bibitem{CDF3791} Abe F, et al (CDF Collaboration).
CDFNOTE 3791, Preliminary (1996) 

\bibitem{Delphi97-1} Abreu P, et al (DELPHI Collaboration).
{\em Z. Phys. C} 76:579 (1997) 

\bibitem{L398} Acciarri M, et al (L3 Collaboration).
{\em Eur. Phys. J. C} 5:195 (1998) 

\bibitem{Opal97-2} Ackerstaff K, et al (OPAL Collaboration).
{\em Z. Phys. C} 76:417 (1997) 

\bibitem{Opal97-1} Ackerstaff K, et al (OPAL Collaboration).
{\em Z. Phys. C} 76:401 (1997) 

\bibitem{SLD96-1} Abe K, et al (SLD Collaboration).
SLAC-PUB-7228 (1996) 

\bibitem{SLD96-2}  Abe K, et al (SLD Collaboration).
SLAC-PUB-7229 (1996) 


\bibitem{CDF4526} Abe F, et al (CDF Collaboration).
CDFNOTE 4526, Preliminary (1998) 

\bibitem{Opal96} Alexander G, et al (OPAL Collaboration).
{\em Z. Phys. C} 72:377 (1996) 

\bibitem{CDF99-5} Affolder T, et al (CDF Collaboration).
{\em Phys. Rev. D} 60:112004 (1999) 

\bibitem{Aleph97-2} Buskulic D, et al (ALEPH Collaboration).
EPS-HEP Jerusalem, Contribution 596 (1997) 

\bibitem{SLD96-3} Abe K, et al (SLD Collaboration).
SLAC-PUB-7230 (1996) 

\bibitem{CDF99-1} Abe F, et al (CDF Collaboration).
{\em Phys. Rev. D} 59:032001 (1999) 

\bibitem{Cleo91} Fulton R, et al (CLEO Collaboration).
{\em Phys. Rev. D} 43:1 (1991) 

\bibitem{PDG96} Barnett R, et al (Particle Data Group).
{\em Phys. Rev. D} 54:1 (1996) 

\bibitem{BOSC} $B$ Oscillation Working Group.
{\em http://lepbosc.web.cern.ch/LEPBOSC/} (1999) 

\bibitem{Moser97} Moser HG, Roussarie A. 
{\em Nucl. Instr. Meth. A} 384:491 (1997)

\bibitem{Aleph99} Barate R, et al (ALEPH Collaboration).
{\em Eur. Phys. J. C} 7:553 (1999) 

\bibitem{Delphi98-2} Abreu P, et al (DELPHI Collaboration).
DELPHI 98-132 CONF 193 (1998) 

\bibitem{Opal99} Abbiendi G, et al (OPAL Collaboration).
{\em Eur. Phys. J. C} 11:587 (1999) 

\bibitem{SLD99-1}  Moore T, et al (SLD Collaboration).
SLAC-R-551 (1999) 

\bibitem{SLD99-2}  Abe K, et al (SLD Collaboration).
SLAC-PUB-8225 (1999) 

\bibitem{Aleph96-1} Buskulic D, et al (ALEPH Collaboration).
{\em Phys. Lett.} B377:205 (1996) 

\bibitem{Aleph98} Buskulic D, et al (ALEPH Collaboration).
{\em Eur. Phys. J. C} 4:367 (1998) 

\bibitem{Delphi98-1} Abreu P, et al (DELPHI Collaboration)
DELPHI 98-131 CONF 192 (1998) 

\bibitem{Delphi99-1} Abreu P, et al (DELPHI Collaboration)
DELPHI 99-109 CONF 296. 

\bibitem{CDF99-6} Abe F, et al (CDF Collaboration).
{\em Phys. Rev. Lett.} 82:3576 (1999) 


\bibitem{Hagelin81} Hagelin J.{\em Nucl. Phys.} B193:123 (1981); 
Voloshin M, et al {\em Sov. J. Nucl. Phys.}
46:112 (1987) 

\bibitem{CDF99-4} Abe F, et al (CDF Collaboration).
{\em Phys. Rev. D} 59:032004 (1999) 

\bibitem{Delphi99}  Abreu P, et al (DELPHI Collaboration).
DELPHI 99-109 CONF 296 (1999) 

\bibitem{L399} Acciarri M, et al (L3 Collaboration).
{\em Phys. Lett.} B438:417 (1999)

\bibitem{Aleph99-1}  Palla F, et al (ALEPH Collaboration).
hep-ex/9905017 (1999)

\bibitem{CDF96} Abe F, et al (CDF Collaboration).
{\em Phys. Rev. Lett.} 77:1945 (1996) 

\bibitem{CDF4672} Abe F, et al (CDF Collaboration).
CDFNOTE 4672 (1998) 


\end{thebibliography}
\end{document}